%% file: main.tex
\documentclass[%
 reprint,
superscriptaddress,
 amsmath,amssymb,
 aps,
]{revtex4-2}
\usepackage{comment}
\usepackage{graphicx}
\usepackage{dcolumn}
\usepackage{bm}
\usepackage[mathlines]{lineno}
\usepackage{siunitx}
\usepackage{ulem}
\usepackage{todonotes}
\usepackage{physics}
\usepackage{float}
\usepackage{booktabs}
\usepackage{afterpage}
\usepackage{xcolor}
\usepackage[version=4]{mhchem} 
\definecolor{lightgray}{gray}{0.6}
\definecolor{medgray}{gray}{0.4}
\definecolor{mRed}{RGB}{230, 0, 50}
\colorlet{newtextColor}{mRed}
\usepackage{blindtext}
\usepackage{hyperref}
\hypersetup{
colorlinks=true,
urlcolor= blue,
citecolor=blue,
linkcolor= blue,
}
\usepackage{etoolbox}

\newcommand{\addcite}[1]{\textsf{\color{red!80!black}[\#]}}

\newcommand{\bea}{\begin{equation} \begin{aligned}}
\newcommand{\eea}{\end{aligned} \end{equation} }

\newif\ifptitle
\newif\ifpnumber
\newcounter{para}

\makeatletter
\def\maketitle{
\@author@finish
\title@column\titleblock@produce
\suppressfloats[t]}
\makeatother

\usepackage{titletoc}
\usepackage{tocloft}
\newcommand{\beginsupplement}{
      \setcounter{table}{0}
      \renewcommand{\thetable}{S\arabic{table}}
      \renewcommand{\theHtable}{S\arabic{table}}
      \setcounter{figure}{0}
      \renewcommand{\thefigure}{S\arabic{figure}}
      \renewcommand{\theHfigure}{S\arabic{figure}}
      \setcounter{section}{0}
      \renewcommand{\thesection}{S\arabic{section}}
      \renewcommand{\theHsection}{S\arabic{section}}
}

\newcommand{\makesititle}{Supplementary Information: \@title}

\begin{document}
\renewcommand{\appendixname}{}

% =============================================================================
\title{Fermiology and the Candidate Chiral Superconductor in Rhombohedral Tetralayer Graphene}
% =============================================================================

\def\StanfordPhys{Department of Physics, Stanford University, Stanford, CA 94305}
\def\SIMES{Stanford Institute for Materials and Energy Sciences, SLAC National Accelerator Laboratory, Menlo Park, CA 94025}
\def\StanfordMatsci{Department of Materials Science and Engineering, Stanford University, Stanford, CA 94305}
\def\PrincetonPhys{Department of Physics, Princeton University, Princeton, New Jersey 08544, USA}
\def\UCSDPhys{Department of Physics, University of California at San Diego, La Jolla, California 92093, USA}
\def\UTDPhys{Department of Physics, University of Texas at Dallas, Richardson, Texas 75080, USA.}

\author{Sandesh~S.~Kalantre}
\affiliation{\SIMES}
\affiliation{\StanfordPhys}

\author{Ben~H.~Alexander}
\affiliation{\SIMES}
\affiliation{\StanfordPhys}

\author{Julian~May-Mann}
\affiliation{\StanfordPhys}

\author{Jonah~Herzog-Arbeitman}
\affiliation{\PrincetonPhys}

\author{Marisa~Hocking}
\affiliation{\SIMES}
\affiliation{\StanfordMatsci}

\author{Qingrui~Cao}
\affiliation{\SIMES}

\author{Kenji~Watanabe}
\affiliation{Research Center for Electronic and Optical Materials, National Institute for Materials Science, 1-1 Namiki, Tsukuba 305-0044, Japan}

\author{Takashi~Taniguchi}
\affiliation{Research Center for Materials Nanoarchitectonics, National Institute for Materials Science, 1-1 Namiki, Tsukuba 305-0044, Japan}

\author{David~Goldhaber-Gordon}
\affiliation{\SIMES}
\affiliation{\StanfordPhys}

\author{Andrew~J.~Mannix}
\affiliation{\SIMES}
\affiliation{\StanfordMatsci}

\author{Trithep Devakul}
\affiliation{\StanfordPhys}

\author{Yves~H.~Kwan}
\email{yveshon.kwan@utdallas.edu}
\affiliation{\UTDPhys}

\author{Daniel~E.~Parker}
\email{danielericparker@ucsd.edu}
\affiliation{\UCSDPhys}

\author{Aaron~Sharpe}
\email{aaron.sharpe@stanford.edu}
\affiliation{\SIMES}
\affiliation{\StanfordPhys}

\date{\today}

% =============================================================================
\begin{abstract}
% =============================================================================

Chiral superconductivity, in which the phase of the superconducting order parameter winds in momentum space, has long been sought for its close link to topological superconductivity.
Recent work~\cite{han2025signatures} reported a superconductor in rhombohedral multilayer graphene emerging from a time-reversal symmetry broken normal state, suggesting that it could be a chiral superconductor.
However, the possibility of chirality depends on the symmetry and structure of the normal-state Fermi surface, which have not been directly measured.
Here we measure quantum oscillations in rhombohedral tetralayer graphene over a broad range of the phase diagram, including the superconducting region.
At densities well above the onset of superconductivity, we reproduce previously-reported oscillations~\cite{han2025signatures} consistent with a spin- and valley-polarized quarter metal with a single simply-connected Fermi pocket.
As the carrier density is reduced, we find a transition to a complex ``multitone'' state that persists through the superconducting region. 
This state's spectrum of quantum oscillations is incompatible with a simply-connected quarter metal. 
The next-simplest candidate normal states suggested by our microscopic modeling (fully-polarized annular, nematic, and three-pocket states) are inconsistent with our measurements, albeit difficult to rule out entirely.
The normal state is thus seen to be richer than previously envisaged, reshaping the search for the superconducting mechanism and the possible chirality of the pairing channel.
\end{abstract}

\maketitle

% =============================================================================
\section{Introduction}\label{sec:intro}
% =============================================================================

Most known superconductors arise from the pairing of electrons in a time-reversal invariant normal state. 
Among the few exceptions to this pattern are the uranium ferromagnetic superconductors \ce{UGe2}~\cite{saxenaSuperconductivityBorderItinerantelectron2000} and \ce{URhGe}~\cite{aokiCoexistenceSuperconductivityFerromagnetism2001}, where superconductivity develops from an itinerant ferromagnet.
Recent work~\cite{han2025signatures} on rhombohedral tetralayer and pentalayer graphene reports superconductivity emerging from a normal state whose anomalous Hall response marks it as an itinerant \textit{orbital} ferromagnet.
Time-reversal symmetry breaking is further evidenced by magnetic hysteresis within the superconducting state, and by the superconductor's robustness to an out-of-plane magnetic field~\cite{han2025signatures}.
Quantum oscillation measurements at densities somewhat near the superconducting region found a spin- and valley-polarized (SPVP) circular quarter metal (CQM), i.e. a non-degenerate metal whose Fermi pocket has the topology of a disk.
It was conjectured that the SPVP CQM phase extends outside of the region where quantum oscillations were directly measured, and is the superconductor's normal state~\cite{han2025signatures}. 

If the SPVP CQM were indeed the normal state, the superconductor would be expected to be chiral~\cite{wangChiralSuperconductivityParent2024, chouIntravalleySpinpolarizedSuperconductivity2025, parra-martinezBandRenormalizationQuarter2025, yoonQuarterMetalSuperconductivityRhombohedral2026,yoon2026majoranacrystal, parra-martinezBandRenormalizationQuarter2025,geier2025chiral,wangChiralSuperconductivityParent2024,christos2025finite,yang2025topological,kim2025topologicalchiral,yoonQuarterMetalSuperconductivityRhombohedral2026,yoon2026majoranacrystal,chen2025intrinsic,patri2025family,qin2026chiral,may2026pairing,jahin2026enhanced,gaggioli2025vortex,chouIntravalleySpinpolarizedSuperconductivity2025,han2025exactmodelschiralflatband,li2025berry}
(that is, the phase of its gap function winds non-trivially in momentum space~\cite{kallin2016chiral}).
Moreover, chiral pairing on a single Fermi surface can realize topological superconductivity~\footnote{A superconductor is defined as topological when its negative energy Bogoliubov-de Gennes quasiparticle bands have non-zero total Chern number~\cite{satoTopologicalSuperconductorsReview2017}.} with gapless Majorana edge modes, and Majorana zero modes bound to vortex cores~\cite{kallin2016chiral, satoTopologicalSuperconductorsReview2017}. 
However, both the chirality and potential topology of the superconductor are ultimately determined by how the electronic pairing occurs, which in turn is highly sensitive to currently unresolved normal state properties --- chiefly the Fermi surface structure. 
Testing this hypothesis of a SPVP CQM normal state requires a direct measurement in the normal state of the superconductor.

% -----------------------------------------------------------------------------
\begin{figure*}[t]
    \centering
    \includegraphics{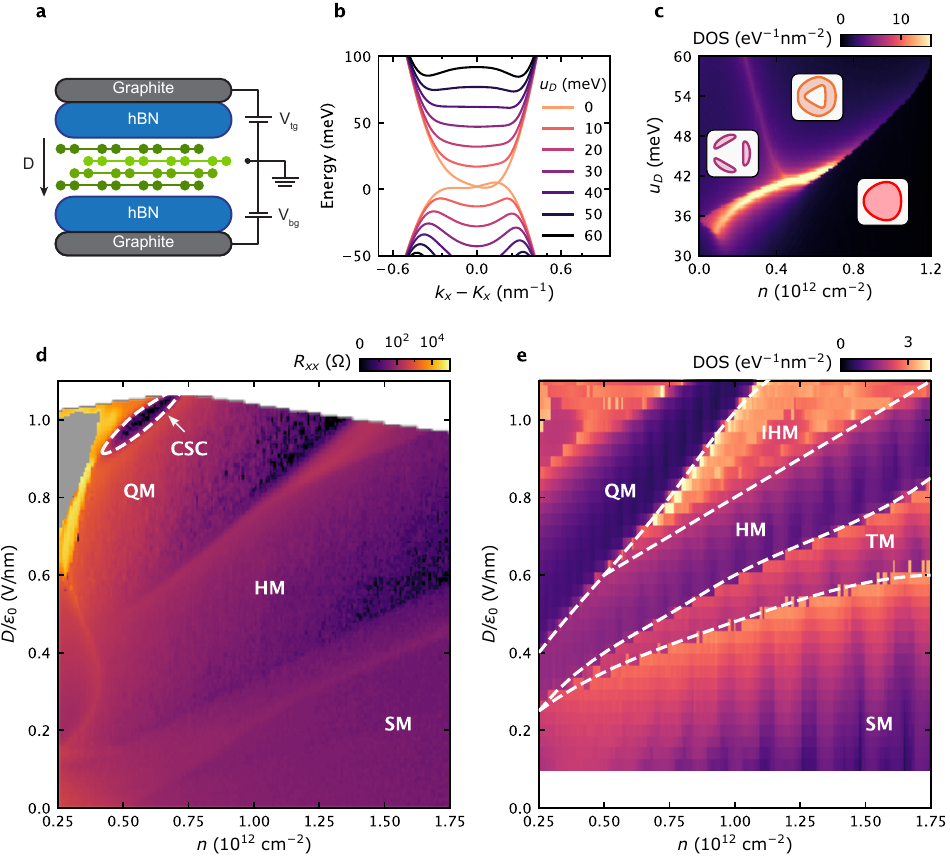}
    \caption{\textbf{Symmetry-broken phases in R4G}:
    (a)~Schematic of the dual-graphite-gated R4G device.
    Top and bottom graphite gates, held at potentials $V_{\mathrm{tg}}$ and $V_{\mathrm{bg}}$, independently control the carrier density $n$ and the displacement field $D$.
    (b)~Low-energy band structure of R4G at representative interlayer potentials $u_D$, showing the opening of a band gap and the evolution into a Mexican hat dispersion with the band minimum at finite momentum, relative to $K$.
    (c)~Single-particle density of states as a function of carrier density $n$ and interlayer potential $u_D$. Sharp features trace vHSs, which in the single-particle picture demarcate Lifshitz boundaries between simply-connected,
    annular, and three-pocket Fermi pocket topologies (insets).
    (d)~Longitudinal resistance $R_{xx}$ as a function of $n$ and $D$ at zero magnetic field and $T = \SI{28}{mK}$. A resistive feature tracks the vHS arc predicted in panel~(c); additional high-resistance features at higher $n$ demarcate boundaries between symmetry-broken phases. 
    The candidate chiral superconducting region (CSC, dashed outline) and representative flavor-polarized phases are annotated: quarter metal (QM), half-metal (HM), and full/symmetric metal (SM). 
    The gray region denotes where the two-terminal resistance of the device is sufficiently large compared to other impedances in the measurement setup that 10\% or more of the expected bias current does not reach the drain.
    (e)~Interacting density of states from self-consistent Hartree-Fock calculations, capturing the principal flavor-polarized phase boundaries observed in transport. TM denotes a triple metal with three equally occupied flavors. IHM denotes an imbalanced half-metal.
    Oscillations seen as a function of $n$ are finite-size artifacts in the calculations.
   }
    \label{fig1}
\end{figure*}
% -----------------------------------------------------------------------------

In this work, we investigate the candidate chiral superconductor (CSC) in rhombohedral tetralayer graphene (R4G).
We resolve oscillations of the critical current of the superconductor with applied magnetic field, consistent with phase-coherent transport.
We then turn to the metallic state surrounding and underlying the candidate CSC, systematically mapping Shubnikov–de Haas (SdH) oscillations across the density/displacement-field phase diagram.
In two dimensions, Onsager quantization relates the frequency of SdH oscillations to the enclosed $k$-space area and degeneracy of the Fermi surfaces~\cite{leeb_field_2025}.
At densities well above the superconducting transition, we observe a single SdH oscillation frequency, consistent with the previously-reported CQM state~\cite{han2025signatures}.

Upon reducing the carrier density, we observe a previously-unreported discontinuous change in the quantum-oscillation spectrum at or above the boundary of the zero-field superconducting region.
Below this transition, we observe striking quantum oscillations with multiple frequencies. 
The most prominent oscillation frequencies (``tones'') depend only weakly on the nominal gate-defined carrier density, and exceed the frequency expected for a singly-degenerate Fermi surface.
This multitone pattern persists continuously across the density and displacement field range of the superconducting dome, provided that superconductivity is suppressed by either $T>T_c$ or $B>B_c$. 
We therefore identify the multitone state as the normal state from which the superconductor emerges. 
The measured quantum oscillations of the multitone state are incompatible with a CQM.
They are also inconsistent with the next-simplest candidates: metals that are still spin- and valley-polarized but have more complex nematic, three-pocket, or annular Fermi pockets.
The unexpected complexity of this normal state raises fundamental questions about the mechanism of pairing and the possibility of a chiral superconducting order parameter.

% -----------------------------------------------------------------------------
\begin{figure*}[t]
    \centering
    \includegraphics{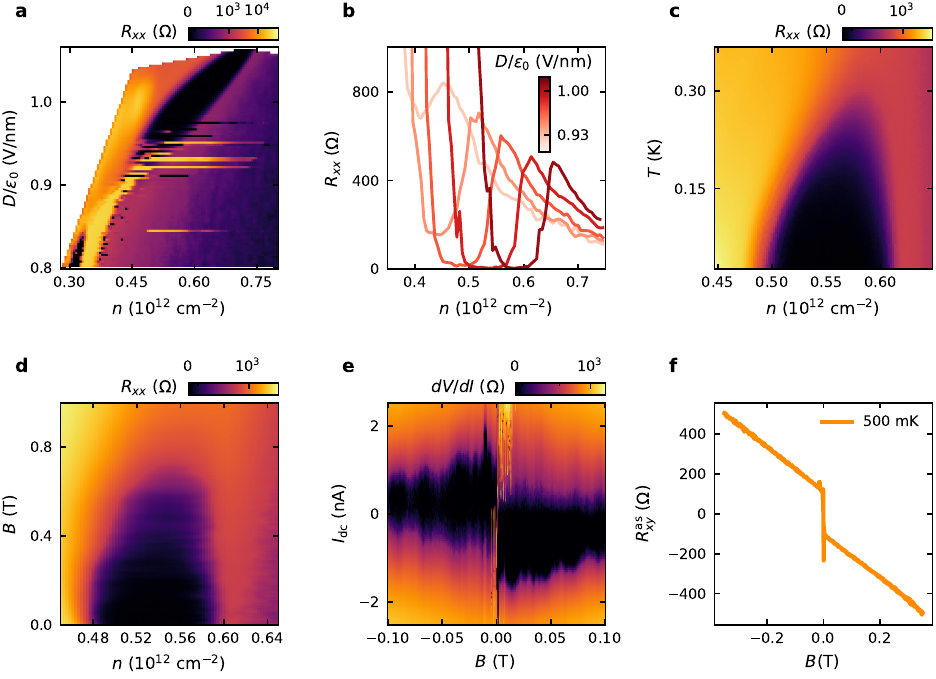}
    \caption{\textbf{Transport characterization of the superconducting state}:
    (a)~Longitudinal resistance $R_{xx}$ as a function of carrier density $n$ and displacement field $D$ in the vicinity of the candidate chiral superconducting (CSC) region at $T = \SI{28}{mK}$ and zero magnetic field. The dark region of vanishing resistance traces the superconducting dome, emerging near $D/\epsilon_0 = \SI{0.9}{V/nm}$ and extending to $D/\epsilon_0 \approx \SI{1.05}{V/nm}$.
    (b)~Line cuts of $R_{xx}$ vs.\ $n$ with $D/\epsilon_0$ between $\SI{0.9}{V/nm}$ and $\SI{1.05}{V/nm}$ (color scale). 
    (c)~$R_{xx}$ as a function of $n$ and temperature $T$ at $D/\epsilon_0 = \SI{1}{V/nm}$, showing the superconducting dome with optimal doping near $n =  \SI{0.6e12}{\per\centi\meter\squared}$ and a critical temperature of $T_c \approx \SI{175}{mK}$.
    (d)~$R_{xx}$ as a function of $n$ and perpendicular magnetic field $B$ at $D/\epsilon_0 = \SI{1}{V/nm}$ and $T = \SI{28}{mK}$. The superconducting region is suppressed by $B \approx \SI{0.4}{T}$. The high-density boundary remains pinned near $n =  \SI{0.65e12}{\per\centi\meter\squared}$, while the low-density boundary shifts to higher density with increasing field.
    (e)~Differential resistance $dV/dI$ as a function of d.c.\ bias current $I_{\mathrm{dc}}$ and perpendicular magnetic field $B$ at $D/\epsilon_0 = \SI{1}{V/nm}$ and optimal doping. Periodic modulations of the critical current with $B$ are consistent with phase-coherent transport, with an extracted period corresponding to a length scale of \SI{260}{nm}.
    (f) Magnetic hysteresis of anti-symmetrized Hall resistance $R_{xy}^{\mathrm{as}}$ as a function of $B$ above $T_c$, indicating time-reversal symmetry breaking in the normal state of the CSC state.}
    \label{fig2}
\end{figure*}
% -----------------------------------------------------------------------------

% =============================================================================
\section{Flavor-Polarized Phases}\label{sec:flavors}
% =============================================================================

Neglecting trigonal warping, the low-energy electronic bands of R4G at zero interlayer potential difference $u_D$ disperse approximately as $k^4$ in the $K$ and $K'$ valleys~\cite{minChiralDecompositionElectronic2008, koshinoTrigonalWarpingBerry2009}. 
Upon increasing $u_D$ (which is experimentally set by an applied displacement field $D$), a band gap opens, and the conduction band minimum evolves from $U$-shaped to Mexican-hat-shaped (Fig.~\ref{fig1}b).
Trigonal warping further splits the Mexican hat's annulus of minima into three pockets at the lowest densities, creating van Hove singularities (vHS).
As shown in Fig.~\ref{fig1}c, the single-particle density of states as a function of $u_D$ and the electron density $n$ exhibits two arcs of vHSs that meet at a higher-order vHS~\cite{shtyk2017monkey,2019NatCo..10.5769Y} (see Sec.~\ref{secapp:continuum_model} for additional discussion). 
For the $u_D$ where these vHSs occur, the band is nearly flat out to a density of $n \approx \SI{0.6e12}{\centi\meter^{-2}}$~\cite{bernevig2025berrytrashcanmodelinteracting}; strong electronic correlations are accordingly expected below this density.

% -----------------------------------------------------------------------------
\begin{figure*}[th]
    \centering
    \includegraphics[width=\textwidth]{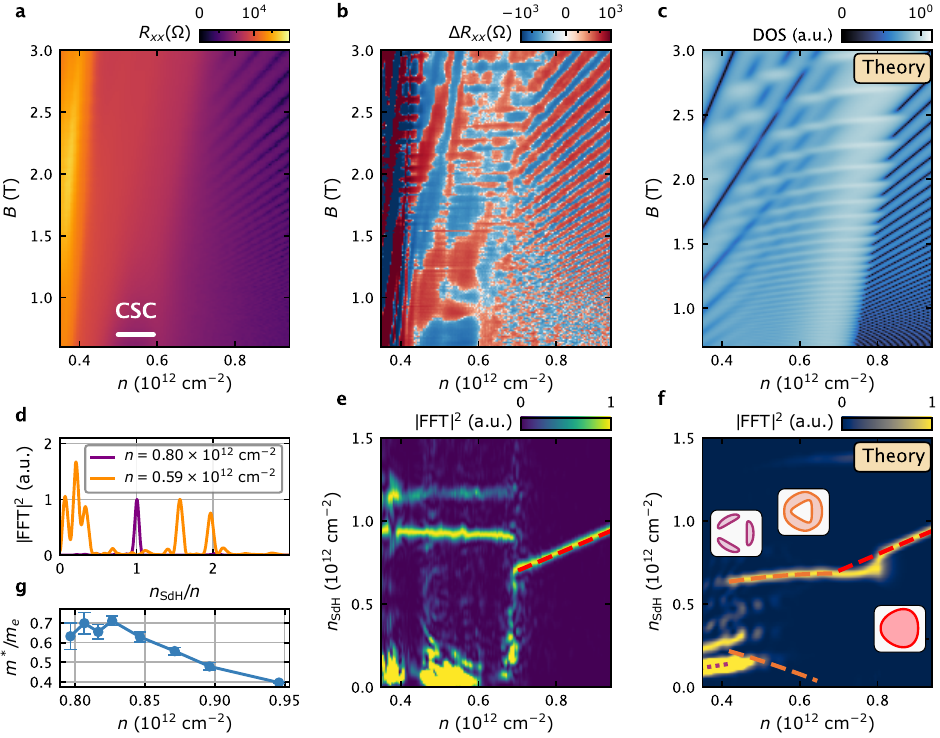}
    \caption{
    \textbf{Quantum oscillations near the superconducting state}:
    (a)~Longitudinal resistance $R_{xx}$ as a function of carrier density $n$ and perpendicular magnetic field $B$ at $D/\epsilon_0 = \SI{1}{V/nm}$ and $T = \SI{28}{mK}$.
    The white line segment indicates the density extent of the superconducting dome at zero field.
    (b)~Oscillatory component $\Delta R_{xx}$ after subtracting a smooth density. 
    Two distinct regions are visible: a Landau fan emanating from the circular quarter metal at higher density, and a region at lower density where oscillations disperse weakly with $n$.
    (c)~Single-particle Landau level calculation at $u_D = \SI{43}{meV}$, reproducing the weak dispersion of LLs with density at low $n$ through level repulsion in the Mexican-hat band.
    (d)~FFT of $\Delta R_{xx}$ in $1/B$ at $n=\SI{0.80e12}{\per\centi\meter\squared}$ (circular quarter metal) and $n=\SI{0.59e12}{\per\centi\meter\squared}$ (multitone state), plotted as a function of normalized effective density $n_{\rm SdH}/n$, Eq.~\eqref{eq:onsager}. 
    A single tone at $n_{\rm SdH}/n=1$ corresponds to a non-degenerate simply-connected Fermi pocket.
    (e)~FFT of $\Delta R_{xx}$ as a function of $n$, plotted as a function of effective density $n_{\rm SdH}$ such that $n_{\mathrm{SdH}} = n$ corresponds to a single simply-connected Fermi pocket (red dashed line). 
    Above $n \approx \SI{0.67e12}{\per\centi\meter\squared}$, a single tone at $n_{\mathrm{SdH}} = n$ indicates a simply-connected quarter metal.
    At lower densities, the spectrum splits into multiple tones, with two nearly-flat tones with $ n_\text{SdH}> n$, signaling a Lifshitz transition. The superconducting region lies entirely within this multitone region.
    (f)~FFT of the simulated quantum oscillations from panel~(c), showing a single tone at $n_{\mathrm{SdH}} = n$ at higher density that splits into multiple tones below the Lifshitz transition.
    Dashed lines indicate SdH frequencies expected from Onsager quantization, color-coded by the zero-field Fermi surface topology (insets). 
    The theoretical FFT spectrum below the transition does not capture the experimental features in (e).
    (g)~ Effective mass as a function of density $n$ extracted from Lifshitz-Kosevich theory in the CQM region at $D/\epsilon_0 =\SI{1}{V/nm}$. Error bars represent standard deviations from fitting.
    }
    \label{fig3}
\end{figure*}
% -----------------------------------------------------------------------------

We experimentally investigate the charge density $n$ and displacement field $D$ phase diagram through magneto-transport in a dual-graphite-gated R4G device (Fig.~\ref{fig1}d and Sec.~\ref{supp_sec:device}).
The core of the device is a $\SI{1.5}{\mu m}$ wide, $\SI{11}{\mu m}$ long rectangle. 
Four contacts along each side are spaced by a center-to-center distance of $\SI{1.5}{\mu m}$ (Fig.~\ref{fig-ed-device}).
Unless explicitly noted, the measured behavior is qualitatively consistent between different contact pairs.
The phase diagram exhibits broad resistive features indicative of a sequence of Stoner-like ferromagnetic transitions among the four isospin (spin and valley) flavors, driven by the high density of states (Fig.~\ref{fig1}d). 
Under applied magnetic field, the quantum oscillations change across these resistive features (Fig.~\ref{fig-ed-qo-field}), identifying transitions from metals with four isospin flavors (dubbed ``symmetric metals'' or SM) to two isospin flavors (``half metals'' or HM) to one isospin flavor (``quarter metal'' or QM).
HM and QM are broad classes that subsume multiple different isospin-polarized and intervalley-coherent states (see Sec.~\ref{supp_sec:SCHF} for the zoology of such states). 
The primary sequence of observed phases and transitions is qualitatively captured by self-consistent Hartree-Fock calculations (SCHF) shown in Fig.~\ref{fig1}e using a standard R4G continuum model~\cite{auerbachIsospinMagneticTexture2025} and screened Coulomb interactions.

% =============================================================================
\section{Superconductivity}\label{sec:superconductivity}
% =============================================================================

We observe superconductivity over a range of $n$-$D$ space similar to that of the superconducting state denoted SC1 in Ref.~\cite{han2025signatures} (Fig.~\ref{fig2}a) and broadly recover the same salient features. 
The gate range accessible within this device is insufficient to reach other superconducting states denoted SC2 and SC3 \cite{han2025signatures}. 
At $D/\epsilon_0=\SI{1}{V/nm}$ and optimal doping, the critical temperature is $T_c = \SI{175}{mK}$, extracted from the temperature at which $R_{xx}$ reaches $5\%$ of the normal state resistance (Fig.~\ref{fig2}c).
Away from optimal doping, $T_c$ falls off, forming a weakly asymmetric dome.
Linecuts of $R_{xx}$ as a function of density at $T = \SI{28}{mK}$ confirm that the resistance of the CSC drops to zero to within experimental noise (Fig.~\ref{fig2}b).

Similar to Ref.~\cite{han2025signatures}, we observe streaks in $R_{xx}$ in both the superconducting region and the surrounding area of $n$-$D$ space (Fig.~\ref{fig2}a). 
Because these streaks occur along the $n$-axis (the fast axis of this 2D map), and can be biased or even completely suppressed by a small magnetic field (Fig.~\ref{fig-ed-switching} and Sec.~\ref{secapp:switching_training}), we attribute them to time-reversal-symmetry-broken domains with slow dynamics. 
A finite anomalous Hall response is observed above $T_c$ in the superconducting region and in the adjacent non-superconducting region, further substantiating time-reversal symmetry breaking in the normal state (Fig.~\ref{fig2}f). 

Since many phenomena can cause sharp reductions in measured $R_{xx}$, it is useful to have other indications that an apparently zero-resistance state is a superconductor. 
A distinctive sign of superconductivity is roughly periodic oscillations in critical current with magnetic field applied normal to the plane of the sample.
We observe oscillations (Fig.~\ref{fig2}e) with a most prominent period of $\approx\SI{30}{mT}$ and additional higher frequency components (Fig.~\ref{fig-fran_fft}).
Assuming a charge-$2e$ condensate, this corresponds to an area of $\SI{7e-2}{\micro \meter\squared}$, for example a square $\SI{260}{\nano\meter}$ on a side.
This length scale is several times smaller than the channel width $\SI{1.5}{\micro\meter}$, perhaps reflecting disorder-induced spatial variations in supercurrent flow.
Indeed, the pattern of oscillations varies between contact pairs (Fig.~\ref{fig-ed-csc-fraun}).
Superconductivity remains stable up to an upper critical magnetic field $\approx \SI{0.4}{T}$ (Fig.~\ref{fig2}d), comparable to the previous report~\cite{han2025signatures}.
On theoretical grounds, a large upper critical field is expected for a chiral superconductor, as the orbital magnetic moment~\cite{zhu2026microscopicoriginorbitalmagnetization} of a chiral superconductor can gain energy in a magnetic field, compensating for the loss of condensation energy. 

We observe additional superconductivity with $T_c~<~\SI{200}{\milli\kelvin}$ from hole doping, similar to observations in rhombohedral trilayer graphene~\cite{zhouSuperconductivityRhombohedralTrilayer2021a}; the study that reported the CSC in R4G did not examine the hole side of the phase diagram.
The consistent behavior between contact pairs in this superconductor (Figs.~\ref{fig-ed-holesc-frauns} and~\ref{fig-si-hole-sc}) suggests the variations between contact pairs in the CSC (Fig.~\ref{fig-ed-csc-fraun}) are likely not due to contact quality but instead due to enhanced sensitivity to mesoscopic disorder.

% =============================================================================
\section{Quantum Oscillations}\label{sec:sdh}
% =============================================================================

To investigate the origin of the CSC, we now turn to the metallic states in the same region of the $n$-$D$ phase diagram, measuring their Shubnikov-de Haas (SdH) oscillations as a function of $1/B$.
The frequencies of such oscillations are a sensitive probe of the Fermiology, with any sudden change in the spectrum as a function of $n$ or $D$ reflecting either a Lifshitz transition or a phase transition. 
The semiclassical Onsager relation~\cite{shoenberg1984magnetic} relates the SdH oscillation frequency $f_{\rm SdH}$ to the enclosed $k$-space area $A_k$ of each Fermi surface orbit, which can be expressed as an effective charge density:
\begin{equation}
    n_{\mathrm{SdH}} \equiv \frac{ef_{\mathrm{SdH}}}{h} = \frac{A_k}{(2\pi)^2},
    \label{eq:onsager}
\end{equation}
where $e$ is the elementary charge and $h$ is Planck's constant.
In these units, a non-degenerate simply-connected Fermi pocket exhibits $n_{\mathrm{SdH}}(n) = n$.
A state with $S$ Fermi surfaces has $S$ effective densities $n_{\mathrm{SdH}}^j$.
Accounting for their degeneracies $g_j$ and carrier signs $\theta_j$, these satisfy 
\begin{equation}
\label{eq:multi_pocket_orbit_sum_rule}
    \sum_{j=1}^S \theta_j g_j n_{\mathrm{SdH}}^j = n.
\end{equation}
This sum rule follows from Luttinger's theorem and is valid even in the presence of interactions~\cite{luttinger1960ground, oshikawa2000topo}, provided one only sums over the elementary frequencies (i.e., not harmonics or breakdown orbits).
Therefore, Onsager quantization analysis of quantum oscillations can provide a distinctive fingerprint of many Fermi surface topologies~\cite{alexandradinataGeometricPhaseOrbital2017, alexandradinataRevealingTopologyFermiSurface2018}.
However, SdH can exhibit frequencies not captured by semiclassical Onsager analysis.
This most commonly happens when electrons tunnel between Fermi surfaces separated on the scale of the inverse magnetic length, forming magnetic breakdown orbits~\cite{cohen_magnetic_1961, alexandradinataGeometricPhaseOrbital2017, alexandradinataSemiclassicalTheoryLandau2018,leeb_field_2025}.

Fig.~\ref{fig3}a shows the magnetoresistance across the superconducting dome and into the surrounding regions at $D/\epsilon_0 = \SI{1}{\volt/nm}$.
For densities above \SI{0.72e12}{\per\centi\meter\squared}, we observe clear oscillations in $R_{xx}$ as a function of $B$.
Upon decreasing density, the average value of $R_{xx}$ abruptly rises, and the oscillations fade.
The density $n_c$ at the transition ranges from \SI{0.72e12}{\per\centi\meter\squared} at \SI{3}{T} down to \SI{0.6e12}{\per\centi\meter\squared} in the zero field limit (Fig.~\ref{fig3}a).
Previous work reported that SdH oscillations were unobservable at densities below the transition, due to a large effective mass~\cite{han2025signatures}.
However, we find that background subtraction reveals clear quantum oscillations over the full displayed range of density (Fig.~\ref{fig3}b). 
The resulting features are insensitive to the details of the background subtraction, and are qualitatively consistent across all measured contact pairs (see Sec.~\ref{supp_sec:fft}).

To extract the tones of these quantum oscillations, we compute the Fourier transform at each density of the background-subtracted resistance $\Delta R_{xx}$ in $1/B$ (Fig.~\ref{fig3}e). 
For densities above $n_c$, we observe a single tone $n_{\mathrm{SdH}} \approx n$, implying a non-degenerate simply-connected Fermi pocket, likely the previously-reported CQM.
Measuring the temperature and magnetic field dependence of the SdH oscillations, we find that the effective mass of the CQM increases from $m^* \approx 0.39 m_e$ to $m^* \approx 0.7m_e$ as the density is reduced toward $n_c$ (Fig.~\ref{fig3}g).

Below $n_c$, the single-tone CQM abruptly vanishes and is replaced by a ``multitone'' state (Fig.~\ref{fig3}e). 
The multitone pattern persists over the entire density range where the CSC is observed (\SIrange{0.47e12}{0.58e12}{\per\centi\meter\squared}), and beyond to \SI{0.35e12}{\per\centi\meter\squared}.

The clearest features of the multitone state are two tones above $n_{\mathrm{SdH}} = n$ (above the red dashed line in Fig.~\ref{fig3}e), which are stable to at least $\SI{670}{mK}$, significantly above $T_c$ (Fig.~\ref{fig:temp-two-tone-both}). 
Both tones are approximately \textit{constant} as a function of density: neither tone shifts by more than \SI{0.09e12}{\per\centi\meter\squared} over the displayed range of densities (Fig.~\ref{fig:two-tone-diff}). 
Their frequencies increase slightly with increasing $D$, but the difference $\approx \SI{0.2e12}{\per\centi\meter\squared}$ between the tones is nearly independent of $D$ (Fig.~\ref{fig:two-tone-diff}).
These tones do not appear to be simple harmonics of any lower-frequency tones.
So within a semiclassical picture the orbit corresponding to each tone would enclose an area in momentum space larger than that of a single fully-polarized Fermi surface (a CQM).

At low effective densities $n_{\mathrm{SdH}} < n$ (Fig.~\ref{fig3}e), we also observe substantial spectral weight at low frequencies. 
This finding, which is robust to different methods of background subtraction, is inconsistent with a non-degenerate simply-connected Fermi pocket.
It suggests instead the presence of small Fermi pockets or breakdown orbits.
The amplitudes of the low tones decay rapidly with temperature, vanishing by $\SI{280}{mK}$.

\begin{figure*}[th]
    \centering
    \includegraphics{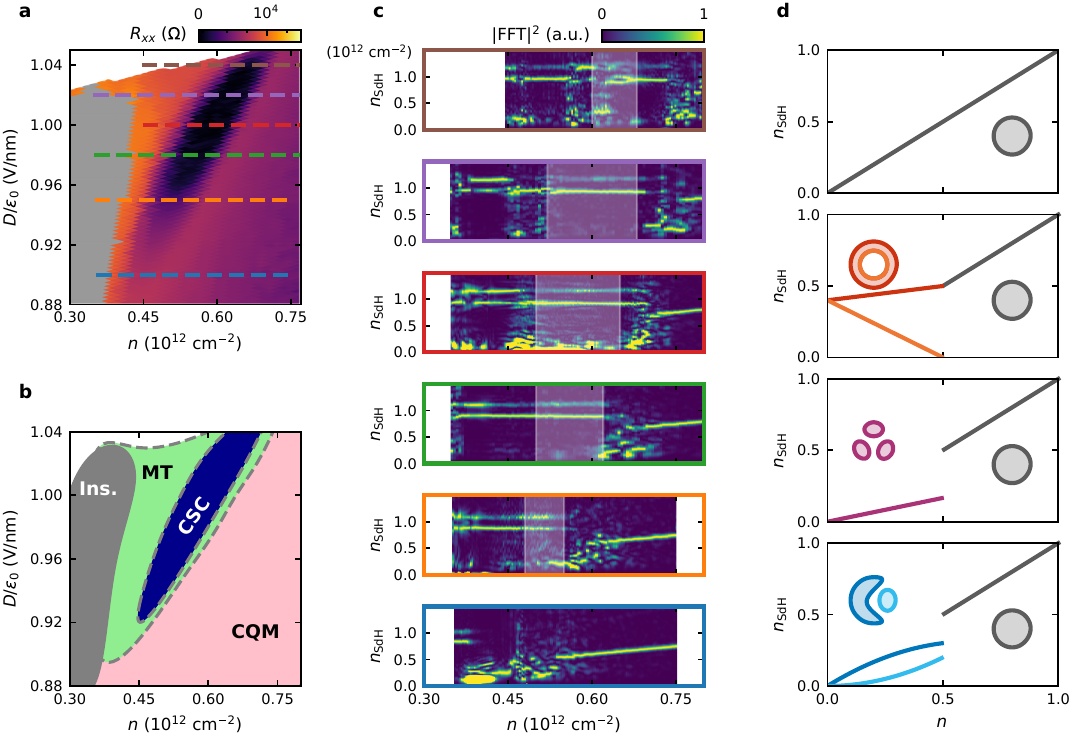}
    \caption{\textbf{Phase diagram near the candidate chiral superconducting state}:
    (a) $R_{xx}$ vs $n$ and $D$ near the CSC state. 
    The gray region denotes where the two-terminal resistance of the device is sufficiently large compared to other impedances in the measurement setup that 10\% or more of the expected bias current does not reach the drain.
    The horizontal dashed lines show the density ranges used for the FFT analysis in (c).
    (b) Schematic phase diagram inferred from the FFT analysis, showing the superconducting region (SC, blue) embedded within a multitone state (MT, green) which is separated from the simply-connected circular quarter metal (CQM, pink) by a transition. 
    (c) FFTs of the oscillatory component $\Delta R_{xx}$ as a function of $n$ at the corresponding $D$ values (color-matched to the dashed lines in (a)), with frequency normalized to $n_\text{SdH}$. 
    Across all values of displacement field, at higher density the spectrum exhibits a single tone at $n_\text{SdH} = n$ that is replaced at lower density by two tones with $n_\text{SdH} > n$. 
    In each panel, the shaded region denotes the extent of the superconducting state for the corresponding $D$. 
    (d) Schematics of the expected quantum oscillation spectra from Onsager quantization of candidate states. From top to bottom, the state at low density is: circular, annular, three-pocket, nematic ``boomerang and bean''. All are assumed to be spin- and valley-polarized.
    }
    \label{fig4}
\end{figure*}

To map the phase boundary between the CQM and the multitone state as a function of $D$, we measure quantum oscillations over the linecuts indicated in Fig.~\ref{fig4}a.
For each sampled value of $D$ we observe a transition from a single tone associated with a CQM to the multitone state (Fig.~\ref{fig4}c).
This transition consistently occurs at a density at or above the onset of superconductivity, and the high tones then depend only weakly on $n$. 
As depicted in the schematic phase diagram of Fig.~\ref{fig4}b (Sec.~\ref{supp_sec:schematic}), our results suggest that CSC emerges from the multitone state, rather than from the CQM.

% -----------------------------------------------------------------------------
% -----------------------------------------------------------------------------

% =============================================================================
\section{Fermiology}\label{sec:Fermiology}
% =============================================================================

We now investigate theoretical models for the SdH oscillations. 
To understand the multitone state, we consider a series of candidate phases ranging from semiclassical phenomenology to microscopic self-consistent Hartree-Fock. 

First we consider a simple one-band model of a trigonally-warped Mexican-hat dispersion that captures the minimal Fermi surface topology of SPVP rhombohedral graphene:
\begin{equation}
\label{eq:1band_model}
    h(\bm{k}) = a_2 |\bm{k}|^2 + a_3 (3 k_xk_y^2 - k_x^3) + a_4 |\bm{k}|^4.
\end{equation}
Depending on the values of the parameters, this model can display a simply-connected Fermi pocket, an annular Fermi pocket, or $C_{3z}$-symmetric three- and four-pocket states. 
The experimental fingerprint of quantum oscillations with two high tones whose frequencies are nearly constant in density is not captured by Onsager-quantized orbits (Eq.~\eqref{eq:onsager}) in any regime of this model (Fig.~\ref{fig-onsager-model}.) 

At $n = \SI{0.5e12}{\per\centi\meter\squared}$ a circular Fermi pocket has $k_F \approx \SI{0.25}{nm^{-1}}$, while annular or three pocket topologies involve even smaller scales. Magnetic breakdown then becomes relevant when the inverse magnetic length reaches $\SI{0.05}{nm^{-1}}$ at $B=\SI{1.5}{\tesla}$. Since breakdown orbits could contribute to the observed oscillations, we next examine the Landau level spectrum from a non-interacting microscopic model of R4G at $u_D = \SI{43}{meV}$ (details in Sec.~\ref{subsec:magnetic_ham}).
The microscopic Landau fan is shown in Fig.~\ref{fig3}c, and the corresponding Fourier transform in Fig.~\ref{fig3}f. 
At densities above \SI{0.7e12}{\per\centi\meter\squared}, the model's simply-connected Fermi pocket produces oscillations with a single tone, matching the experiment closely. 
A Lifshitz transition to an annular Fermi pocket occurs near $n = \SI{0.7e12}{\per\centi\meter\squared}$; the transition density increases slightly with increasing $B$. At densities below the transition, the hole-like Fermi surface generates branches of Landau levels with both negative and positive slopes in the Landau fan.
The intersecting electron-like and hole-like branches combine to create oscillations with weak dispersion in density, giving a single almost-flat tone.
The pattern of frequencies in Fig.~\ref{fig3}f deviates from Eq.~\eqref{eq:onsager}, showing that an Onsager picture is insufficient to describe this regime.

However, regardless of parameter values chosen (Sec.~\ref{sec:additional_LL_calcs}) the Landau level calculation fails to capture two key aspects of the experimental data: (i) the second flat tone, and (ii) the fact that both tones have frequencies \textit{greater} than that of the single tone seen at densities just above the transition.
We have also verified that neither the opposite valley Landau fan nor lower $u_D$ values qualitatively match the experiment (Sec.~\ref{sec:additional_LL_calcs}).
The discrepancies between the Landau level calculation and the observed tones disfavors the hypothesis that the multitone state is a SPVP phase with an annular Fermi pocket.

Given the proximity of the multitone state to a single-particle vHS, interactions could strongly reshape this state.
Symmetry-breaking is the simplest scenario, which we examine using self-consistent Hartree-Fock (details in Sec.~\ref{supp_sec:SCHF}).
We find the SPVP ground state can break $C_3$ rotation symmetry to form a nematic metal featuring a ``horseshoe'' with one pocket or a ``boomerang and bean'' with two Fermi pockets (Fig.~\ref{fig4}d and \ref{HF_single_screened_set1_eps20_FS}).
Neither state produces two nearly-flat tones within Onsager quantization (Fig.~\ref{fig4}d).
As detailed in Sec.~\ref{subsection:HF_results}, we also find that spin-polarized intervalley coherent (SIVC) states are essentially degenerate with SPVP states. The SIVC states have annular Fermi pockets at the relevant densities and displacement fields (Fig.~\ref{IVC_FS_uD52_eps20}).
As discussed in Sec.~\ref{subsection:HF_results}, naively applying Eq.~\eqref{eq:onsager} to SIVC states does not yield two nearly-flat tones. 
That said, predicting SdH oscillations from SIVC states is non-trivial, since such time-reversal-invariant states may be destabilized by magnetic fields on the scale applied for the SdH measurements.

% =============================================================================
\section{Discussion}\label{sec:discussion}
% =============================================================================

In this work we characterize the candidate CSC in rhombohedral tetralayer graphene, and its associated normal state.
Below we summarize our findings on the superconductor, then discuss candidates for the normal state phase, and conclude with implications for the superconducting mechanism and possible pairing channels.

We replicate the key signatures of the candidate CSC: $T_c\approx\SI{175}{mK}$, an out-of-plane critical field $\approx\SI{0.4}{T}$, and an anomalous Hall response above $T_c$~\cite{han2025signatures}.
We also observed previously-unreported oscillations in the critical current as a function of magnetic field, consistent with phase-coherent transport.

Using SdH oscillations, we then characterized the metallic states in and around the CSC in the $n$ and $D$ phase diagram.
At higher densities than the CSC, we observed a single SdH frequency (``tone'') from a circular quarter metal, whose associated effective mass increases nearly twofold as its density decreases towards the CSC.
At or above the onset density of the CSC, the SdH oscillations undergo a transition into a pattern with multiple frequencies that we dub the ``multitone state.''
The multitone state is present throughout the entire $n$ and $D$ region of the CSC at $B > B_c$, for temperatures both above and below $T_c$.
From these findings, we can draw two concrete inferences:
(I) The multitone state is likely the normal state on top of which pairing occurs. 
Its Fermiology therefore sets the allowed pairing channels.
We cannot fully exclude the possibility that the state represented by the multitone spectrum is not the normal state of the CSC, but rather is induced by an out-of-plane magnetic field less than \SI{0.5}{\tesla}, however we see no evidence of such a transition in the data.
(II) The existence of a transition out of the circular quarter metal implies that the multitone state cannot also be a circular quarter metal, but must be more complex.

We will now use features of the multitone SdH spectrum to constrain the viability of candidate normal states. 
Here are the salient features we identify.
Two tones appear at high frequencies $n_{\mathrm{SdH}}>n$ that are nearly constant as a function of density.
The $n_{\mathrm{SdH}}<n/2$ range has high spectral weight with numerous low-frequency tones.
In the intermediate range $n/2 <n_{\mathrm{SdH}}<n$, the spectral weight is low; occasional low-amplitude peaks are consistent with harmonics of lower-frequency tones.
We now analyze normal state candidates in the context of these data.

We begin by examining the viability of spin- and valley-polarized metallic states. 
The presence of two high tones na\"ively suggests an annular Fermi surface with inner and outer radii corresponding to $\SI{0.9e12}{\per\centi\meter\squared}$ and $\SI{1.2e12}{\per\centi\meter\squared}$, which correspond to momenta with high kinetic energy in the band structure, pushing them outside the flat region of the band.
Such a large-radius annulus would incur a severe kinetic penalty at the band structure level (Fig.~\ref{figS:R4G_dispersion_density_orbital_zeeman}) and an even larger exchange penalty in SCHF calculations relative to the ground state candidates we identify on energetic grounds (see Sec.~\ref{secapp:Ann_from_Mag} for extended discussion).
Furthermore, with our assumed spin and valley polarization, applying the Onsager relation to these tones would indicate that the annular Fermi pocket contains a roughly constant density $\Delta n \approx \SI{0.2e12}{\per\centi\meter\squared}$, considerably below the gate-defined density.
We found no way to combine the measured high- and/or low-frequency tones to satisfy the Luttinger sum rule,  Eq.~\eqref{eq:multi_pocket_orbit_sum_rule}. 
However, the missing density could in principle lie in additional Fermi pockets which are not reflected in observed tones, or in localized states as discussed below.

Among the ground state candidates identified through SCHF calculations, only the outer boundary of an annular state could produce a tone relatively insensitive to gate-tuned density in an Onsager picture.
But each such fully-polarized state would produce only a single high tone, not two, and its frequency would be far lower than we observe.
In an Onsager picture, tones from a nematic or three-pocket ground state candidate would not be flat.
Given that the high-frequency tones do not appear to correspond to Fermi surfaces, they may instead arise from breakdown orbits that enclose multiple smaller Fermi pockets such as those in the three-pocket state.
However, calculations of the microscopic R4G Landau level spectrum at a range of displacements with spin and valley polarization imposed (Sec.~\ref{sec:Fermiology} and Sec~\ref{sec:additional_LL_calcs}) do not produce a pair of flat tones, despite including all breakdown effects.
This suggests beyond-Onsager effects do not alter the above conclusions.
In sum, the annular, nematic, and three-pocket spin- and valley-polarized states all fail to explain the multitone pattern, and particularly its two high, gate-voltage-insensitive tones.

The next-simplest possibility is that the multitone state is a time-reversal symmetry breaking metal that is not spin- and valley-polarized.
Intervalley coherent states (IVCs), found in moir\'e graphene~\cite{kwan2021kekule, PhysRevB.108.235128,nuckolls2023quantum, kim2023imaging} and ABC trilayer graphene~\cite{arp2024intervalley, liu2024visualizingincommensurateintervalleycoherent}, are one class of such states.
Our Hartree-Fock calculations show that IVCs are essentially degenerate with the spin- and valley-polarized phases over the relevant region of $n$ and $D$ (Sec.~\ref{subsection:HF_results}).
Though the anomalous Hall effect in the normal state rules out a time-reversal symmetric IVC state, a valley-imbalanced IVC state with finite orbital magnetization could account for the anomalous Hall effect. 
Whether an IVC is compatible with the experimentally-observed SdH pattern warrants further theoretical examination.

The difficulty of accounting for the multitone pattern with any of the above candidate phases leads us to consider alternative explanations.
The linear relation between the gate voltage and itinerant carrier density could fail in this region of $n$ and $D$. For example, charge could enter a reservoir of localized states, which could explain the weak dependence of SdH frequency on gate voltage.
Such a localized reservoir would need a large capacity, as the two high tones persist nearly unchanged over a broad density range of \SI{0.25e12}{\per\centi\meter\squared}, 
while not imparting
strong disorder to the itinerant carriers, given the clean quantum oscillations and sharp phase transitions observed in this region of gate voltage. These requirements would be hard to satisfy through disorder localization.
An intriguing source for a localized charge reservoir which could satisfy the requirements is a metallic Wigner crystal~\cite{han2026evidencemetallicwignercrystal, dong2026crystalscaughtdopingmetallic, feng2026selfdopedcrystalpreemptedbandinversion}, i.e.,~an electron crystal coexisting with itinerant carriers.
If additional carriers tended to enter as part of the Wigner crystal instead of as itinerant carriers, this could produce tones with weak dependence on the total carrier density. 
Further theoretical work is required to determine if the observed multitone pattern, including the puzzlingly-high pair of tones that do not evolve with gate voltage, could be produced by a state that breaks translation symmetry. 

We close with the implications for the nature of the superconductor.
The time-reversal symmetry-breaking in the normal state and the high upper critical field of the superconducting state remain suggestive of a chiral superconductor.
This is corroborated by theoretical studies that have shown dominant $p_x \pm i p_y$ pairing (i.e., the $E$ and $\bar{E}$ representations of $C_{3z}$) starting from a SPVP normal state with various Fermiologies~\cite{parra-martinezBandRenormalizationQuarter2025, christos2025finite, chen2025intrinsic, patri2025family, qin2026chiral, may2026pairing}.
However, due to trigonal warping, non-chiral $f$-wave pairing (i.e., the $A$ representation of $C_{3z}$) cannot be ruled out~\cite{yoonQuarterMetalSuperconductivityRhombohedral2026, chouIntravalleySpinpolarizedSuperconductivity2025,han2025exactmodelschiralflatband}.
Such $f$-wave pairing can be nodal or fully-gapped depending on the Fermi surface(s) of the normal state. Should the normal state be a partially valley-imbalanced IVC, time-reversal symmetry-breaking effects would be weaker and $f$-wave superconductivity could become favorable.
Normal states that break $C_{3z}$ symmetry~\cite{qin2026extremeanisotropymetallicsuperconducting, nguyen2025hierarchysuperconductivitytopologicalcharge}, due to either intrinsic nematicity or extrinsic strain, allow yet other non-chiral pairing channels.

Finally, we address the prospect that the superconductor is topological.
It is theoretically established that chiral pairing of a circular Fermi surface leads to topological superconductivity~\cite{satoTopologicalSuperconductorsReview2017, read2000paired}~\footnote{This holds within the weak coupling picture of chiral superconductivity. For strong pairing, it is possible to have topologically-trivial chiral superconductivity, even when the normal state has a single Fermi surface.}.
The applicability of this statement to the CSC is challenged by our finding that this superconductor does not emerge from the CQM.
Further experimental and theoretical effort will be required to resolve the nature of the multitone normal state, with fundamental implications for pairing in the associated superconductor.

\textit{Note added}: 
While this manuscript was in the final stages of preparation, the authors became aware of Ref.~\cite{dutta2026reconfigurablechiralsuperconductivity}, which reports orbital ferromagnetism in the candidate chiral superconductor of R5G and reconfigurable low-current switching between states of opposite chirality. 
Within the CSC, we observe that our measured pairs are qualitatively consistent at zero d.c. bias but behave differently under finite d.c.~bias, consistent with that picture.
We also note that the same R4G sample analyzed in the present manuscript was examined using nanoSQUID magnetometry in
Ref.~\cite{sheekey2026visualizingorbitalmagnetismelectron}.

% =============================================================================
\begin{acknowledgments}
% =============================================================================

We thank Sayak Bhattacharjee, Gal Shavit, Ben Foutty, Ruoxi Zhang, Junkai Dong, Yifan Li, Tomohiro Soejima, Andrea Young, Ben Feldman, and Steve Kivelson for fruitful discussions.
Fabrication, measurement, and analysis performed by S.S.K., B.H.A., M.H., Q.C., and A.S. was supported by the Department of Energy, Office of Science, Basic Energy Sciences, Materials Sciences and Engineering Division, under Contract DE-AC02-76SF00515. 
M.H. acknowledges partial support from the U.S. Department of Defense through the Graduate Fellowship in STEM Diversity program. 
S.S.K. acknowledges graduate financial support from the Knight-Hennessy fellowship.
K.W. and T.T. acknowledge support from the Japan Society for the Promotion of Science KAKENHI (Grant Numbers 21H05233 and 23H02052) and World Premier International Research Center Initiative, Ministry of Education, Culture, Sports, Science, and Technology, Japan. 
Analysis performed by D.E.P. was supported under NSF CAREER award no. DMR-2542485.
Analysis performed by T.D. was supported by the Air Force Office of Scientific Research under award number FA9550-25-1-0343.
Part of this work was performed at the Stanford Nano Shared Facilities (SNSF), supported by the National Science Foundation under Award No. ECCS-2026822, as well as in the nano@Stanford labs, which are supported by the National Science Foundation as part of the National Nanotechnology Coordinated Infrastructure under Award No.ECCS-1542152. 

\end{acknowledgments}

\section*{Author Contributions}
S.K., M.H., and Q.C. assembled the heterostructure and fabricated the sample.
S.K., Q.C., and A.S. performed low temperature transport measurements. 
S.K. and B.A. analyzed the data with input from all authors. 
J.M.M., J.H-A., Y.K., and D.E.P. completed the theoretical calculations.
T.T. and K.W. provided hBN crystals. 
S.K., J.M.M., Y.K., D.E.P., and A.S. wrote the paper with input from all authors. 
A.S. conceived of and supervised the project.

\section*{Competing interests}
The authors declare no competing interests.

\section*{Data and Code availability}
The data and code from this study are available at the Stanford Digital Repository~\cite{data_repo}.

\bibliography{refs}

% =============================================================================
%\section*{Methods}
% =============================================================================

%\setcounter{figure}{0}
%\renewcommand{\figurename}{}
%\renewcommand{\thefigure}{Extended Data Fig.\arabic{figure}}
%\makeatletter
%\renewcommand{\fnum@figure}{\thefigure}
%\makeatother  

\include{si}

\end{document}

%% file: si.tex
\title{Supplementary Information: Fermiology and the Candidate Chiral Superconductor in Rhombohedral Tetralayer Graphene}
\maketitle

\beginsupplement
\onecolumngrid

\makeatletter
\pretocmd{\section}{\clearpage}{}{}
\makeatother

% =============================================================================

% Where you want the SI TOC:
\startcontents[SI]
\printcontents[SI]{}{1}{\setcounter{tocdepth}{2}}

% =============================================================================
\section{Stack assembly and device fabrication}\label{supp_sec:device}
% =============================================================================

\begin{figure*}[h]
    \centering
    \includegraphics[width=0.9\linewidth]{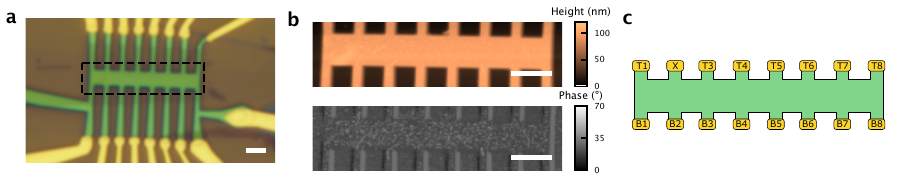}
    \caption{
    \textbf{Device morphology and measurement schematic.}
    (a) Optical micrograph of the fully fabricated dual-graphite-gated R4G Hall bar. The black dashed box indicates the region mapped by AFM.
    (b) AFM topography (top) and phase (bottom) images corresponding to the area highlighted in (a).
    (c) Schematic of the Hall bar enumerating contacts. Pins B1 and B7 serve as the source and drain, respectively, while contact pairs B2-B3, B3–B4, and B4–B5 denote the three measurement configurations extensively investigated in this study. All scale bars are 2\,\si{\micro\meter}.
    }
    \label{fig-ed-device}
\end{figure*}

\begin{figure*}[h]
    \centering
    \includegraphics[width=0.9\linewidth]{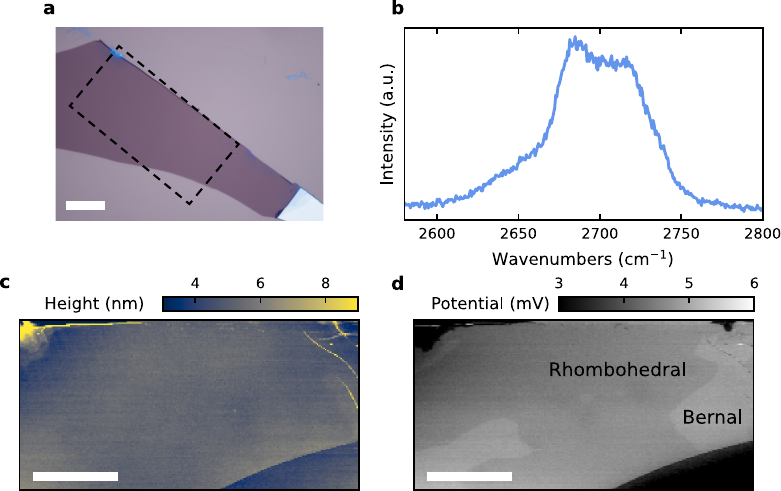}
    \caption{
    \textbf{Identification of rhombohedral stacking order.}
    (a) Optical micrograph of the R4G flake. The black dashed box indicates the region mapped by AM-KPFM. (b) Raman spectra of the 2D peak of the rhombohedral flake after release onto the bottom hBN and bottom graphite gate.
    (c) AFM topography and (d) relative surface potential images corresponding to the area highlighted in (a). All scale bars are 15\,\si{\micro\meter}.
    }
    \label{fig-rhomboid}
\end{figure*}

Rhombohedral regions in tetralayer graphene flakes were identified using Raman spectroscopy (Horiba XPlora, 532\,nm excitation) alongside amplitude-modulated Kelvin probe force microscopy (AM-KPFM) (Bruker IconIR atomic force microscope (AFM)) with MikroMasch NSC18-Pt probes as seen in Fig.~\ref{fig-rhomboid}.
The crystallographic orientation was discerned using atomically-resolved torsional force microscopy (TFM) on a Bruker Icon AFM with ARROW-NCPt probes.
The tips were mounted in a DTRCH-AM torsional probe holder to enable excitation of torsional modes. 
The torsional resonance used for imaging occurs near $\SI{1.6}{MHz}$. 
Further details of the imaging procedure can be found in the supplemental materials of Ref.~\cite{pendharkarTorsionalForceMicroscopy2024}.
The rhombohedral regions were subsequently isolated from Bernal regions via anodic oxidation using a separate ARROW-NCPt probe.

A dual graphite-gated stack was fabricated in three assembly steps using standard dry-transfer techniques: the bottom hBN and bottom graphite, R4G with a top hBN, and top graphite with a carrier hBN.
Large graphite flakes (thicknesses $\lesssim$ 5 nm) were patterned into strips using anodic oxidation, and excess graphite between the strips was removed using sacrificial hBN flakes. 
These were then used as the top and bottom gates.
A standard poly(bisphenol A carbonate)/polydimethylsiloxane (PC/PDMS) stamp was used to pick up an hBN flake followed by one of the graphite strips.
This bottom stack was subsequently released onto a $\SI{285}{nm}$ SiO$_2$/Si substrate (doped Si with dry thermal oxide, Nova wafer) by melting the PC film. 
The film was then dissolved via chloroform.
As a final precaution to ensure the cleanliness of the top surface of the hBN, AFM tip cleaning was performed in Contact Mode on a Bruker IconIR AFM with an NSC18-Pt probe.

Another hBN flake and the R4G flake were picked up in the same way. 
Special care was taken to roll on to the R4G stamp along the zigzag direction to prevent relaxation.
This stack was subsequently aligned and released onto the bottom gate structure by melting the PC film. 
Similarly to before, the majority of the film was dissolved in chloroform before AFM tip cleaning was performed using Contact Mode of a Bruker IconIR AFM with NSC18-Pt probe. The rhombohedral nature of the flake was confirmed using Raman spectroscopy (Horiba XPlora, 532 nm excitation), shown in Fig.~\ref{fig-rhomboid}(b).

Finally, an hBN flake and another patterned graphite strip were picked up using a third PC/PDMS stamp. This stack was aligned and released on to the other portion of the completed heterostructure. The film was dissolved using chloroform, but no subsequent AFM cleaning was performed.

Tapping AFM was used to identify a bubble-free region of the completed heterostructure.
We patterned the stack into a Hall bar geometry using selective CHF$_3$/O$_2$ plasma etching in an Oxford Plasma Pro 80 reactive ion etcher.
Ohmic contacts were made by electron-beam deposition of Cr/Au (5\,\si{\nano\meter}/75\,\si{\nano\meter}) using a Kurt J. Lesker LAB18 electron-beam evaporator.
Patterns for defining the device geometry were made using a Raith EBPG 5200+ electron-beam lithography system. 
See Fig.~\ref{fig-ed-device} for optical images of the device and Fig.~\ref{fig1}(a) for a schematic of the gate geometry. The top and bottom hBN thicknesses were 17\,\si{\nano\meter} and 26\,\si{\nano\meter} respectively.

The dual gated geometry of the device allows for independent control of the electron density, $n$, and the perpendicular electric displacement field, $D$. 
Within a parallel-plate capacitor model: $n=(\epsilon_\mathrm{BN}\epsilon_0/e$)($V_\mathrm{bg}$/$d_\mathrm{bg} + V_\mathrm{tg}$/$d_\mathrm{tg}$) and $D=-(\epsilon_\mathrm{BN}\epsilon_0/2$)($V_\mathrm{bg}$/$d_\mathrm{bg} - V_\mathrm{tg}$/$d_\mathrm{tg}$).
Here, $V_{bg}$ ($V_{tg}$) is the voltage applied to the bottom (top) gate, $\epsilon_\mathrm{BN}=3$ is the relative dielectric constant of hBN, $\epsilon_0$ is the vacuum permittivity, $e$ is the elementary charge, and $d_\mathrm{bg}$ ($d_\mathrm{tg}$) is the thickness of the bottom (top) hBN.

% =============================================================================
\section{Electron transport}\label{supp_sec:transport}
% =============================================================================

We performed transport measurements in a Leiden Cryogenics CF-900 dilution refrigerator using a custom probe. 
The probe consists of coaxial wiring with custom filtering.
The wires first pass through a cured mixture of epoxy and bronze powder to filter GHz frequencies, then through low-pass RC filters mounted on sapphire plates to filter MHz frequencies.
Samples were mounted using a Kyocera custom 32-contact ceramic leadless chip carrier (drawing PB-44567-Mod with no nickel adhesion layer under gold, to reduce spurious magnetic effects).
A radiation shield around the lower part of the probe protects the sample from high-frequency blackbody radiation coming from the 50\,\si{\milli\kelvin} stage of the cryostat. 
This radiation shield was not used in the presented measurements due to damage to the assembly fixing the shield to the probe.
Subsequent calibrations of electron temperature $T_e$ with mesoscopic metal wires~\cite{pierre2003dephasing}, suggest a higher $T_e \approx 80$--$100\,\si{\milli\kelvin}$ than the $25\,\si{\milli\kelvin}$ base temperature of the cryostat.
Because the quantum oscillation measurements probe the normal state of the superconductor, the higher electron temperature does not affect the conclusions of our measurements.

Stanford Research Systems SR830/SR860 lock-in amplifiers paired with NF Corporation LI-75A or Basel Precision Instruments SP1004 voltage preamplifiers were used to perform four-terminal resistance measurements.
A \SI{1}{\giga\ohm} bias resistor was used to apply an AC bias current of up to \SI{5}{nA} RMS at a frequency of \SI{6.451}{Hz}.
Keithley 2400 source-measure units or Stanford Research Systems DC205 precision voltage sources were used to apply voltages to the gates.
The current through the device was simultaneously measured using a DL Instruments DL-1211 current preamplifier.

% =============================================================================
\section{Schematic phase diagram}\label{supp_sec:schematic}
% =============================================================================

\begin{figure*}[h]
     \centering
    \includegraphics{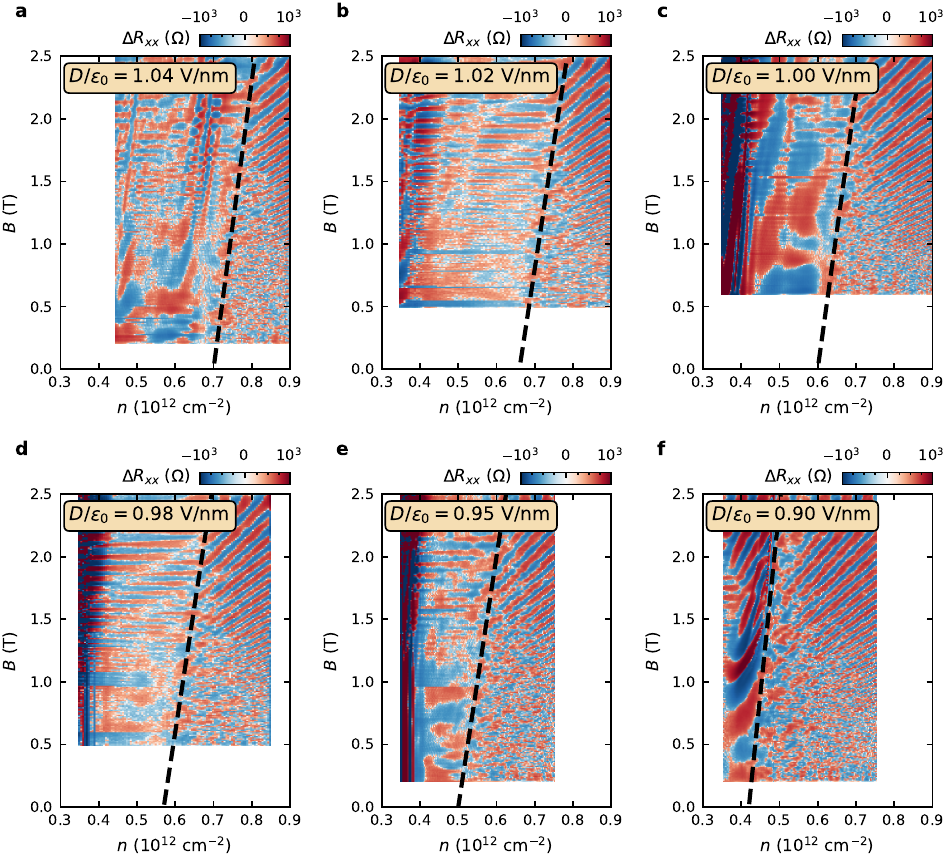}
    \caption{\textbf{Extraction of zero field phase boundary between MT and CQM state} (a-f) Background-subtracted longitudinal resistance $\Delta R_{xx}$ 
as a function of carrier density $n$ and perpendicular magnetic field $B$, 
measured at displacement fields $D/\varepsilon_0$ ranging from 0.90 to 
1.04~V/nm. The background subtraction removes the slowly varying 
magnetoresistance, revealing the onset of quantum oscillations 
associated with the CQM state. The dashed black line 
marks the approximate phase boundary of the CQM state, determined by the 
onset of these oscillations. Extrapolating this boundary to $B = 0$ yields 
the zero-field density at which the system transitions from the 
multitone state into the CQM.
    }
    \label{fig-detrend-vs-D}
\end{figure*}

The boundary between the multitone state and the circular quater metal (CQM) is constrained by the zero-field intercept of a linear fit to the interface between their quantum oscillations (Fig.~\ref{fig-detrend-vs-D}). In Fig.~\ref{fig4}b, we determine the MT-CQM boundary at a given displacement field as the average between the densities for the onset of CQM quantum oscillation as determined by two different methods: the zero-field intercept (Fig.~\ref{fig-detrend-vs-D}) and the FFT (Fig.~\ref{fig4}c).
The boundary of insulator is defined as the values of $n$ and $D$ where the device's two-terminal resistance is sufficiently large that 10\% or more of the expected $\SI{1}{nA}$ bias current is not recovered at the drain. 
The boundary of the superconductor is schematically determined by the shape of the superconducting dome. 

% =============================================================================
\section{Streaks and field training}\label{secapp:switching_training}
% =============================================================================

\begin{figure*}[h]
     \centering
    \includegraphics{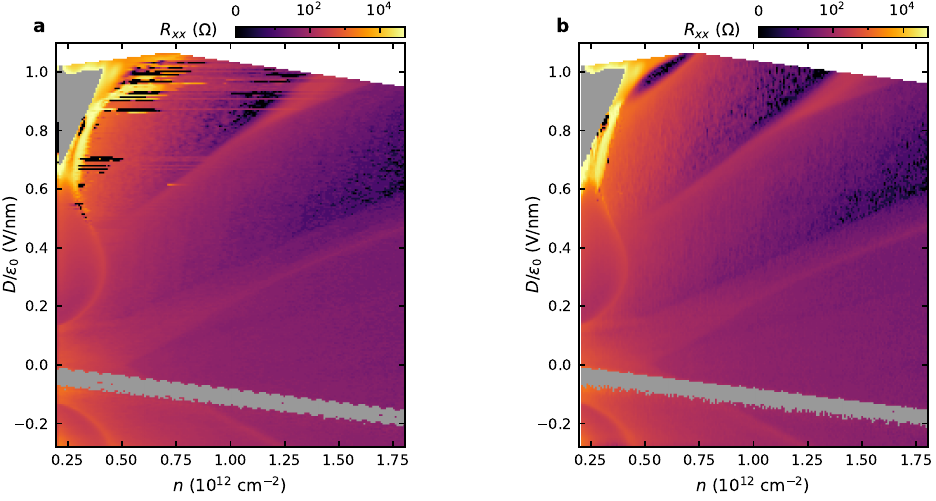}
    \caption{\textbf{Field cycling of metastable states in the vicinity of the CSC region.}
    (a) Longitudinal resistance $R_{xx}$ as a function of $n$ and $D$ at $T = 28$\,mK after a nominal zero-field cooldown to base temperature.  The gray region denotes where the two-terminal resistance of the device is sufficiently large that 10\% or more of the expected 1\,\si{\nano\ampere} bias current is not recovered at the drain.
    The map exhibits pronounced streaking artifacts (horizontal striations and pixel-scale jumps), most visible in the high-$D$ region near the candidate chiral superconducting state and along phase boundaries.
    (b) The same $R_{xx}$ map measured after cycling the perpendicular magnetic field $0 \to +0.5$\,T $\to -0.5$\,T $\to 0$ prior to the $n$-$D$ scan. The streaks are largely suppressed and the underlying phase boundaries become clearly resolved. 
    We attribute the streaks in~(a) to low-energy metastable states near the CSC region and phase boundaries that compete with the true ground states; field cycling in~(b) appears to relax the system towards the ground-state configuration.}
    \label{fig-ed-switching}
\end{figure*}

$R_{xx}$ exhibits streaks in the region of $n$–$D$ space surrounding the candidate superconducting state that depend on the sweep direction in gate space (Fig.~\ref{fig2}a).
A small magnetic field of $\mathcal{O}(\si{mT})$ can bias them.
After quenching the superconducting magnet to reduce the amount of trapped flux, the device exhibits dramatically more streaking (Fig.~\ref{fig-ed-switching}a).
Subsequently training our $\SI{14}{T}$ superconducting magnet by cycling to finite field and returning to zero results in trapped magnetic flux typically $\mathcal{O}(\si{mT})$.
This trapped flux is enough to entirely suppress these streaks (Fig.~\ref{fig-ed-switching}b). Note that after field cycling, we take measurements at nominal zero field as set by the magnet controller.
For both Fig.~\ref{fig2}a and Fig.~\ref{fig-ed-switching}a, $n$ is swept at fixed $D$; for Fig.~\ref{fig-ed-switching}b, $D$ is swept at fixed $n$.

% =============================================================================
\section{Comparison of the phase diagram, quantum oscillations and Fraunhofer oscillations across contact pairs}\label{secapp:contact-comp}
% =============================================================================

\begin{figure*}[h]
    \centering
    \includegraphics{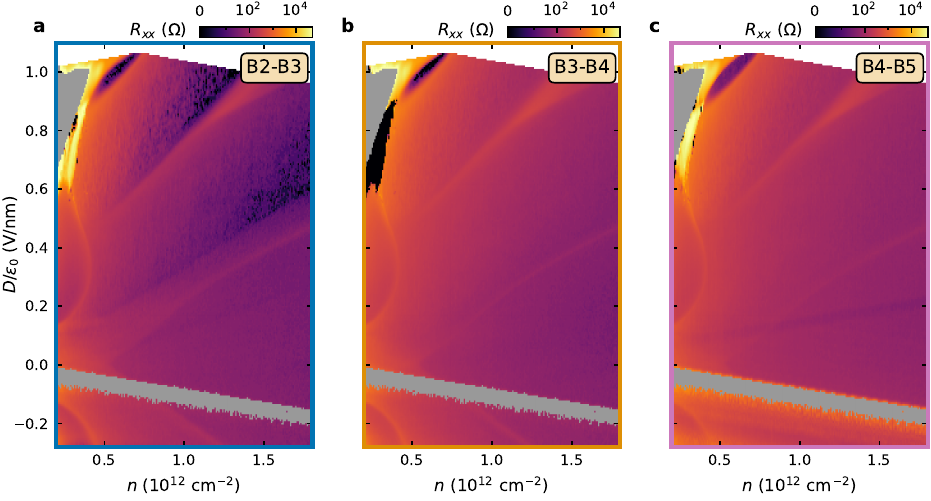}
    \caption{\textbf{Reproducibility of the $n$-$D$ phase diagram across contact pairs.}
    Longitudinal resistance $R_{xx}$ as a function of carrier density $n$ and displacement field $D$ at $T = 28$\,mK and $B = 0$\,T, measured using three different voltage probe pairs: (a) B2-B3, (b) B3-B4, and (c) B4-B5. The same source and drain contacts (pins B1 and B7) are used in all three measurements.  The gray region denotes where the two terminal resistance of the device is sufficiently large that 10\% or more of the expected 1\,\si{\nano\ampere} bias current is not recovered at the drain. The principal features of the phase diagram, including the candidate chiral superconducting region, the arc tracking the van Hove singularity, and the phase boundaries corresponding to flavor transitions, are reproduced across all three contact pairs. This confirms that these features reflect intrinsic properties of the device rather than contact-specific artifacts.}
    \label{fig-ed3}
\end{figure*}

\begin{figure*}[h]
    \centering
    \includegraphics{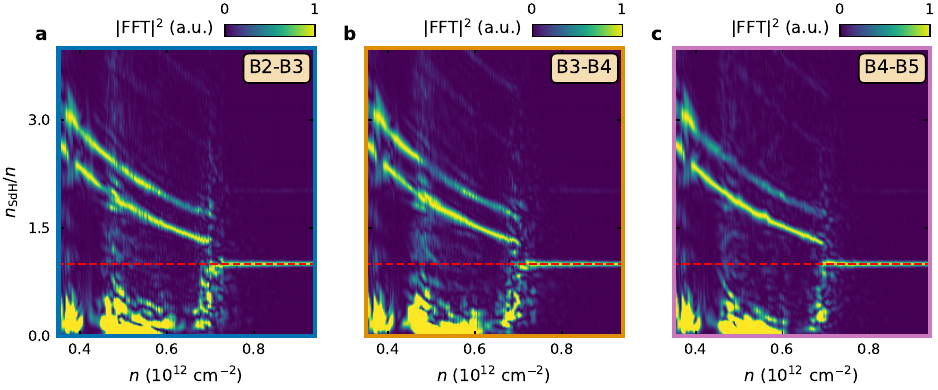}
    \caption{\textbf{Normalized quantum oscillation frequency as a function of $n$ across contact pairs.}
    (a,b,c) Quantum oscillation frequency normalized to density at $D/\epsilon_0 = 1\,\si{V/nm}$ extracted from three different voltage probe pairs: (a) B2-B3, (b) B3-B4, and (c) B4-B5. The same source and drain contacts (pins B1 and B7) are used in all three measurements. At densities above $n = 0.7 \times 10^{12} \si{\centi\meter^{-2}}$, a quarter metal state with $n_{\mathrm{SdH}}/n = 1 $ (red line) is seen which abruptly transitions into the multitone state at lower densities.}
    \label{fig-ed3_normalized_oscillation}
\end{figure*}

\begin{figure*}
    \centering 
    \includegraphics[width=\textwidth]{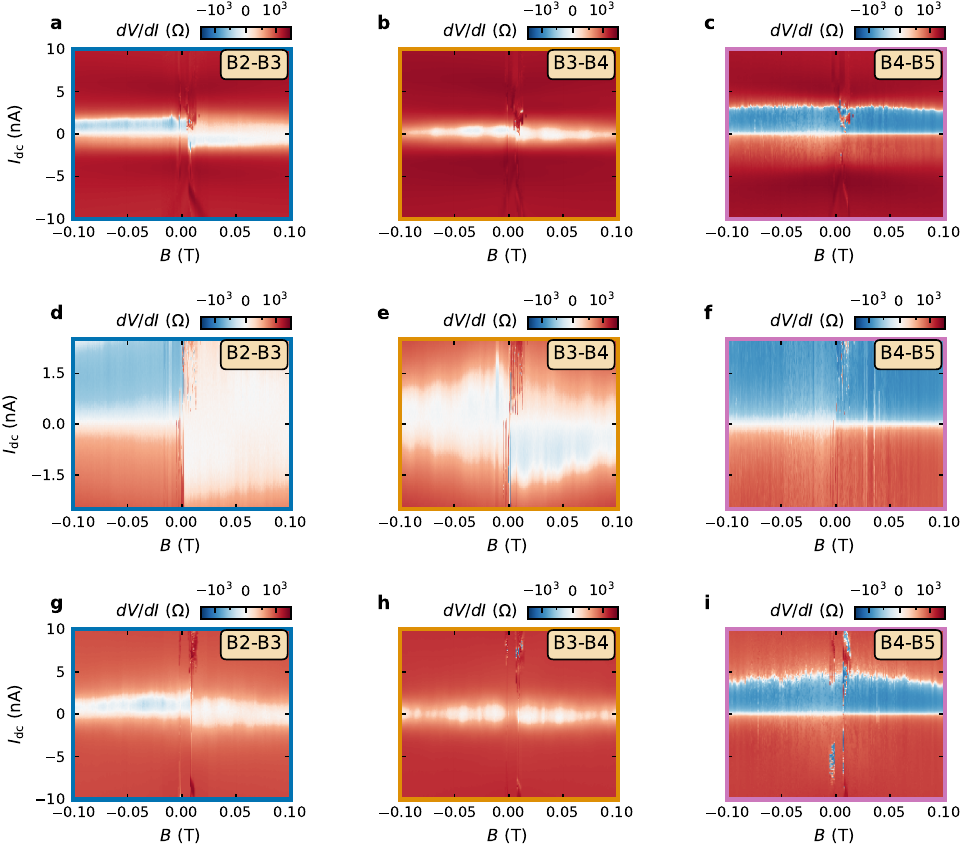}
    \caption{
    \textbf{Varying response to d.c. bias between contact pairs in the CSC.}
    Differential resistance $dV/dI$ as a function of d.c.\ bias current $I_{\mathrm{dc}}$ and perpendicular magnetic field $B$ at $T = \SI{28}{\milli\kelvin}$, measured at three densities within the superconducting dome and at $D/\epsilon_0  = \SI{1.01}{V/nm}$. Each row corresponds to a fixed $(n, D/\epsilon_0 )$, and the three columns show measurements taken simultaneously across the three voltage probe pairs.
    (a-c) $(n, D/\epsilon_0 ) = (\SI{0.58e12}{\per\centi\meter\squared},\, \SI{1.01}{V/nm})$.
    (d-f) $(n, D/\epsilon_0 ) = (\SI{0.60e12}{\per\centi\meter\squared},\, \SI{1.01}{V/nm})$.
    (g-i) $(n, D/\epsilon_0 ) = (\SI{0.64e12}{\per\centi\meter\squared},\, \SI{1.01}{V/nm})$.
    The shape and extent of the zero-resistance region differ substantially across the three contact pairs at each density, with some pairs showing a well-defined supercurrent window and others showing strongly suppressed or absent superconductivity at small bias currents. This contact-pair dependence indicates that the superconducting state is spatially inhomogeneous on the scale of the device channel. We attribute this inhomogeneity to competition between the CSC phase with other competing phases and not to the spatial disorder in the channel.
    }
   \label{fig-ed-csc-fraun}
\end{figure*}

\begin{figure*}
   \centering
   \includegraphics[width=0.95\textwidth]{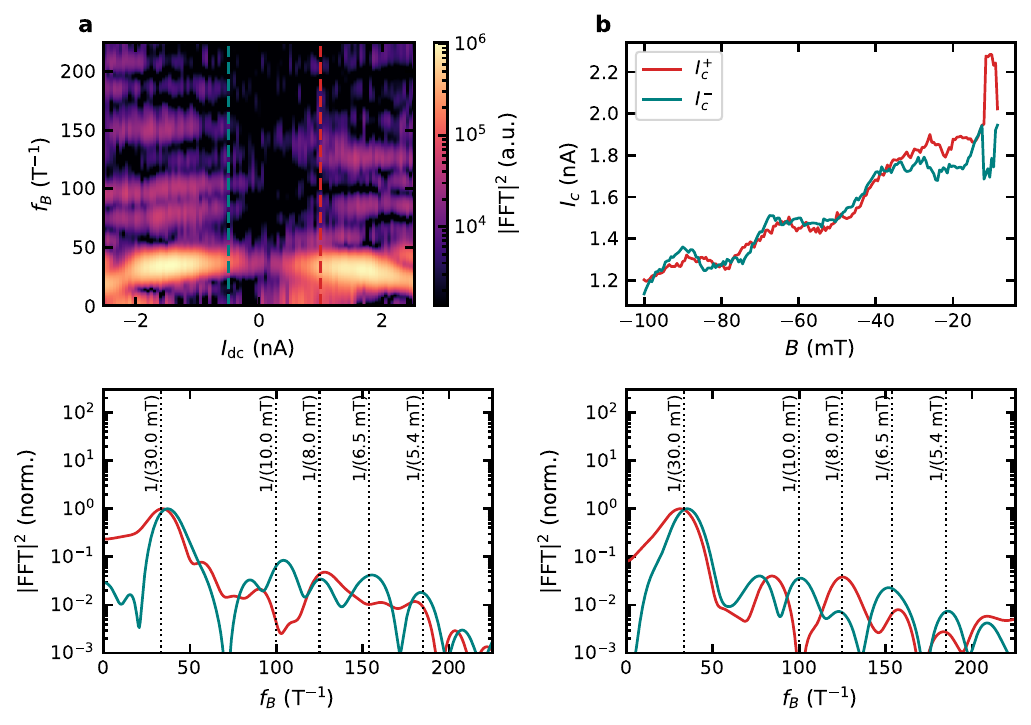}
   \caption{\textbf{Critical-current oscillations in the CSC.} All data are from Fig.~\ref{fig-ed-csc-fraun}e (equivalently Fig.~\ref{fig2}e) 
   (a) Power spectrum $|\mathrm{FFT}|^2$ of the longitudinal resistance $R_{xx}(B)$ versus d.c. bias $I_{dc}$ and frequency $f_B$. For each $I_{dc}$, $R_{xx}(B)$ over the window $B \in [-100, -8]$ mT is high-pass filtered to remove the slow magnetoresistance background, Hann-windowed, and Fourier transformed. Dashed lines mark the bias values of the line cuts in (c). 
   (b) Positive- and negative-bias critical currents $I_c^{+}$, $I_c^{-}$ extracted from $R_{xx}$ using a threshold of $200\ \Omega$, referenced to the supercurrent center $I_0 = 0.18$ nA. 
   (c) Line cuts of (a) at $I_{dc} = 1.0$ and $-0.50$ nA. 
   (d) Power spectrum of $I_c^{\pm}(B)$ from (b) on the same $f_B$ axis as (c).}
   \label{fig-fran_fft}
\end{figure*}

\begin{figure*}
   \centering
   \includegraphics[width=0.95\textwidth]{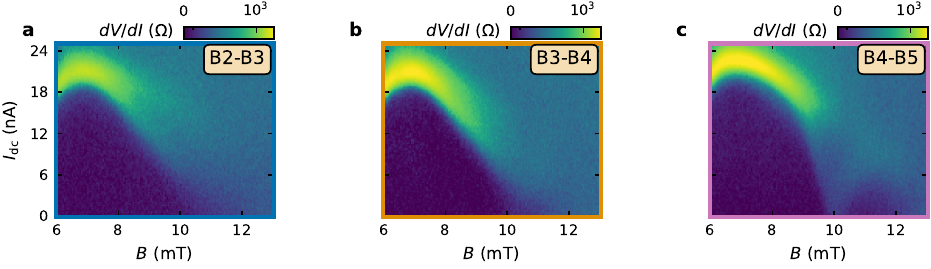}
   \caption{\textbf{Fraunhofer interference in the hole-side superconducting state across contact pairs.}
    Differential resistance $dV/dI$ as a function of d.c.\ bias current $I_{\mathrm{dc}}$ and perpendicular magnetic field $B$ at $T = \SI{28}{\milli\kelvin}$, measured at the hole-side superconducting state. Current is sourced at T1 and drained at B6. Panels (a-c) show measurements across the three voltage probe pairs: B2-B3, B3-B4 and B4-B5 respectively. All three contact pairs exhibit Fraunhofer-like modulations of the critical current with a similar period and envelope shape, in marked contrast to the strong contact-pair dependence observed in the candidate chiral superconducting state on the electron side. The consistency of the Fraunhofer pattern across contact pairs indicates that the hole-side superconductor is spatially homogeneous on the scale of the device channel.}
   \label{fig-ed-holesc-frauns}
\end{figure*}

In Figs.~\ref{fig-ed3} to \ref{fig-ed-holesc-frauns}, we compare results using three different voltage probe pairs B2-B3, B3-B4, and B4-B5. Fig.~\ref{fig-ed3} demonstrates that the principal features of the $n$-$D$ phase diagram, namely the region of the candidate CSC and the flavor phase transitions, are robust across different contact pairs. Fig.~\ref{fig-ed3_normalized_oscillation} shows that the frequency spectrum of SdH oscillations at $D/\epsilon_0=\SI{1.}{V/nm}$ is consistent across contact pairs in both the multitone state and the circular quarter metal. Fig.~\ref{fig-ed-csc-fraun} reveals that transport as a function of finite d.c. bias current $I_\text{dc}$ in the CSC does show appreciable variation across the contact pairs, demonstrating that the CSC is sensitivity to mesoscopic disorder in the device channel. By contrast, the $I_\text{dc}$ vs $B$ transport maps in Fig.~\ref{fig-ed-holesc-frauns} show that the hole-side superconductor is less sensitive to disorder in the device channel.

We also observe an asymmetric dependence of the magnitude of the critical current on the sign of $I_{\mathrm{dc}}$ (Fig.~\ref{fig-ed-csc-fraun}).
In systems such as rhombohedral graphene with broken $C_{2z}$ symmetry, the Josephson diode effect can give rise to such an asymmetric critical current~\cite{nadeem2023superconducting}. 
However, such asymmetries must be carefully investigated to rule out spurious mechanisms~\cite{chirolliDiodeEffectFraunhofer2025, rashidiSelfFieldInducedJosephsonDiode2025}.
Though all contact pairs behave consistently under $I_{\mathrm{dc}}=0$, their behavior under finite $I_{\mathrm{dc}}$ is different (Fig.~\ref{fig-ed-csc-fraun}).
If $I_{\mathrm{dc}}$ nucleates domains of competing states or drives domain wall motion, the differing behavior between contact pairs and asymmetric $I_{\mathrm{dc}}$ dependence may arise from varied domain structure and pinning in the channel, instead of a Josephson diode effect.

% =============================================================================
\section{Comparison between single particle DOS and experimental $n$-$D$ phase diagram}
% =============================================================================

\begin{figure*}[h]
   \centering 
   \includegraphics[width=\textwidth]{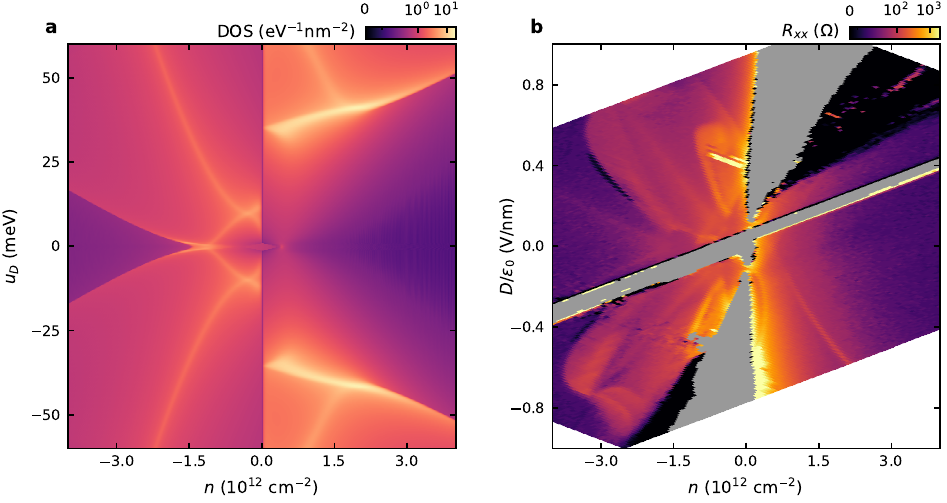}
   \caption{\textbf{Comparison between single particle DOS and $R_{xx}$ over the $n$-$D$ phase diagram.} (a) Single particle DOS calculated using a continuum model (Set 1 parameters, see Sec.~\ref{secapp:continuum_model}) as a function of density and interlayer potential, $u_D$. (b) $R_{xx}$ measured as a function of density and displacement field with a measurement configuration optimized for hole side measurements. The gray region denotes where the two terminal resistance of the device is sufficiently large that 10\% or more of the expected 1\,\si{\nano\ampere} bias current is not recovered at the drain. Between panels (a) and (b) we see a correlation between the calculated DOS and observed magnitude of $R_{xx}$, especially on the hole side. Good agreement between DOS and $R_{xx}$ validates the magnitudes of parameters used in the model.}
   \label{fig-ed-sdos-exp}
\end{figure*}

In Fig.~\ref{fig-ed-sdos-exp}, we compare the single particle DOS calculated using a continuum model (Sec.~\ref{secapp:continuum_model}) as a function of density $n$ and interlayer potential $u_D$ to the longitudinal resistance $R_{xx}$ mapped over density $n$ and displacement field $D$. While the single particle DOS does not capture Stoner transitions observed in the experiment, we see a correlation between low DOS regions in the phase diagram and the observed low resistance regions in the experiment. 

\section{Quantum oscillations as a function of $n$ and $D$ at fixed magnetic fields}
\begin{figure*}[h]
   \centering 
   \includegraphics[width=\textwidth]{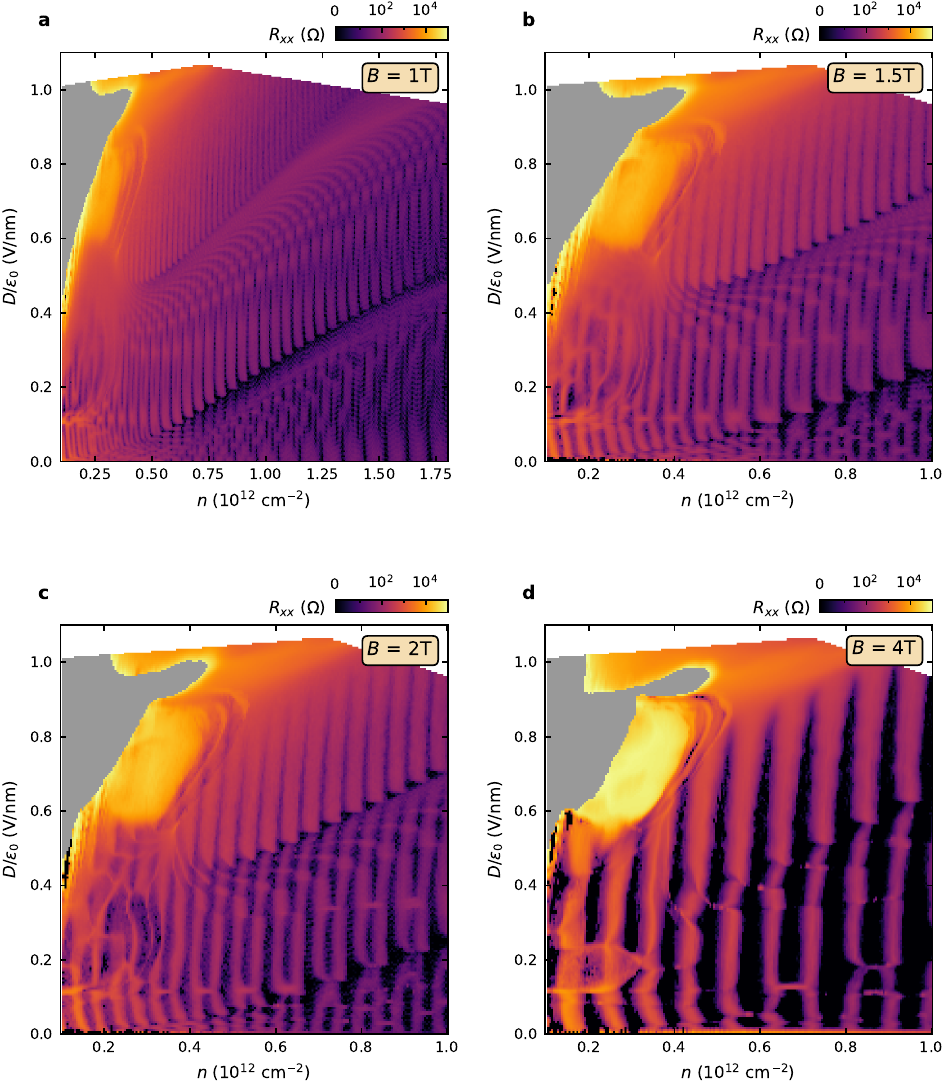}
   \caption{\textbf{$n$-$D$ phase diagram as a function of magnetic field.} Longitudinal resistance $R_{xx}$ as function of density $n$ and displacement field $D$ at a fixed magnetic fields: (a) 1\,\si{\tesla}, (b) 1.5\,\si{\tesla}, (c) 2\,\si{\tesla} and (d) 4\,\si{\tesla}. The gray region denotes where the two terminal resistance of the device is sufficiently large that 10\% or more of the expected 1\,\si{\nano\ampere} bias current is not recovered at the drain.}
   \label{fig-ed-qo-field}
\end{figure*}

Fig.~\ref{fig-ed-qo-field} shows $R_{xx}$ mapped across the $n$-$D$ plane at fixed perpendicular magnetic fields of 1, 1.5, 2, and 4\,T. At low fields (1--1.5\,T), quantum oscillations are visible as rapid oscillatory features across a wide range of densities, reflecting the underlying Fermi surface structure of the correlated parent states. As the field increases to 2 and 4\,T, individual Landau level gaps begin to be resolved and the oscillations evolve into well-separated resistance minima. The highly resistive region (shaded in gray) at large $D$ and low $n$ reflects an insulating regime, while the low-resistance regions at intermediate densities trace the boundaries of the isospin-ordered phases identified in the zero-field phase diagram.

% =============================================================================
\section{Anomalous Hall near the CSC state}\label{sec-ah-hall-map}
% =============================================================================

\begin{figure*}
   \centering 
   \includegraphics[width=0.9\textwidth]{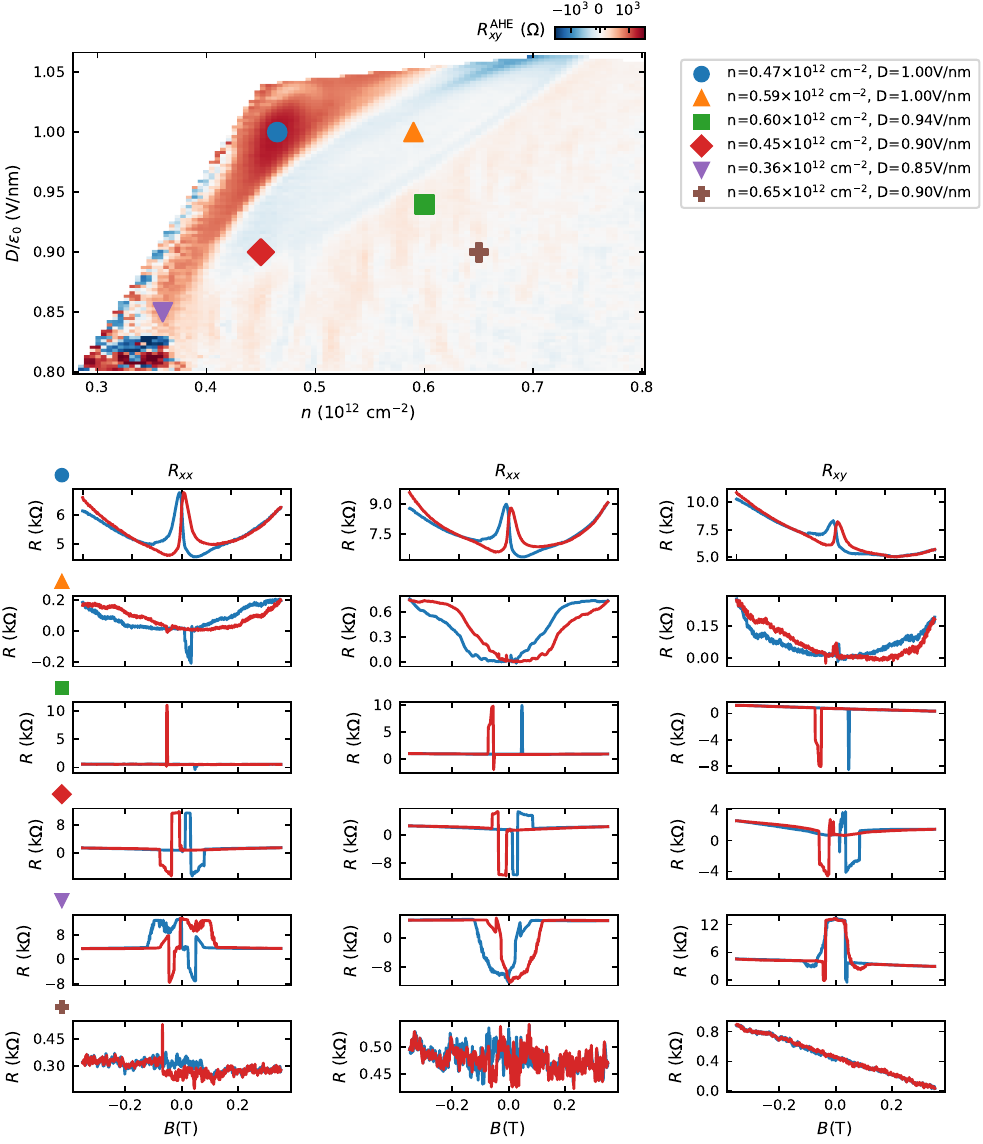}
   \caption{\textbf{Anomalous Hall response near the CSC state at 28\,\si{\milli\kelvin}.} Top panel shows anomalous Hall resistance as a function of density $n$ and displacement field $D$ taken at base temperature of 28\,\si{\milli\kelvin}. Each row below the $n$-$D$ map shows hysteresis loops in two longitudinal resistance pairs denoted by $R_{xx}$, and one non-local Hall pair denoted by $R_{xy}$, at fixed value of $n$ and $D$ as marked by symbols in the top panel. Blue curve denotes field swept from negative to positive values while the red curve denotes field swept from positive to negative values.}
   \label{fig-ed-ah}
\end{figure*}

\begin{figure*}
   \centering 
   \includegraphics[width=0.9\textwidth]{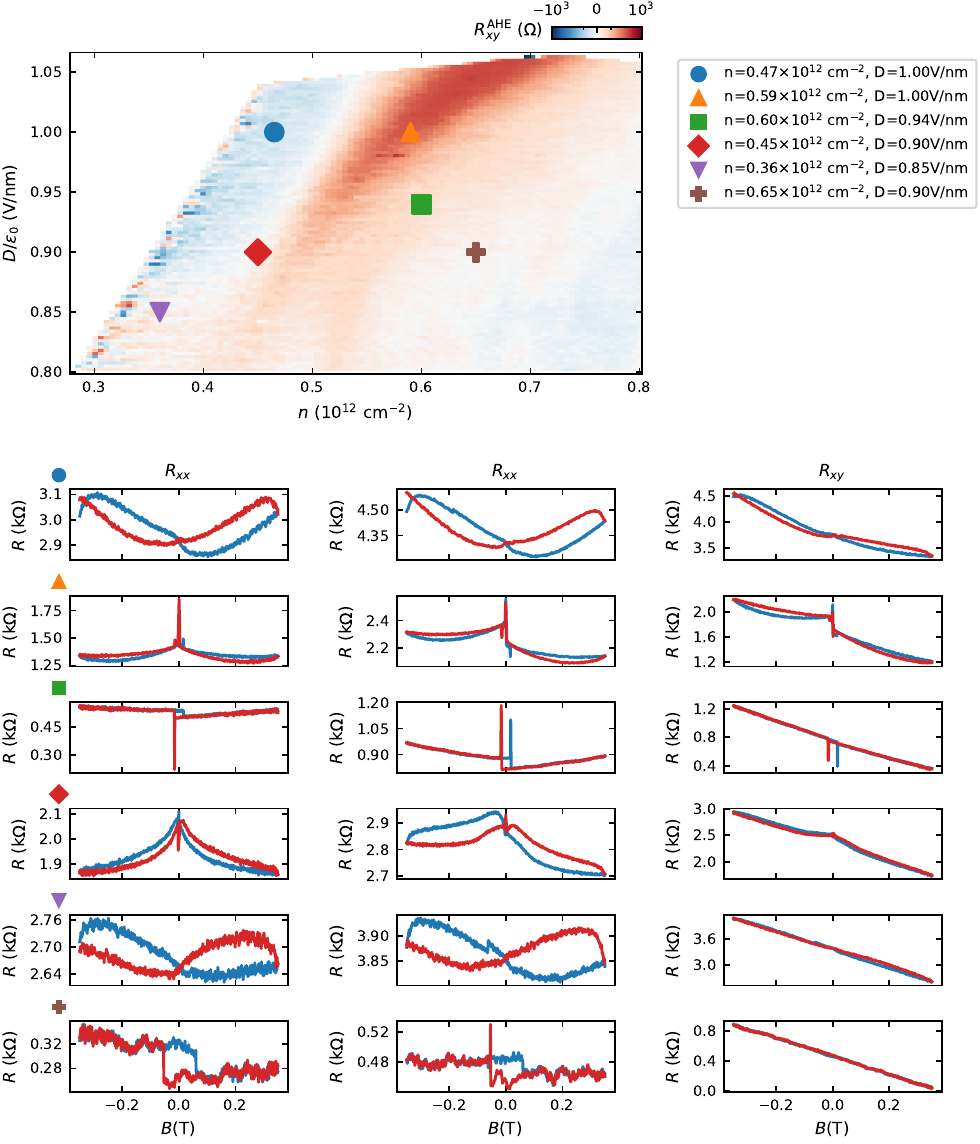}
   \caption{\textbf{Anomalous Hall response near the CSC state at 500\,\si{\milli\kelvin}.} Top panel shows anomalous Hall resistance as a function of density $n$ and displacement field $D$ taken at temperature of 500\,\si{\milli\kelvin}. Each row below the $n$-$D$ map shows hysteresis loops in two longitudinal resistance pairs denoted by $R_{xx}$, and one non-local Hall pair denoted by $R_{xy}$, at fixed value of $n$ and $D$ as marked by symbols in the top panel. Blue curve denotes field swept from negative to positive values while the red curve denotes field swept from positive to negative values.}
   \label{fig-ed-ah-500mK}
\end{figure*}

To map the anomalous Hall response shown in Fig.~\ref{fig-ed-ah} and Fig.~\ref{fig-ed-ah-500mK} and minimize the contribution from both the regular Hall effect and $R_{xx}$ mixing due to anisotropy, we measured $R_{yx}$ at $B=\pm\SI{100}{mT}$ and $\pm\SI{200}{mT}$.
The anomalous Hall response is then calculated according to $R_{yx}^\mathrm{AHE}=(R_{yx}^{\SI{100}{mT}}-R_{yx}^{-\SI{100}{mT}})-(R_{yx}^{\SI{200}{mT}}-R_{yx}^{-\SI{200}{mT}})/2$.
This procedure removes the regular Hall component when its contribution to $R_{yx}$ has a linear dependence on $B$, which is in general true for the small $B$ considered here.

% =============================================================================
\section{High-field Landau fans across the CSC state}
% =============================================================================

\begin{figure*}[h]
   \centering 
   \includegraphics{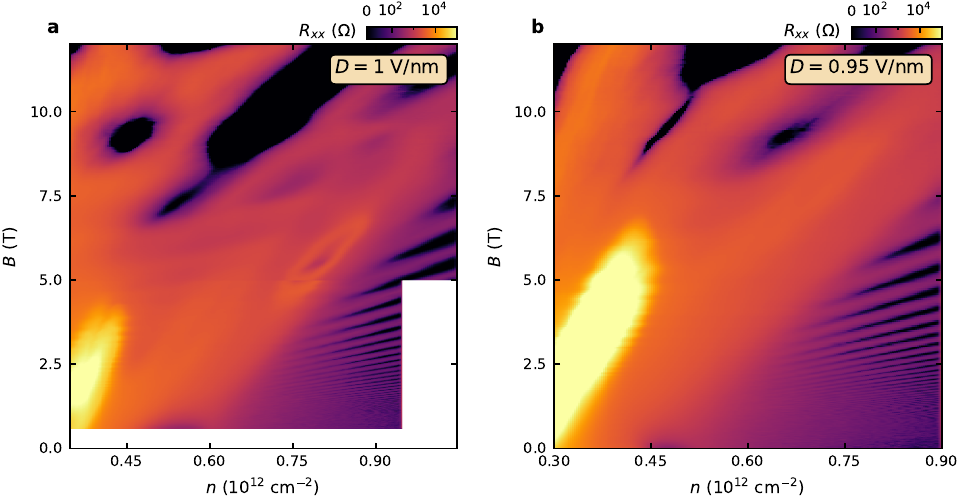}
   \caption{\textbf{High field Landau fans across the CSC.} Landau fans with an extended field range up to $12\,\si{\tesla}$ at displacement fields (a) $1\,\si{V/nm}$ and (b) $0.95\,\si{V/nm}$. The density axis at these displacement fields crosses the CSC region in the $n$-$D$ phase diagram.}
   \label{fig-si-highfield-fan}
\end{figure*}

% =============================================================================
\section{Analysis of quantum oscillations using the Lifshitz-Kosevich theory}
% =============================================================================

The observed quantum oscillations may be fit to the Lifshitz-Kosevich formula \cite{lifshitsTheoryShubnikovHaas1958, shoenberg1984magnetic}. For each tone, we expect
\begin{equation}
    R_{xx} \propto R_{\mathrm{T}} R_{\mathrm{D}} \cos( 2 \pi \qty(\frac{ h n_{\mathrm{SdH} }}{e B} + \gamma)),
\end{equation}
where,
\begin{align}
    R_{\mathrm{T}} = \frac{\alpha T}{B \sinh(\alpha T/B)}, && R_{\mathrm{D}} = \exp(-\alpha T_{\mathrm{D}}/B), &&
    \alpha = \frac{2 \pi^2 k_B }{\hbar e} m^{\ast}, && T_{\mathrm{D}} = \frac{\hbar}{2 \pi k_B \tau_q},
\end{align}
and $\gamma$ is a phase offset. To practically fit to this formula, we first fit the effective mass $m^{\ast}$ using only the temperature dependence of the amplitude of SdH oscillations (see Sec.~\ref{supp_sec:effective-mass}). Compared to the simply-connected CQM, this approach is more complex in the multitone state, limiting the applicability of this straightforward model there. In the CQM, it turns out that $\alpha < \SI{10}{\tesla/\kelvin}$, so $\sinh(\alpha T/B) \approx \alpha T/B$ at base temperature and $R_{\mathrm{T}} \approx 1$ can be neglected during fitting. In the CQM, where the effective mass is straightforward to characterize, this fit leads to a lifetime of $\tau_q = \SI{4}{\pico\second}$.

\begin{figure*}[h]
    \centering
    \includegraphics{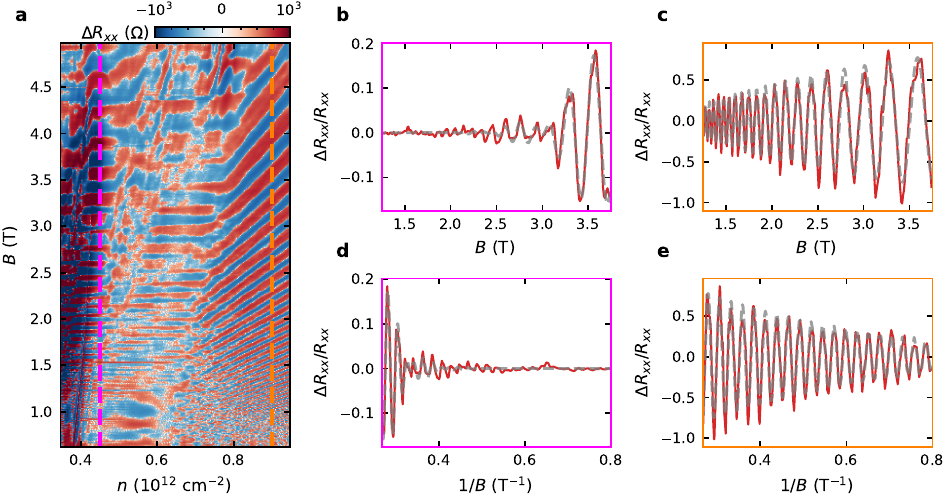}
    \caption{\textbf{Lifshitz-Kosevich fit to quantum oscillations in the multitone and CQM regime.} (a) Background subtracted $R_{xx}$ with linecuts indicated, which are color coded to the borders in panels (b)-(e). Solid red lines in (b-c) and (d-e) show background subtracted $R_{xx}$ plotted against field and inverse field, respectively. Grey dashed line indicates fit to the Lifshitz-Kosevich formula. The extracted frequencies agree with those obtained from the FFT analysis. Note the beating pattern visible in the multitone regime in panel (d), indicating the presence of at least two frequencies. }
    \label{fig:fft-compare-B1Blinecuts}
\end{figure*}

In the following table, we list the extracted parameters for the multitone state at low density (purple line in Fig.~\ref{fig:fft-compare-B1Blinecuts}a) and the CQM at high density (orange line in Fig.~\ref{fig:fft-compare-B1Blinecuts}a).
\begin{table}[h]
    \centering
    \begin{tabular}{cccc}
    \toprule
        Fit & $n_\text{SdH}$ & $\gamma$ & $\alpha T_{\mathrm{D}}$ \\[0.3em] \hline\\[-1.0em]
        CQM & \SI{0.91e12}{\centi\meter^{-2}} & \SI{27}{\degree} & \SI{2.5}{\tesla} \\
        multitone state, lower freq. & \SI{0.89e12}{\centi\meter^{-2}} & \SI{44}{\degree} & \SI{26}{\tesla} \\
        multitone state, higher freq. & \SI{1.19e12}{\centi\meter^{-2}} & \SI{-48}{\degree} & \SI{23}{\tesla}\\
    \bottomrule
    \end{tabular}
    \caption{Lifshitz-Kosevich fit parameters.}
    \label{tab:placeholder}
\end{table}

% =============================================================================
\section{Extraction of effective mass in the circular quarter metal state from temperature dependent quantum oscillations}\label{supp_sec:effective-mass}
% =============================================================================

\begin{figure*}[h]
   \centering 
    \includegraphics[width=0.95\linewidth]{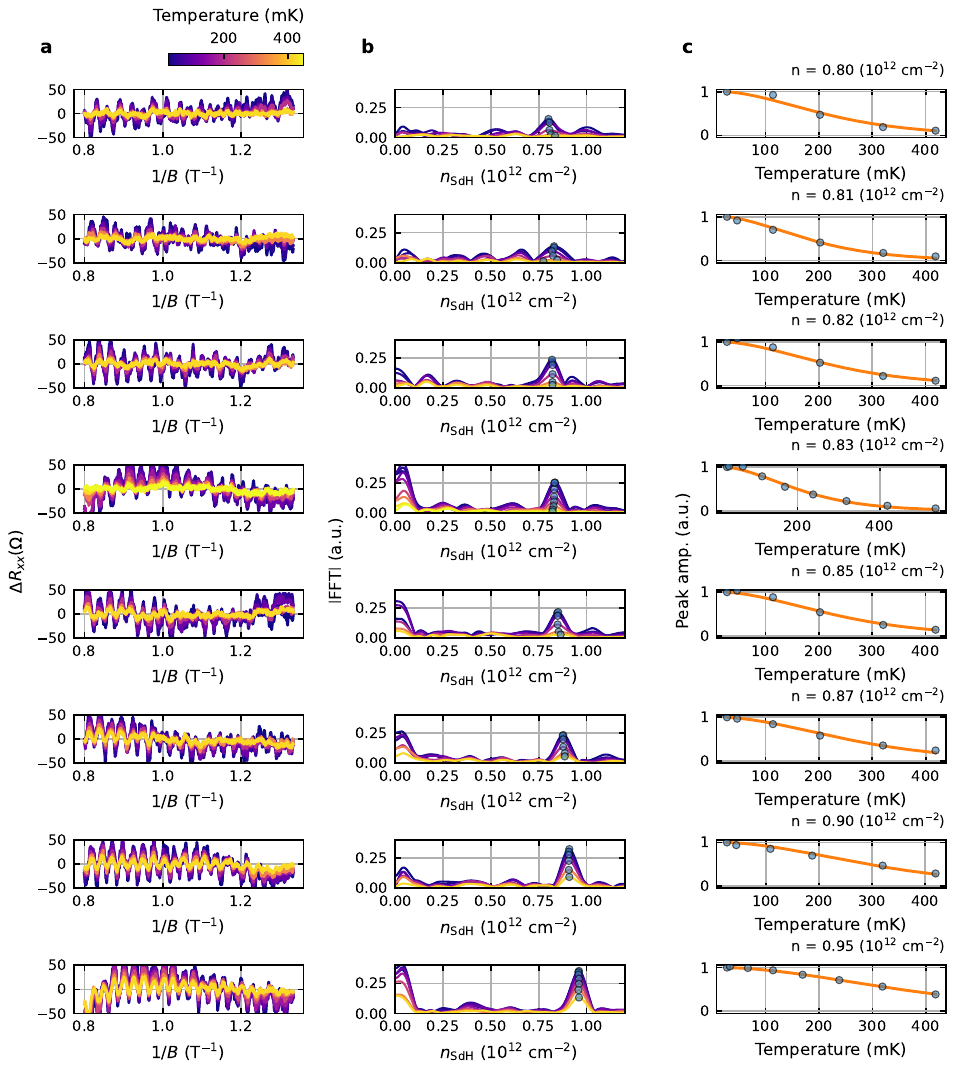}
    \caption{\textbf{Temperature dependence of quantum oscillations in the circular quarter metal at $D/\epsilon_0 = \SI{1}{V/nm}$ for contacts B3-B4.} (a) Background subtracted $R_{xx}$; (b) Fourier transform; and (c) fit of Fourier transform peak amplitudes to expected Lifshitz-Kosevich temperature dependence (orange curve). }
   \label{fig-ed_effectivemass}
\end{figure*}

% =============================================================================
\section{Temperature dependence of quantum oscillations in multitone state}
% =============================================================================

\begin{figure*}[h]
   \centering 
   \includegraphics[width=0.95\textwidth]{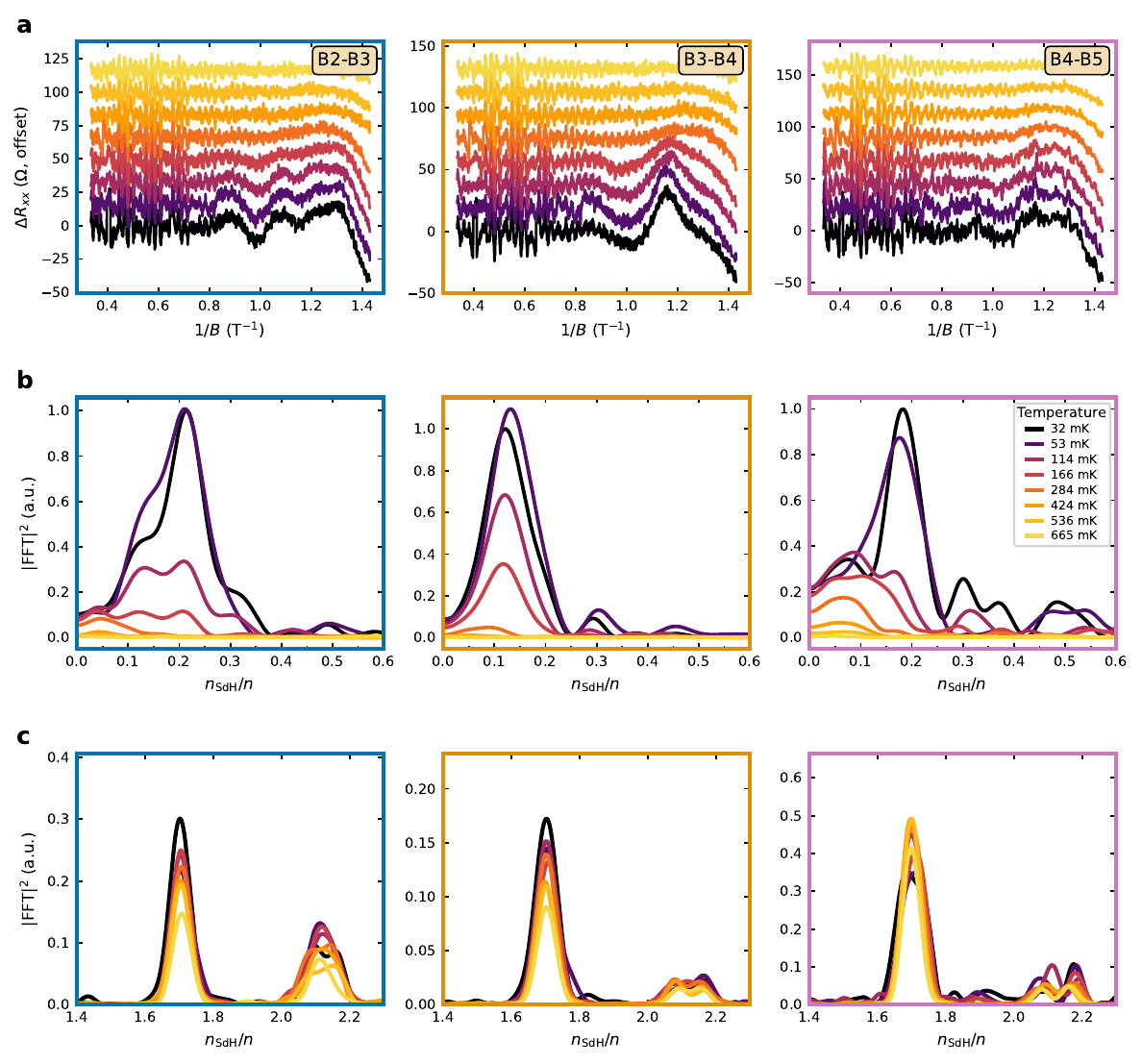}
   \caption{\textbf{Temperature dependence of the Shubnikov--de Haas spectrum in the multitone state.}
      All data are taken at fixed carrier density $n = 0.54\times10^{12}\,\mathrm{cm^{-2}}$ and displacement field
      $D = \SI{1}{V/nm}$ over a magnetic-field window $B = 0.7$--$3\,\mathrm{T}$.
      Columns correspond to three longitudinal contact pairs (B2--B3, B3--B4, B4--B5), and line color encodes temperature from $\sim\SI{32}{mK}$ to $\sim\SI{665}{mK}$.
      (a) Background-subtracted longitudinal resistance $\Delta R_{xx}$ versus inverse field $1/B$, offset vertically for clarity.
      The smooth magnetoresistance background is removed with a Butterworth high-pass filter applied in $B$.
      (b) Corresponding power spectra $|\mathrm{FFT}|^2$ as a function of the normalized frequency $n_{\mathrm{SdH}}/n$ showing the low-frequency tones ($n_{\mathrm{SdH}}/n < 0.6$).
      (c) The same spectra near the two high-frequency tones.}
   \label{fig:temp-two-tone-both}
\end{figure*}

\begin{figure*}[h]
   \centering 
   \includegraphics[width=0.95\textwidth]{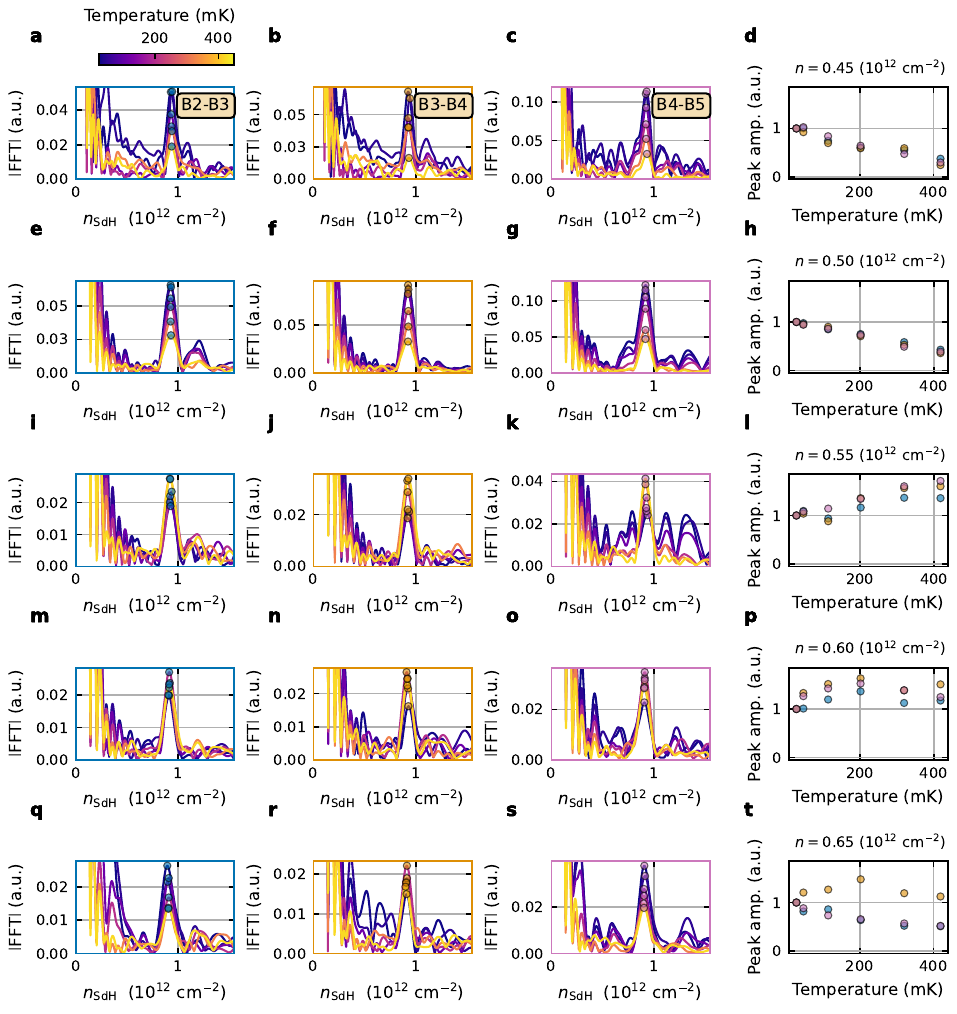}
   \caption{\textbf{Temperature dependence of high frequency quantum oscillations in the multitone regime.} (a-t) Fourier transform amplitude of $R_{xx}$ with respect to $1/B$ at fixed displacement $D/\epsilon_0 = 1 \si{V/nm}$.  Each row corresponds to a fixed density within the multitone region. The first three columns correspond to different contact pairs, while the right-most column shows temperature dependence of the indicated peak amplitude (with colors indicating contact pairs). The peak tracked here  is the lower of the two high-frequency tones, which is non-monotonic in temperature for some contacts and densities. While the higher tone is robust at low temperature, it becomes more difficult to resolve as temperature increases.     
   }
   \label{fig:temp-two-tone-high}
\end{figure*}

\begin{figure*}
   \centering 
   \includegraphics[width=0.95\textwidth]{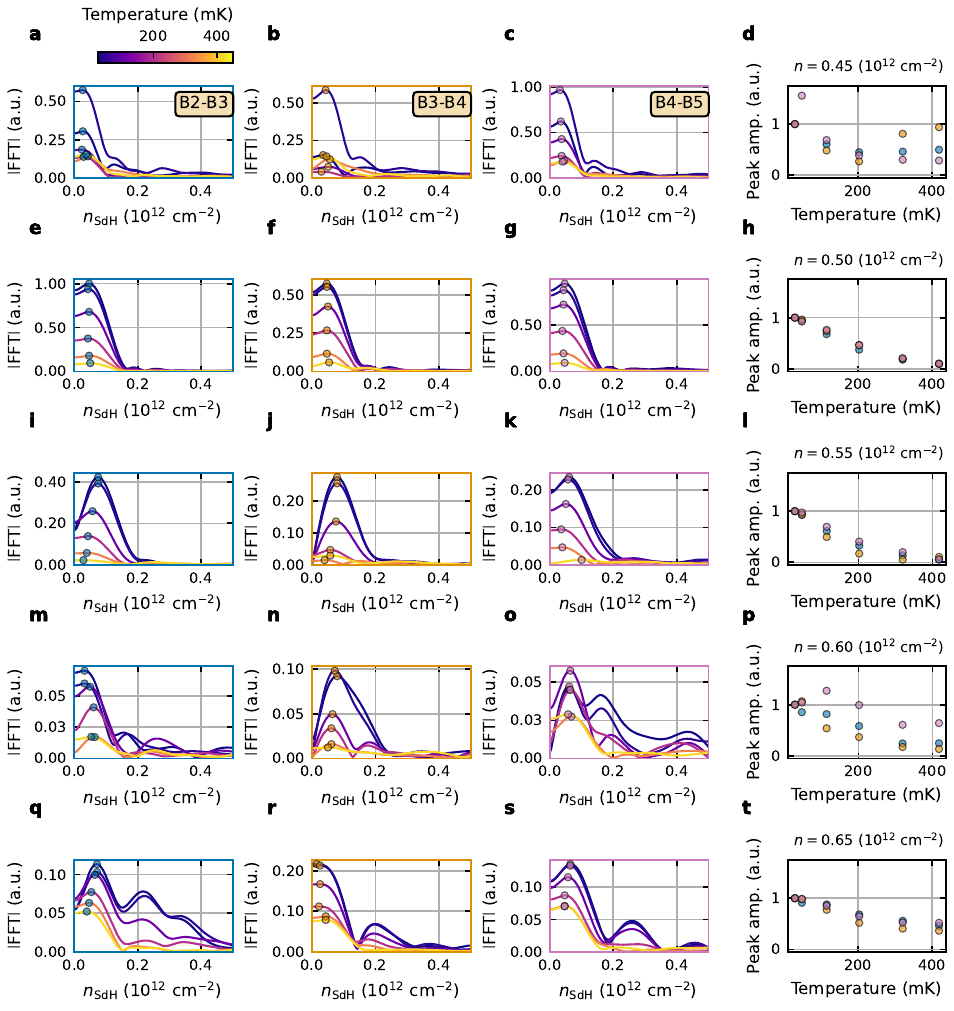}
   \caption{\textbf{Temperature dependence of low frequency quantum oscillations in the multitone regime.} (a-t) Fourier transform amplitude of $R_{xx}$ with respect to $1/B$ at fixed displacement $D/\epsilon_0 = 1 \si{V/nm}$, focusing on the low-frequency tones. A linear background subtraction is applied to $R_{xx}$.  Each row corresponds to a fixed density within the multitone region. The first three columns correspond to different contact pairs, while the right-most column shows temperature dependence of the indicated peak amplitude  (with colors indicating contact pairs).}
   \label{fig:temp-two-tone-low}
\end{figure*}

% =============================================================================
\section{Details of FFT analysis of SdH oscillations}\label{supp_sec:fft}
% =============================================================================

In this section, we provide additional details of the pipeline used to process the Shubnikov-de Haas (SdH) oscillations and perform the frequency analysis using the fast Fourier transform (FFT).
In particular, we establish the robustness of the SdH oscillations in the multitone state to different processing techniques.

\begin{figure*}[h]
    \centering
    \includegraphics[width=\textwidth]{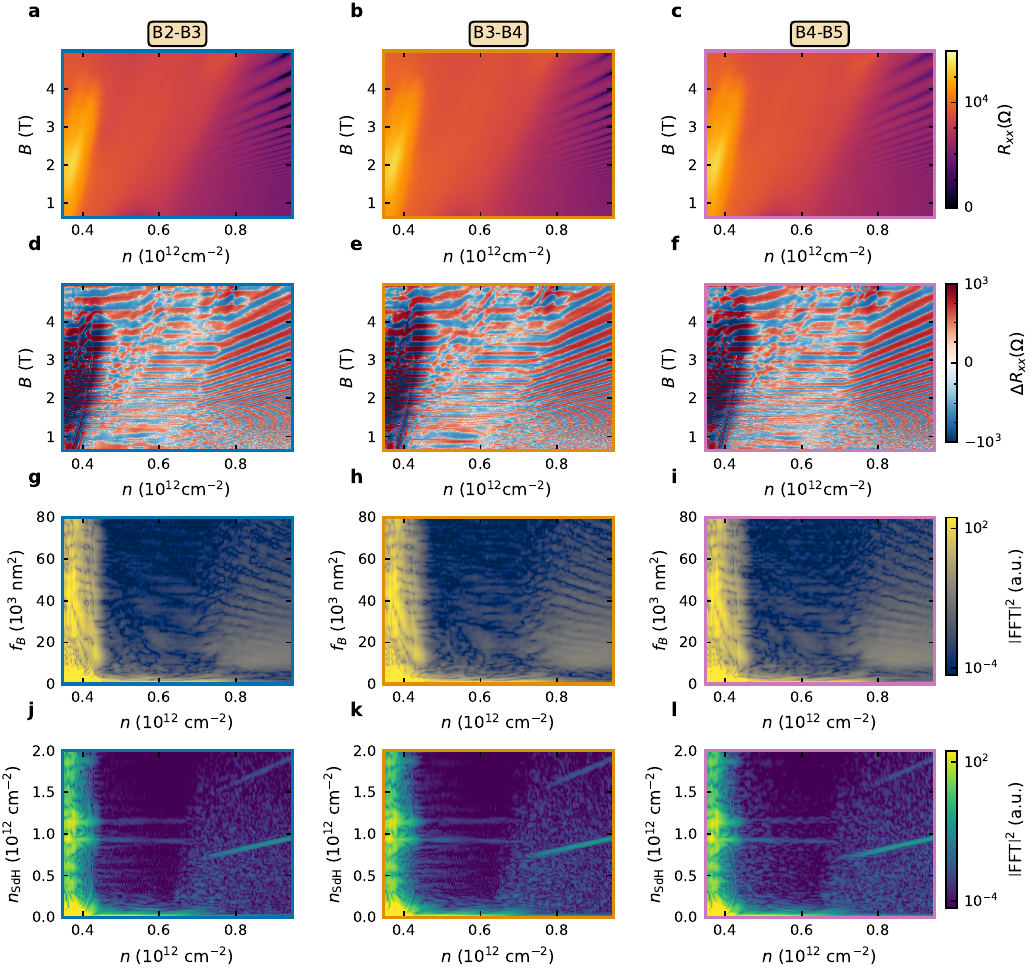}
    \caption{\textbf{Fourier analysis of quantum oscillations versus $B$ and $1/B$ across contact pairs.} Landau fans at  $D/\epsilon_0=1 \si{\volt/\nano\meter}$ (a-c). Background subtraction (d-f) reveals oscillatory features. Fourier transforms in $B$ (g-i) and $1/B$ (j-l) illustrate that the two high-frequency features in the multitone state are periodic in $1/B$ and robust to contact pair choice. }
    \label{fig:fft-B-1_B-contact-pair}
\end{figure*}

As a preliminary check, we first verify in Fig.~\ref{fig:fft-B-1_B-contact-pair} that the quantum oscillations are periodic in $1/B$ rather than $B$. In the high density range $n>0.7\times 10^{12}\,\text{cm}^{-2}$, the $1/B$ Fourier transform for all three contact pairs (panels j,k,l) reveals a clear tone with $n_{\text{SdH}}=n$, corresponding to expected dominant frequency of the circular quarter metal (CQM). We also observe a harmonic at $n_\text{SdH}=2n$. By contrast, the Fourier transform with respect to $B$ (panels g,h,i) does not yield clear tones in the same density regime. For the multitone state at low densities $n<0.7\times 10^{12}\,\text{cm}^{-2}$, we observe two nearly-flat high-frequency tones in the $1/B$ Fourier transform, while no clear spectral features are present in the Fourier transform with respect to $B$. 

\begin{figure*}[h]
    \centering
     \includegraphics[width=\textwidth]{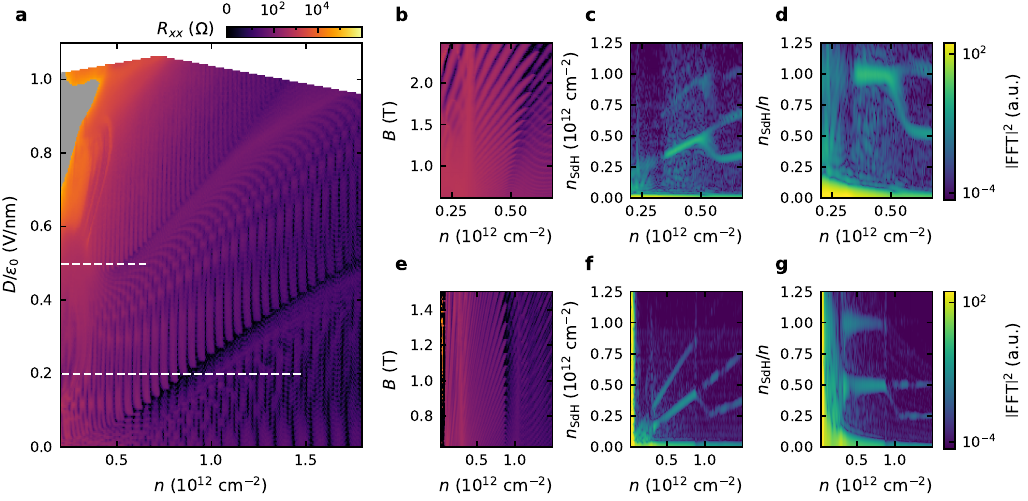}
    \caption{\textbf{Quantum oscillations in the quarter metal, half metal, and full metal regimes.}  (a) Gate map at $B = \SI{1}{\tesla}$ with locations of Landau fans from panels (b) and (e) indicated. We plot frequency spectra normalized to $n_{\mathrm{SdH}}$ in panels (c,f) and $f_{\nu} = n_{\mathrm{SdH}}/n$ in panels (d,g). Panel (d) shows a transition from the quarter metal ($f_{\nu}=1$) to half metal ($f_{\nu}=1/2$) regimes, while panel (g) shows a transition from the the half metal to symmetric metal ($f_{\nu}=1/4$) regimes. }
    \label{fig:fft-QO-in-metalic_regimes}
\end{figure*}

In Fig.~\ref{fig:fft-QO-in-metalic_regimes}, we also show well-resolved quantum oscillations that are periodic in $1/B$ in the half metal and full metal regimes. For the linecuts shown, the half (full) metal is characterized by a dominant $n_\text{SdH}=n/2$ ($n_\text{SdH}=n/4$) tone, consistent with equal-sized simply-connected Fermi pockets in two (four) isospin flavors.

To demonstrate the robustness of the high-frequency tones in the multitone state, we present the characterization process in more detail. The latter can be broken down into three primary parts: (1) optional background subtraction; (2) interpolation to even spacing in $1/B$; (3) Fourier transformation. We now discuss important aspects of each of these.

\begin{figure*}[h]
   \centering 
   \includegraphics{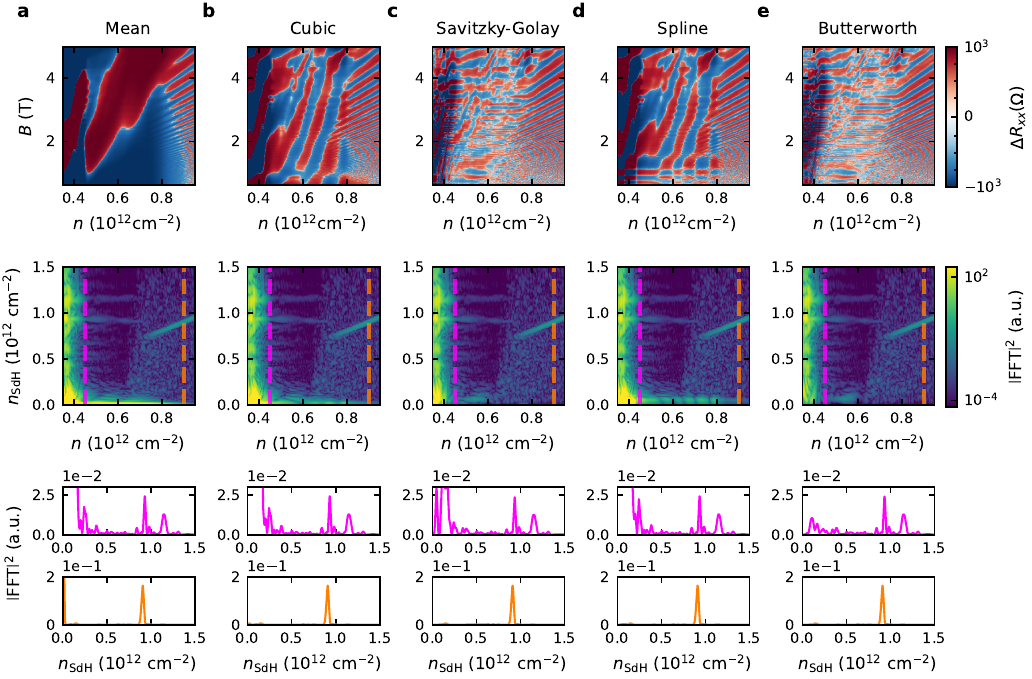}
   \caption{\textbf{Comparison of background subtraction methods}. Background subtracted $\Delta R_{xx}$ (top) and Fourier transform (middle, with dashed lines corresponding to linecuts below) at $D/\epsilon_0=1 \si{\volt/\nano\meter}$ using (a) mean subtraction, (b) cubic polynomial subtraction, (c) Savitzky–Golay filter, (d) spline fitting, (e) Butterworth filter. The data are obtained using contacts B3-B4.}
   \label{fig:fft-compare-bgsub-methods}
\end{figure*}

\begin{figure*}[h]
   \centering 
   \includegraphics[width=0.95\textwidth]{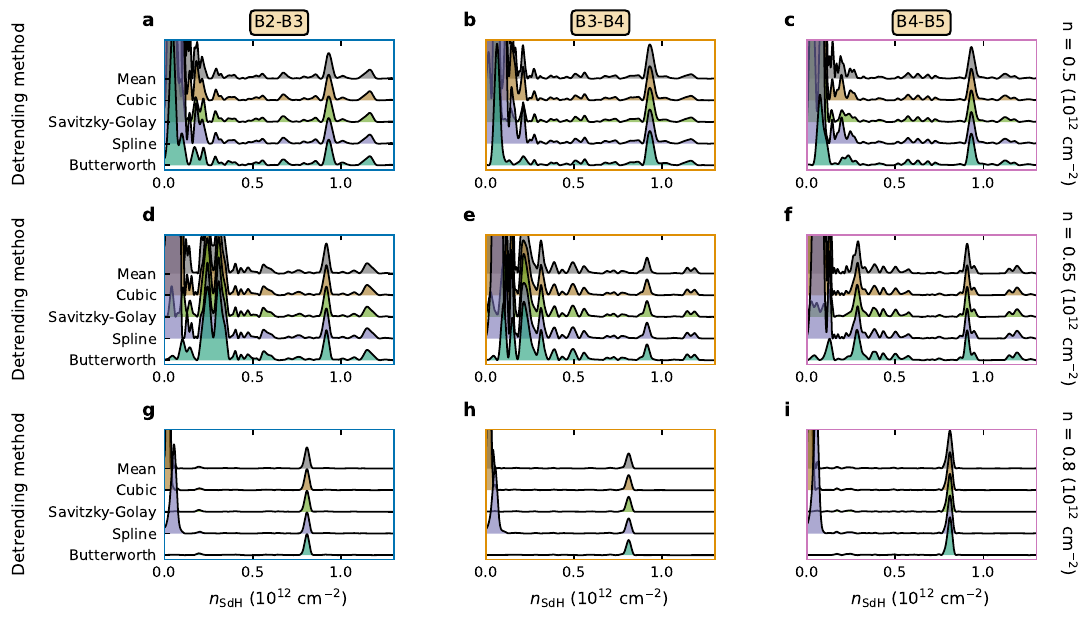}
   \caption{\textbf{Comparison of linecuts using different background subtraction methods}. The results from three contact pairs are shown at $n = \SI{0.5e12}{\centi\meter^{-2}}$ (a-c),  $n = \SI{0.65e12}{\centi\meter^{-2}}$ (d-f) and  $n = \SI{0.8e12}{\centi\meter^{-2}}$ (g-i), all at $D/\epsilon_0=1 \si{\volt/\nano\meter}$. Almost all features at frequencies above approximately $n_\text{SdH}=\,$\SI{0.3e12}{\centi\meter^{-2}} vary minimally with respect to background subtraction techniques, but show some slight variations across contact pairs.}
   \label{fig:fft-compare-bgsub-cuts}
\end{figure*}

While not necessary for observation of high frequency tones, background subtraction is particularly useful to enhance their clarity. In the multitone state, there is significant magnetoresistance with low frequency components, which means it is helpful to perform background subtraction when examining the high frequency tones, especially when fitting to the form of oscillations as in Figs.~\ref{fig:fft-compare-B1Blinecuts}. Fig.~\ref{fig:fft-compare-bgsub-methods} and \ref{fig:fft-compare-bgsub-cuts} show that the high frequency tones are robust against a variety of background subtraction methods (mean, cubic, Savitzky-Golay, spline, Butterworth). Of these, Butterworth filtering (which explicitly suppresses low frequencies in $B$ below some cutoff) was chosen for results where background subtraction is required.

\begin{figure*}[h]
    \centering
    \includegraphics{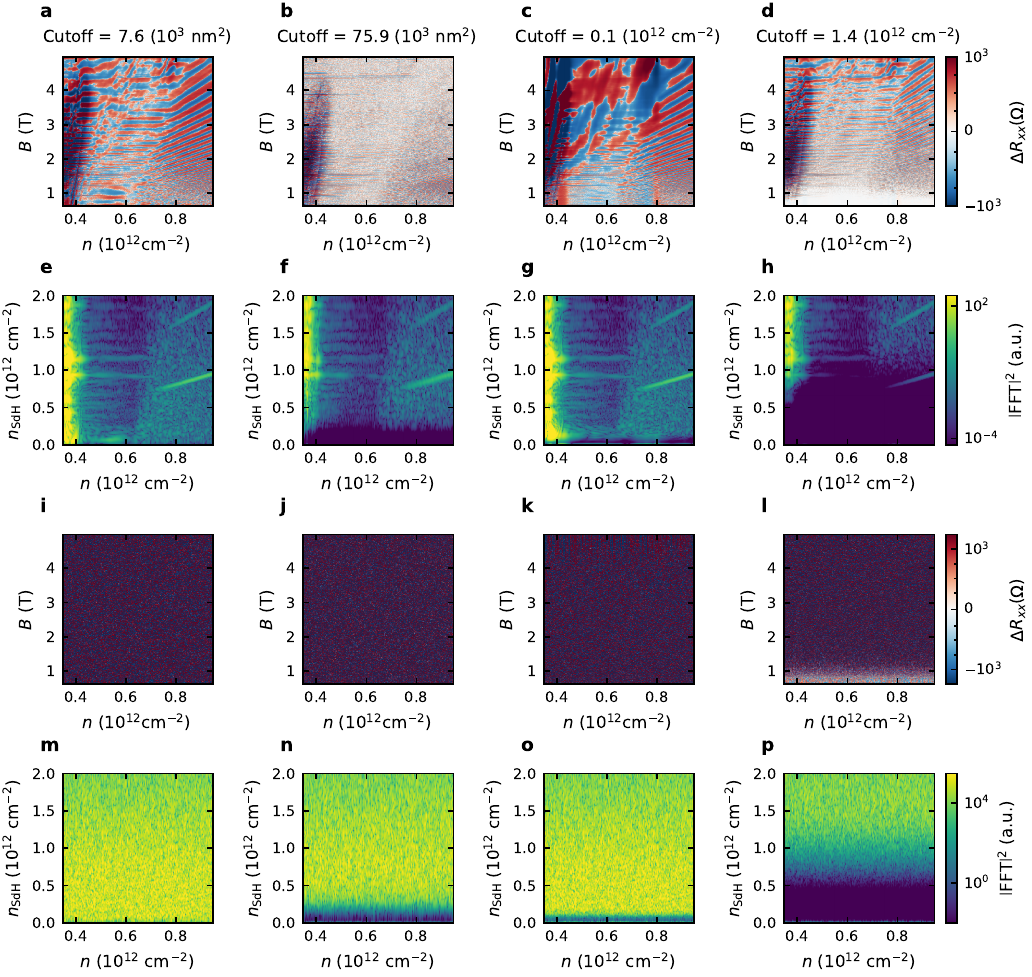}
    \caption{\textbf{Analysis of Butterworth filtering}. Data are taken using contacts B3-B4 at $D/\epsilon_0=1 \si{\volt/\nano\meter}$. Filtering applied in $B$ (a,b) and $1/B$ (c,d) with bandwidths of 1/8 (a,c) and 1/80 (b,d) of the full data range in $B$ and $1/B$, respectively. Corresponding Fourier transforms are shown in (e-h). Panels (i-p) show the same analysis applied to Gaussian noise with mean and standard deviation matching the original resistance data, in order to diagnose artifacts arising from the analysis procedure. }
    \label{fig:fft-compare-bwfilter-params}
\end{figure*}

\begin{figure*}[h]
    \centering
    \includegraphics{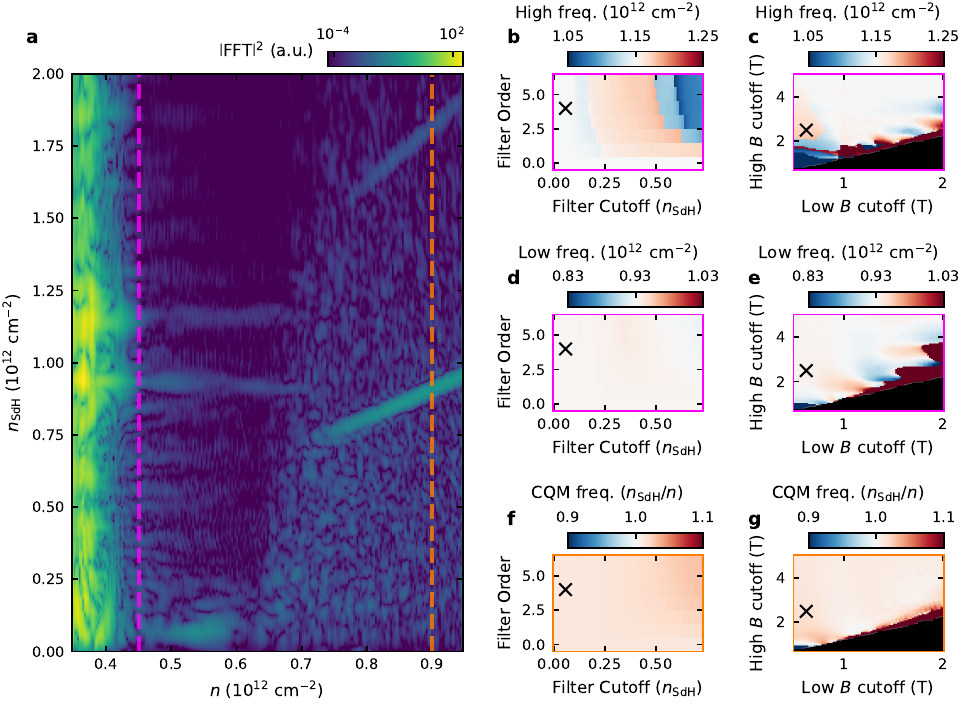}
    \caption{\textbf{Effects of Butterworth filter parameters and field cutoff on extracted frequencies.} Data are taken using contacts B3-B4 at $D/\epsilon_0=1 \si{\volt/\nano\meter}$.
    (b,d,f) Varying filter parameters for the (b) upper and (d) lower high-frequency tone in the multitone regime (density indicated by purple line in (a)), and the primary tone in the CQM (density indicated by orange line in (a)).
    (c,e,g) Same as (b,d,f), expect varying the magnetic field cutoffs.
     Black crosses mark filter/cutoff parameters used in typical analysis.}
    \label{fig:fft-Bcutoff-effects}
\end{figure*}

In Fig.~\ref{fig:fft-compare-bwfilter-params}, we examine various choices of filter cutoff for the Butterworth filter. We also investigate the differences between applying filtering to the original data, taken in constant spacing in $B$, or interpolated data, with constant spacing in $1/B$ (the interpolation procedure will be discussed shortly).
We find that performing background subtraction after interpolating to even spacing in $1/B$ has a tendency to cause artifacts at low frequencies (see Fig.~\ref{fig:fft-compare-bwfilter-params}g). Therefore, in the results presented in this work, we perform background subtraction on data evenly spaced in $B$.
The largest concern when applying background subtraction is the introduction of artifacts that may be falsely interpreted as real physical signals. This is possible with aggressive background subtraction: strong aliasing effects can appear as peaks in the Fourier transform, particularly at low frequencies. Fig.~\ref{fig:fft-compare-bwfilter-params}l illustrates how aggressive background subtraction, particularly when applied to interpolated data, has a tendency to introduce apparent structure even when applied to Gaussian noise. The two high frequency tones in the multitone state are, however, present with modest or no background subtraction, ruling out the possibility that they are spurious artifacts. Additional quantitative details on robustness of the these high frequency features to background subtraction are shown in Fig.~\ref{fig:fft-Bcutoff-effects}.

After performing background subtraction, it is essential to interpolate data to even spacing in $1/B$ prior to the FFT. 
The spacing between points was chosen as the smallest difference between data points in $1/B$. 
Before performing the FFT, we apply a Hann window to suppress spectral leakage from the finite field range.
Additionally, we zero-pad the data to allow better identification of the peak positions.

Fig.~\ref{fig:fft-Bcutoff-effects} shows the detailed effects of modifying the window range over which we take a Fourier transform on the extracted peak frequency. 
While the single tone of the CQM is consistently observed, the low amplitude of oscillations in the multitone regime means that peaks may fail to appear if the data is not taken to sufficiently high fields. 
In particular, data up to at least \SI{2.5}{\tesla} is necessary for a clear extraction of the high-frequency tones.
This is consistent with the expected exponential suppression of these tones at low fields. Above this threshold though, 
we note that the extracted frequencies are relatively insensitive to further changes to the field cutoff parameters. 

\begin{figure*}
    \centering
    \includegraphics{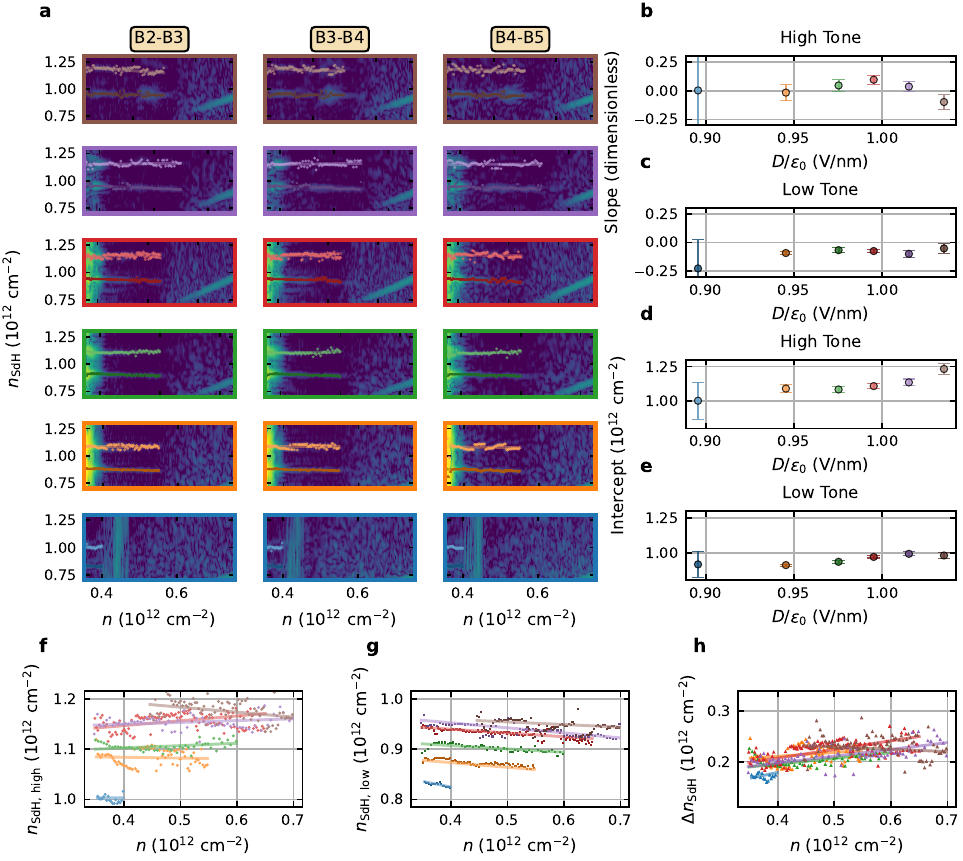}
    \caption{\textbf{Density dependence of high frequency tones in the multitone state.} (a) Two high-frequency tones marked on squared Fourier transform amplitude of $R_{xx}$. The displacement field in each row is color coded according to the colored markers in (b-e). The three columns correspond to the contact pairs B2-B3, B3-B4, and B4-B5. (b,c) Slope of the higher and lower tone, respectively, as a function of $n$. (d,e) Frequency intercept of the higher and lower tone, respectively. High (f) and low (g) tones, along with their difference (h), are shown as a function of density for various displacement fields $D$. }
    \label{fig:two-tone-diff}
\end{figure*}

\begin{figure*}[h]
   \centering 
   \includegraphics[width=0.8\textwidth]{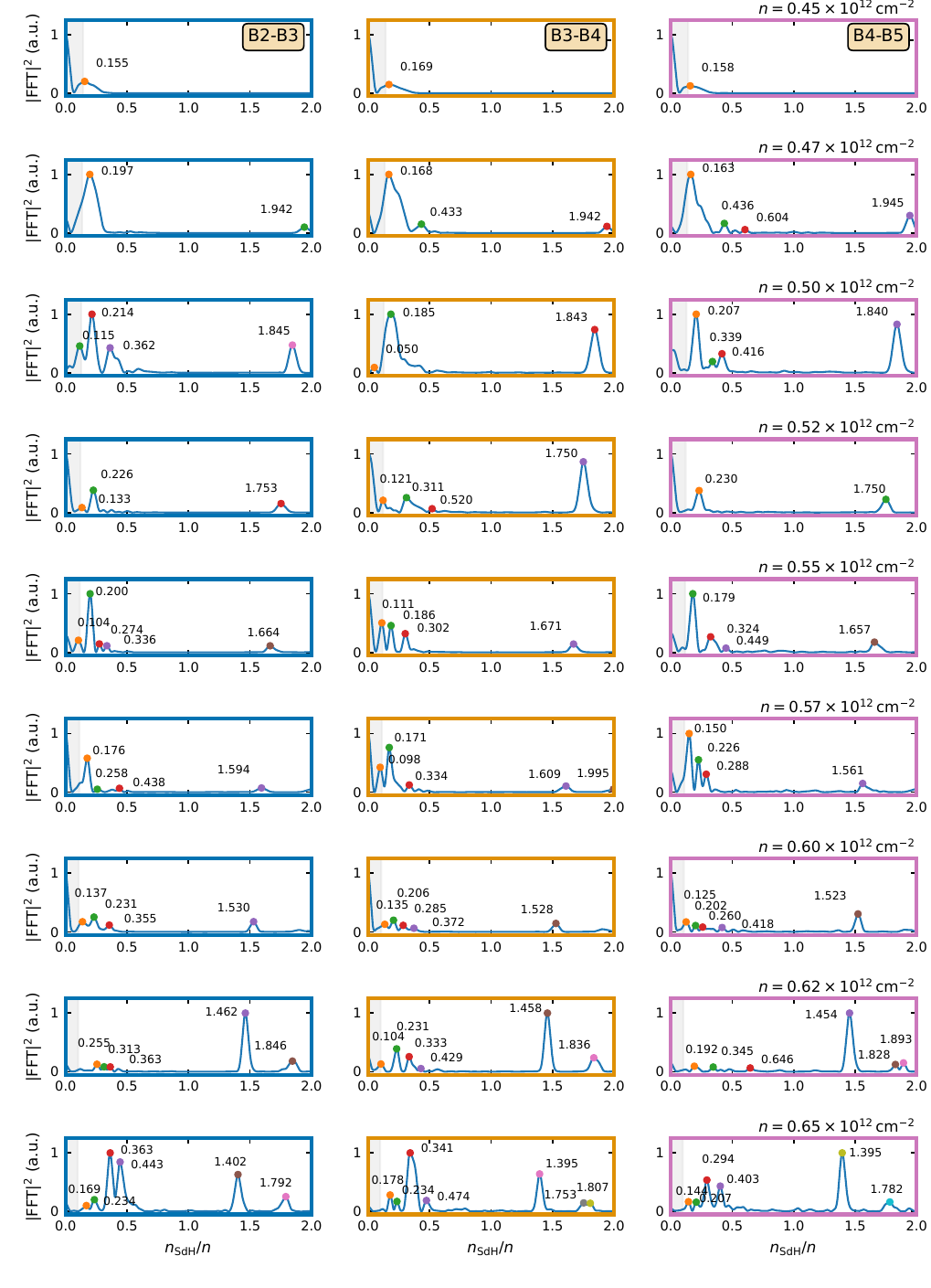}
   \caption{\textbf{Density dependence of the Shubnikov–de Haas spectrum across contact pairs.}
  Normalized power spectra $|\mathrm{FFT}|^2$ of $\Delta R_{xx}$ as a function of the normalized frequency $n_{\mathrm{SdH}}/n$, for a series of fixed carrier densities $n$ (rows, $n = 0.45$–$0.65\times10^{12},\mathrm{cm^{-2}}$) at $D/\epsilon_0 = \SI{1}{V/nm}$. Each column corresponds to one of three voltage-probe pairs (B2--B3, B3--B4, B4--B5; colored frames). Each line cut is high-pass filtered with a density-scaled Butterworth filter to remove the smooth background; each spectrum is normalized to its maximum. Markers denote identified peaks, labeled by their normalized frequency. The shaded region denotes frequencies below a three-cycle resolution limit of the field window, where extracted peaks are unreliable.}
   \label{fig-all_linecuts}
\end{figure*}

In Fig.~\ref{fig:two-tone-diff} we track the high-frequency tones in the multitone state as a function of $n$ and $D$. We find that these tones vary very weakly with $n$. The variation of these tones with $D$ is somewhat stronger. However, the difference in the frequencies of the two high tones does not change appreciably with $D$.
In Fig.~\ref{fig-all_linecuts}, we plot the FFT frequency spectra at fixed $D/\epsilon_0=1\,\text{V}/\text{nm}$ for different values of $n$.

% =============================================================================
\section{Transport in an additional superconducting state on the hole side}
% =============================================================================

\begin{figure*}[h]
   \centering 
   \includegraphics[width=0.8\textwidth]{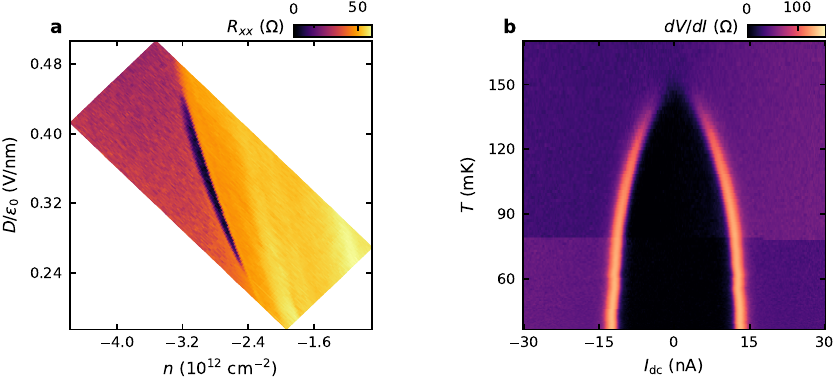}
   \caption{\textbf{Superconducting state on the hole side.} (a) $R_{xx}$ as a function of $n$ and $D$ on the hole side. (b) $dV/dI$ vs d.c. current bias and temperature in the superconducting state. The critical temperature of the superconductor is about 150\,\si{\milli\kelvin}.}
   \label{fig-si-hole-sc}
\end{figure*}

We observe an additional superconducting state on the hole side of the $n$-$D$ phase diagram. Fig.~\ref{fig-si-hole-sc}(a) shows the extent of the superconducting state which occurs at a flavor transition in the phase diagram. As in rhombohedral trilayer graphene~\cite{zhouSuperconductivityRhombohedralTrilayer2021a}, this superconductor emerges from a SM (i.e., a time-reversal symmetric state) near a Stoner-like transition into a partially isospin-polarized half-metal. The critical temperature of this superconducting state is $150\,\si{\milli\kelvin}$. From the uniformity of the critical current as measured by multiple contact pairs, we infer that this particular superconducting state is spatially uniform across the device. Fraunhofer oscillations as a function of magnetic field in this superconducting state are shown in Fig.~\ref{fig-ed-holesc-frauns}.

% =============================================================================
\section{Single-particle continuum model for rhombohedral tetralayer graphene}\label{secapp:continuum_model}
% =============================================================================

The real and reciprocal lattices of R4G are spanned by
\begin{align}
	\bm{R}_1 = \left( a,0 \right), \quad
	\bm{R}_2 = \left( \frac{1}{2}, \frac{\sqrt{3}}{2} \right),\qquad
	\bm{G}_1 = \frac{2\pi}{a}\left(1, -\frac{1}{\sqrt{3}}\right), \quad
	\bm{G}_2 = \frac{2\pi}{a}\left(0, \frac{2}{\sqrt{3}}\right),
\end{align}
where $a \approx \SI{2.46}{\angstrom}$ is lattice constant of graphene (i.e.~$\sqrt{3}$ times larger than the carbon-carbon bond length).
We define standard high-symmetry points $\bm{K} = \frac{2}{3} \bm{G}_1 + \frac{1}{3} \bm{G}_2$ and $\bm{K}' = -\bm{K}$, and index valley by $\tau \in \{K,K'\} = \{1,-1\}$.

In the plane wave basis, the non-interacting continuum model of R4G for a single spin sector in valley $K$ is
\begin{equation}
    \hat{H}^K_\text{s.p.}=\sum_{\bm{k},l,l',\sigma,\sigma'
    }c^\dagger_{\bm{k},l,\sigma}[h^K_0(\bm{k})]_{l\sigma,l'\sigma'}c_{\bm{k},l',\sigma'},
\end{equation}
where $\bm{k}$ is measured relative to $\bm{K}$, and $l=1,2,3,4$ and $\sigma=A,B$ are layer and sublattice indices respectively.
$h_0^K(\bm{k})$ is expressed in terms of $2\times 2$ blocks in sublattice space as~\cite{ghazaryan2023platform}
\begin{equation}
\label{eq:microscopic_RMG_Hamiltonian}
	h_0^K(\bm{k})
	= 
	\begin{bmatrix} 
		\mathcal{D}_{1}(\bm{k}) & \mathcal{V}(\bm{k}) & \mathcal{W} & 0\\
		\mathcal{V}^\dagger(\bm{k}) & \mathcal{D}_{2}(\bm{k}) & \mathcal{V}(\bm{k}) & \mathcal{W}\\
	\mathcal{W}^\dagger & \mathcal{V}^\dagger(\bm{k}) & \mathcal{D}_3(\bm{k}) & \mathcal{V}(\bm{k})\\
			    0 & \mathcal{W}^\dagger & \mathcal{V}^\dagger(\bm{k}) & \mathcal{D}_4(\bm{k})
	\end{bmatrix},
\end{equation}
where 
\begin{equation}
    \mathcal{D}_l(\bm{k})= \begin{pmatrix} 0  & \gamma_{0} \tilde{k}^-\\ \gamma_0 \tilde{k}^+ & 0 \end{pmatrix} + \mathcal{U}_{\ell},\quad \mathcal{V}(\bm{k})=\begin{pmatrix} -\gamma_4 \tilde{k}^- & -\gamma_3 \tilde{k}^+ \\ \gamma_1 & -\gamma_4 \tilde{k}^- \end{pmatrix},\quad \mathcal{W}=\begin{pmatrix} 0 & \frac{\gamma_2}{2}\\ 0 & 0 \end{pmatrix},
\end{equation}
and we have defined dimensionless momenta
\begin{equation}
    \label{eq:scaled_RMG_momenta}
    \tilde{k}^\pm\equiv \frac{\sqrt{3}a}{2}(k_x\pm ik_y).
\end{equation}
The onsite potentials read
\begin{gather}
    \mathcal{U}_1=\text{diag}[\Delta_2+V_1(D),\Delta_2+\delta+V_1(D)]\\
    \mathcal{U}_2=\text{diag}[-\Delta_2+\delta+V_2(D),-\Delta_2+\delta+V_2(D)]\\
    \mathcal{U}_3=\text{diag}[-\Delta_2+\delta+V_3(D),-\Delta_2+\delta+V_3(D)]\\
    \mathcal{U}_4=\text{diag}[\Delta_2+\delta+V_4(D),\Delta_2+V_4(D)]
\end{gather}
where $V_l(D)$ captures the contribution arising from the externally applied displacement field $D$.
In App.~\ref{secapp:interlayer_screening}, we discuss self-consistent Hartree calculations that estimate the internally-screened $V_l(D)$.
For calculations where we neglect such capacitive effects, we simply model the displacement field as a linear interlayer potential $u_D=V_{l+1}-V_l$.
The continuum model for valley $K'$ can be obtained by time-reversal symmetry. The continuum models for other numbers of layers, $L$, are constructed analogously. 

The single-particle parameters that we use are listed in the table below.
We consider two sets of parameters.
Set 1 is taken from Ref.~\cite{auerbachIsospinMagneticTexture2025}, where the parameters were fitted to experimental data. All calculations in the main text use the Set 1 parameters.
In Set 2, we adjust the values of $\gamma_2,\delta,\Delta_2$. The inversion-symmetric potentials $\delta,\Delta_2$ are taken from ab-initio calculations in Ref.~\cite{2023PhRvB.108o5406P} while the enlarged value of $\gamma_2$ is suggested from Ref.~\cite{vituri2025incommensurate}. These sets show similar features in the density of states as discussed below. 
\begin{center}
\begin{tabular}{llllllll}
	\toprule
	&
    $\gamma_0$ (meV) & 
	$\gamma_1$ (meV) & 
	$\gamma_2$ (meV) & 
	$\gamma_3$ (meV) & 
	$\gamma_4$ (meV) & 
	$\delta$ (meV)	&
	$\Delta_2$ (meV) \\ 
	\midrule
	Set 1:\,\, &3100 & 380 & -15 & 290 & 141 & 10.5 & 2\\
    \midrule
    Set 2:\,\, &3100 & 380 & -24 & 290 & 141 & 14.3 & 12.2\\
    \bottomrule
\end{tabular}
\end{center}

% =============================================================================
\subsection{Fermiology and higher order van Hove singularities}
% =============================================================================

To understand the Fermiology of the lowest conduction band analytically, we employ projection to the orthogonal chiral basis of $L$-layer rhombohedral graphene\cite{2024PhRvB.109t5122H}:
\bea
    [\Psi_{A}(\bm{k})]_{\alpha l} &= (1, x, \dots, x^{L-1})_l \delta_{\alpha A} \\
    [\Psi_{B}(\bm{k})]_{\alpha l} &= (\bar{x}^{L-1}, \bar{x}^{L-2}, \dots, 1)_l \delta_{\alpha B} \\
\eea
where $x = - \gamma_0 \tilde{k}^-/\gamma_1$, $\alpha$ is sublattice, and $l$ is the layer number. We collect these vectors into the matrix $\Psi(\bm{k}) = [\Psi_A(\bm{k}) , \Psi_B(\bm{k})] / \mathcal{N}(\bm{k})$ with a normalization factor $\mathcal{N}(\bm{k}) = |\Psi_A| = |\Psi_B|$ such that $\Psi^\dag(\bm{k}) \Psi(\bm{k}) = \mathbf{1}_{2}$. This low-energy basis is accurate at small momenta $|x|<1$. Numerically we check that $\Psi(\bm{k})$ has at least $95\%$ overlap with the actual conduction band wavefunction for $|k| \leq 0.4\text{nm}^{-1}$. 

We now form the effective Hamiltonian $h_{\text{eff}}(\bm{k}) = \Psi^\dag(\bm{k}) h_0^K(\bm{k}) \Psi(\bm{k})$. We consider a linear interlayer potential $u_D$. $h_{\text{eff}}(\bm{k})$ is a $2\times 2$ Hamiltonian which can diagonalized exactly. We take the resulting eigenvalues and expand them in powers of $k$ to obtain the behavior in the vicinity of the $K$ point. For the tetralayer case ($L=4$), we find the conduction eigenvalue has the expansion
\bea
E_{L=4}(\bm{\tilde{k}}) = E_0 + E_2 |\bm{\tilde{k}}|^2 + E_4 |\bm{\tilde{k}}|^4 + E_5 |\bm{\tilde{k}}|^5 \cos 3 \theta + O(|\bm{\tilde{k}}|^6),
\eea
where $\theta$ is the angle that $\bm{\tilde{k}}$ makes with the $x$-axis. The expressions for the leading coefficients are 
\bea
E_0 &= \frac{3u_D}{2} + \Delta_2 \\
E_2 &=\frac{\gamma_0 \left(\gamma_0 \gamma_2^2-3 \gamma_0 u_D^2+3 \gamma_0 u_D (\delta -2 \text{$\Delta _2$})+6 \gamma_1 \gamma_4 u_D\right)}{3 \gamma_1^2 u_D} \\
E_4 &= -\frac{\gamma_0^3 \left(\gamma_0 \gamma_2^4+27 \gamma_0 u_D^4+6 \gamma_2 u_D^2 (2 \gamma_0 \gamma_2-9 \gamma_1 \gamma_3)\right)}{27\gamma_1^4 u_D^3} \\
E_5 &= \frac{2 \gamma_0^5 \gamma_2}{3 \gamma_1^4 u_D} \\
\eea
and we check that the $O(|\bm{\tilde{k}}|^6)$ term has a positive coefficient. Our main observation is that there is a critical displacement field $u_D > 0$ such that
\bea
-3 \gamma_0 u_D^2+ u_D \left( 3 \gamma_0  (\delta -2 \text{$\Delta _2$})+6 \gamma_1 \gamma_4 \right) + \gamma_0 \gamma_2^2 &= 0 
\eea
and hence $E_2 = 0$, thereby generating an anomalously flat low-energy dispersion $E(\bm{\tilde{k}}) = E_4 |\bm{\tilde{k}}|^4 + E_5 |\bm{\tilde{k}}|^5 \cos 3\theta$. This leads to a higher-order van Hove singularity (vHS) with the DOS diverging like $|E_4 \omega|^{-1/2}$, where $\omega$ is the energy deviation from the vHS. Note that the divergence is a power-law in comparison to the usual logarithmic divergence of a 2D vHS near a quadratic saddle point $k_x^2 - k_y^2$. 

We contrast this behavior with multilayer graphenes with other layer numbers $L$. Following the same procedure, we derive that in $L = 2$ Bernal bilayer graphene the conduction band dispersion is
\bea
E_{L=2}(\bm{\tilde{k}}) &= \left(\text{$\Delta_2$}+\frac{u_D}{2}\right) +|\bm{\tilde{k}}|^2 \left(\frac{\gamma_{0} (\gamma_{0} \delta +2 \gamma_{1} \gamma_{4})}{\gamma_{1}^2}-\frac{\gamma_{0}^2 u_D}{\gamma_{1}^2}+\frac{\gamma_{3}^2}{u_D}\right) +  \frac{2 \gamma_{0}^2 \gamma_{3} |\bm{\tilde{k}}|^3 \cos (3 \theta )}{\gamma_{1} u_D} + \dots
\eea
Again, $u_D > 0$ can be tuned to generate an anomalously flat dispersion $E_{L=2}(\bm{\tilde{k}}) \sim O(k^3 \cos 3\theta)$ which has a $|\omega|^{-1/3}$ power-law divergence. This type of dispersion is called a ``monkey-saddle"~\cite{shtyk2017monkey}. Note that the divergence of the DOS at the higher-order $k^4 + k^5 \cos 3\theta$ vHS for R4G is even stronger. It may be called a ``monkey-bucket."

For $L = 3$ trilayer graphene, we find that there is no $u_D$ where $E_2$ can be tuned to zero. Hence there are no divergences in the DOS at the $K,K'$ points (although there are other vHSs elsewhere in the dispersion). For $L= 5$ pentalayer graphene, we find the same $O(|\bm{\tilde{k}}|^4)$ condition can be achieved as in tetralayer graphene. Specifically
\bea
E_{L=5}(\bm{\tilde{k}})&= (\text{$\Delta_2$}+2 u_D) +  \frac{\gamma_{0} (\gamma_{0} (\delta -2 \text{$\Delta_2$})+2 \gamma_{1} \gamma_{4}- u_D \gamma_{0} )}{\gamma_{1}^2} |\bm{\tilde{k}}|^2 + \frac{\gamma_{0}^4 \left(9 \gamma_{2}^2-16 u_D^2\right)}{16 \gamma_{1}^4 u_D}|\bm{\tilde{k}}|^4 + \dots \ .
\eea
We find similar behavior for hexalayer graphene. 

\begin{figure}[h]
	\centering
	\includegraphics[width=\textwidth]{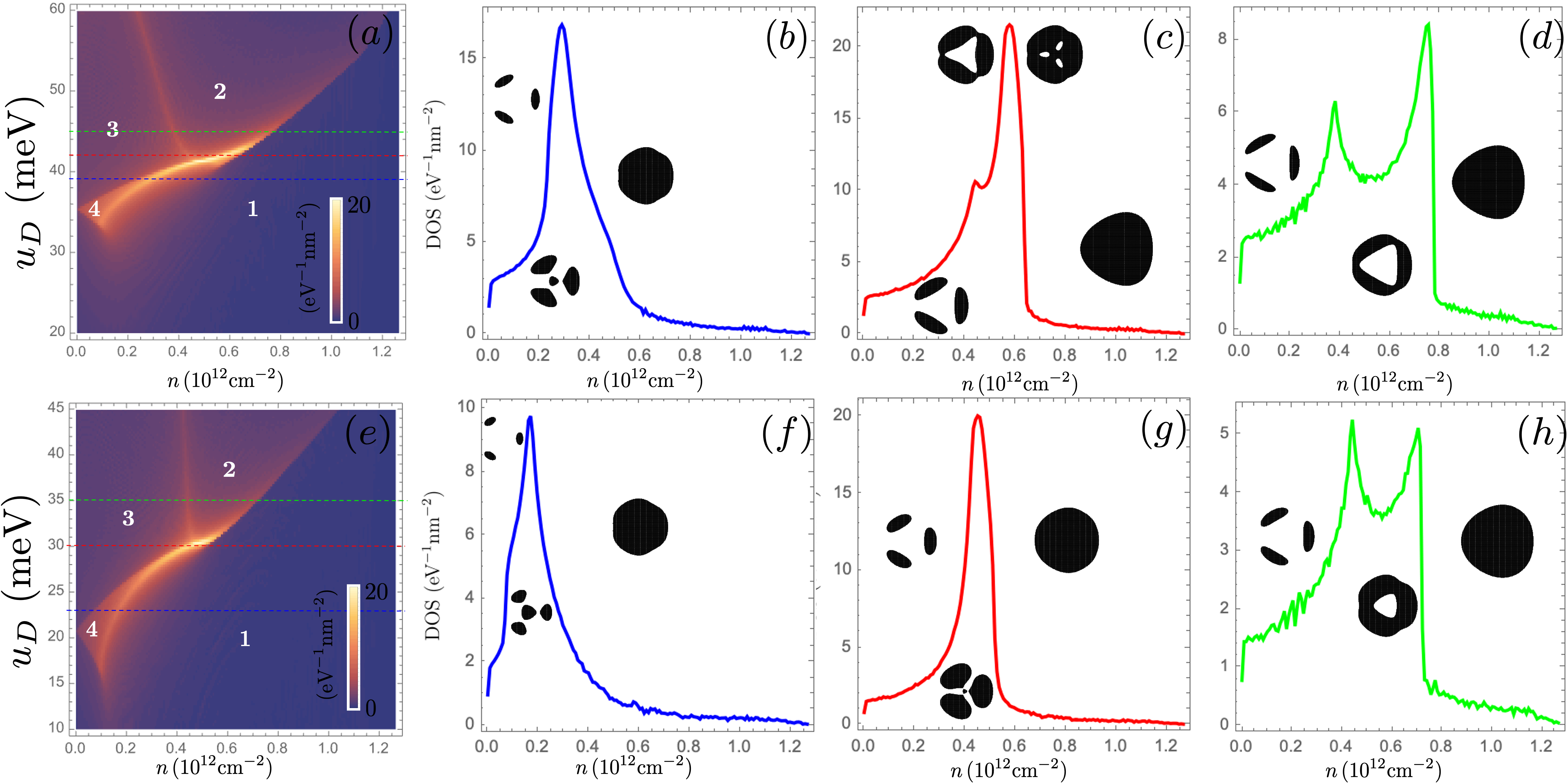}
	\caption{
		(a)  Non-interacting density of states as a function of $n$ and $u_D$  for the Set 1 parameters. We mark different types of Fermi surfaces as: single Fermi surface (1), annular (2), three-pocket  (3), and four-pocket (4). At a critical $u_D$, two logarithmic vHS collide at the higher order vHS at the $K$ point. The vHS line above the critical $u_D$ separates regions 2 and 3, while the step-edge feature separates regions 1 and 2. The lower vHS line separates regions 4 and 1. Another step-edge feature separates regions 3 and 4. (b)-(d) display line-cuts of the density of states at values of $u_D$ corresponding to the colored dashed lines in (a). Sample Fermi seas are shown as insets. %.1, .27, .6,[0.3,0.47,0.6, 1],.2 , .45, 1.
        (e)-(h) show the corresponding results for the Set 2 parameters. 
	}
	\label{fig:R4G_fermiology}
\end{figure}

Fig.~\ref{fig:R4G_fermiology} shows a map of the density of states for the Set 1 parameters with example Fermi surfaces. We observe multiple Fermi surface topologies, with transitions between them marked by singular behaviors in the DOS (either step edges when a new pocket is created, vHS at a saddle point, or higher order vHS at fine-tuned displacement fields as discussed above). We contrast the non-interacting DOS for both Set 1 and Set 2 parameters for R4G in Fig.~\ref{fig:R4G_fermiology}(a) and (e) respectively, where we have indicated with labels $1,2,3,4$ the different types of Fermiology (see caption). We also show fixed-$u_D$ line-cuts of the DOS, as well as illustrations of different types of Fermi seas.

% =============================================================================
\subsection{Magnetic Hamiltonian for rhombohedral graphene}\label{subsec:magnetic_ham}
% =============================================================================

We now define a continuum model for rhombohedral multilayer graphene at finite magnetic field, which has been studied previously by e.g.~Refs.~\cite{zhang2011magnetoelectric,slizovskiy2019films,auerbachIsospinMagneticTexture2025}. We consider an out of plane magnetic field $B$ and apply minimal substitution $k_j \to \hat{\pi}_j$ where $[\hat{\pi}_x, \hat{\pi}_y] = -i / \ell_B^2$ and $\ell_B = \sqrt{\hbar / e |B|}$ is magnetic length. For the $\tau$ valley, we then define ladder operators
\begin{align}
	\hat{a}_\tau &= - \frac{\ell_B}{\sqrt{2}}\left( \tau \hat{\pi}_x - i \hat{\pi}_y \right),  
	 &\hat{a}_\tau^\dagger &= - \frac{\ell_B}{\sqrt{2}}\left( \tau \hat{\pi}_x + i \hat{\pi}_y \right),  
\end{align}
so that $[\hat{a}_\tau, \hat{a}_{\tau}^\dagger] = -\tau$.

With these conventions, the dimensionless momenta of Eq.~\eqref{eq:scaled_RMG_momenta} become
\begin{equation}
    \tilde{k}^\pm\equiv \frac{\sqrt{3}a}{2}(k_x\pm ik_y)
    \to \hat{\pi}^\pm  = \frac{i\sqrt{2}}{\ell_B} \frac{\sqrt{3}a}{2}\hat{a}^{\mp \tau}.
\end{equation}
They behave like ladder operators in the Landau level basis $\ket{n,\tau}$ according to
\begin{align}
	\hat{\pi} \ket{n,K} = i \frac{\sqrt{2}}{\ell_B} \frac{\sqrt{3}a}{2} \sqrt{n+1} \ket{n+1,K}, \quad
	\hat{\pi}^\dagger \ket{n,K} =  (-i) \frac{ \sqrt{2}}{\ell_B} \frac{\sqrt{3}a}{2}  \sqrt{n-1} \ket{n+1,K},\\
	\hat{\pi} \ket{n,K'} = i \frac{\sqrt{2}}{\ell_B} \frac{\sqrt{3}a}{2} \sqrt{n} \ket{n-1,K'}, \quad
	\hat{\pi}^\dagger \ket{n,K'} =  (-i) \frac{ \sqrt{2}}{\ell_B} \frac{\sqrt{3}a}{2}  \sqrt{n+1} \ket{n+1,K'}.
\end{align}
With this redefinition of $\tilde{k}^\pm \to \hat{\pi}^\pm$, we may again use $\hat{h}_{0}^\tau$ from Eq.~\eqref{eq:microscopic_RMG_Hamiltonian}, which now acts on the basis $	\ket{n,\sigma,\ell,\tau}$ To avoid spurious quasi-zero modes from the chiral anomaly, we take a $(\sigma,\ell,\tau)$-dependent cutoff for the Landau level basis of
\begin{equation}
	N(\sigma,\ell,\tau) = \begin{cases}
		N_L + \ell -1 & (\sigma,\tau) = (A,K)\\
		N_L + \ell & (\sigma,\tau) = (B,K)\\
		N_L + \ell & (\sigma,\tau) = (A,K')\\
		N_L + \ell -1 & (\sigma,\tau) = (B,K').
	\end{cases}
\end{equation}
In practice we take $N_{L} = 150-250$ to accurately resolve Landau levels down to $\approx \SI{0.5}{\tesla}$.

% =============================================================================
\subsection{Landau level simulations}\label{sec:additional_LL_calcs}
% =============================================================================

\begin{figure*}[h]
   \centering 
   \includegraphics{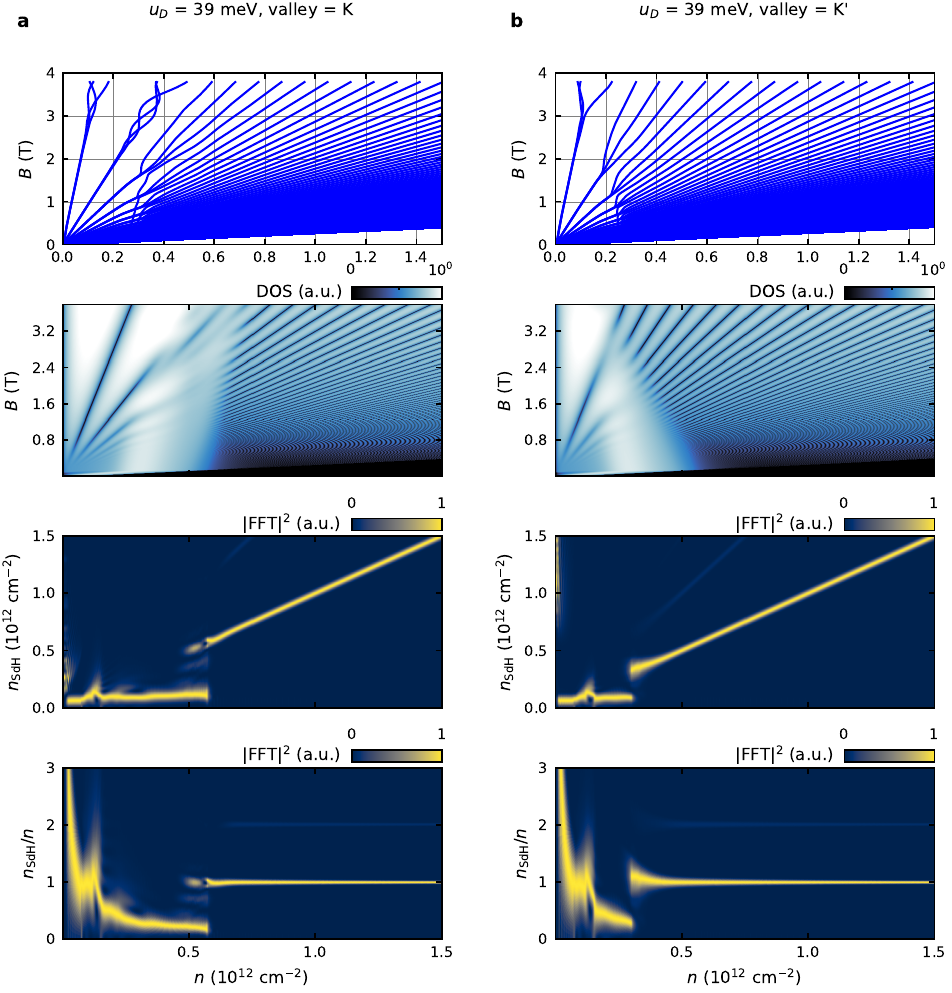}
   \caption{Theoretical Landau levels and corresponding FFTs as a function of density for $u_D = 39\,\si{\milli\electronvolt}$}
   \label{fig-theory-LL-39}
\end{figure*}

\begin{figure*}[h]
   \centering 
   \includegraphics{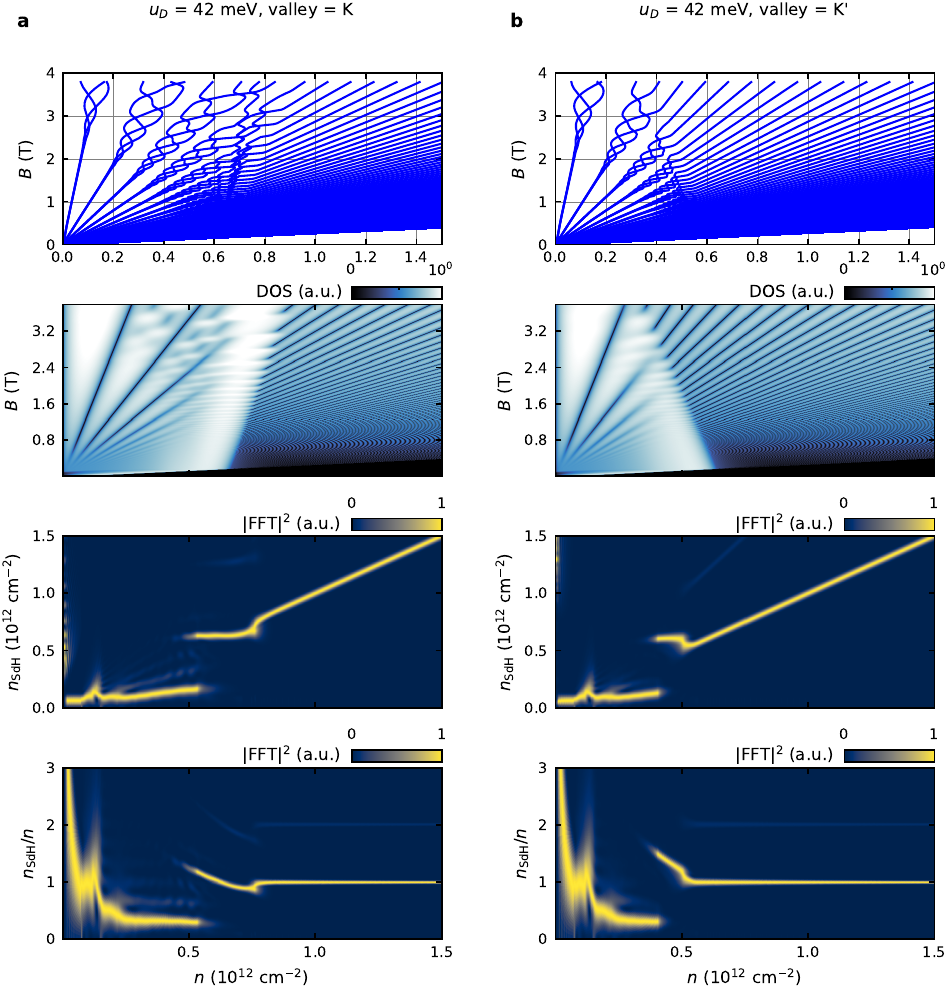}
   \caption{Theoretical Landau levels and corresponding FFTs as a function of density for $u_D = 42\,\si{\milli\electronvolt}$}
   \label{fig-theory-LL-42}
\end{figure*}

\begin{figure*}[h]
   \centering 
   \includegraphics{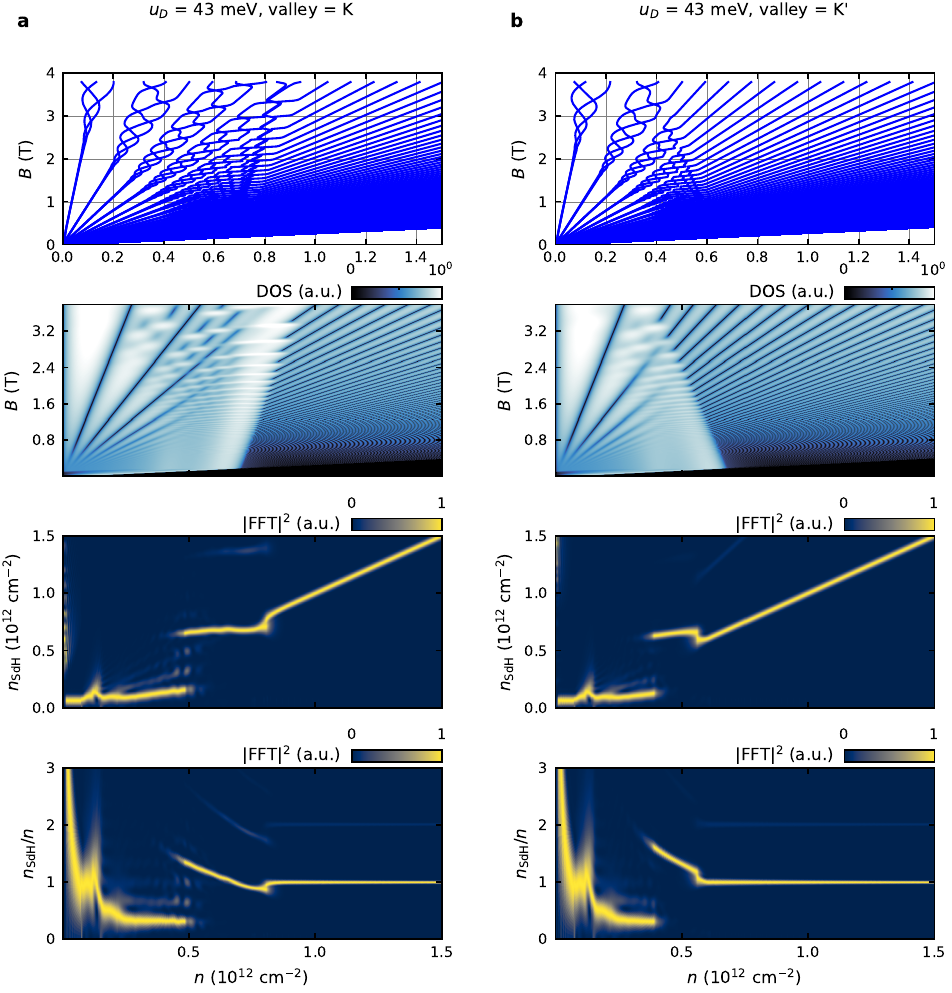}
   \caption{Theoretical Landau levels and corresponding FFTs as a function of density for $u_D = 43\,\si{\milli\electronvolt}$}
   \label{fig-theory-LL-43}
\end{figure*}

\begin{figure*}[h]
   \centering 
   \includegraphics{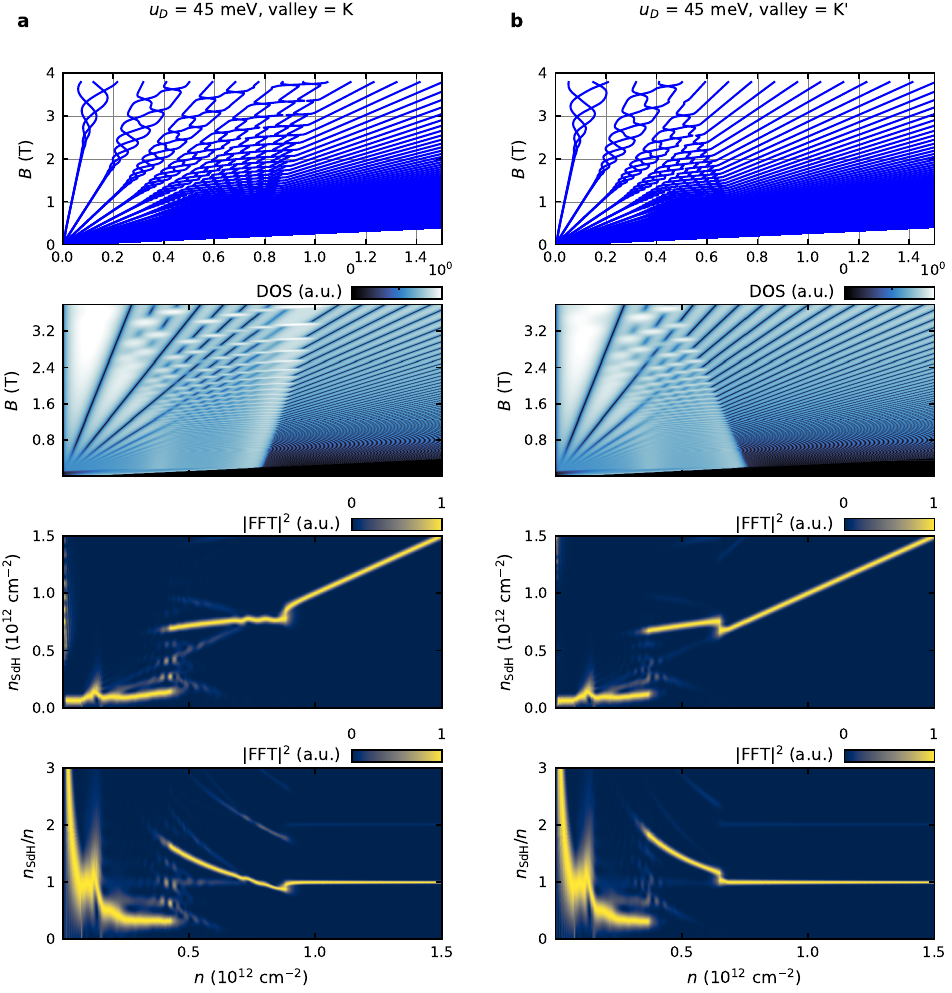}
   \caption{Theoretical Landau levels and corresponding FFTs as a function of density for $u_D = 45\,\si{\milli\electronvolt}$}
   \label{fig-theory-LL-45}
\end{figure*}

\begin{figure*}[h]
   \centering 
   \includegraphics{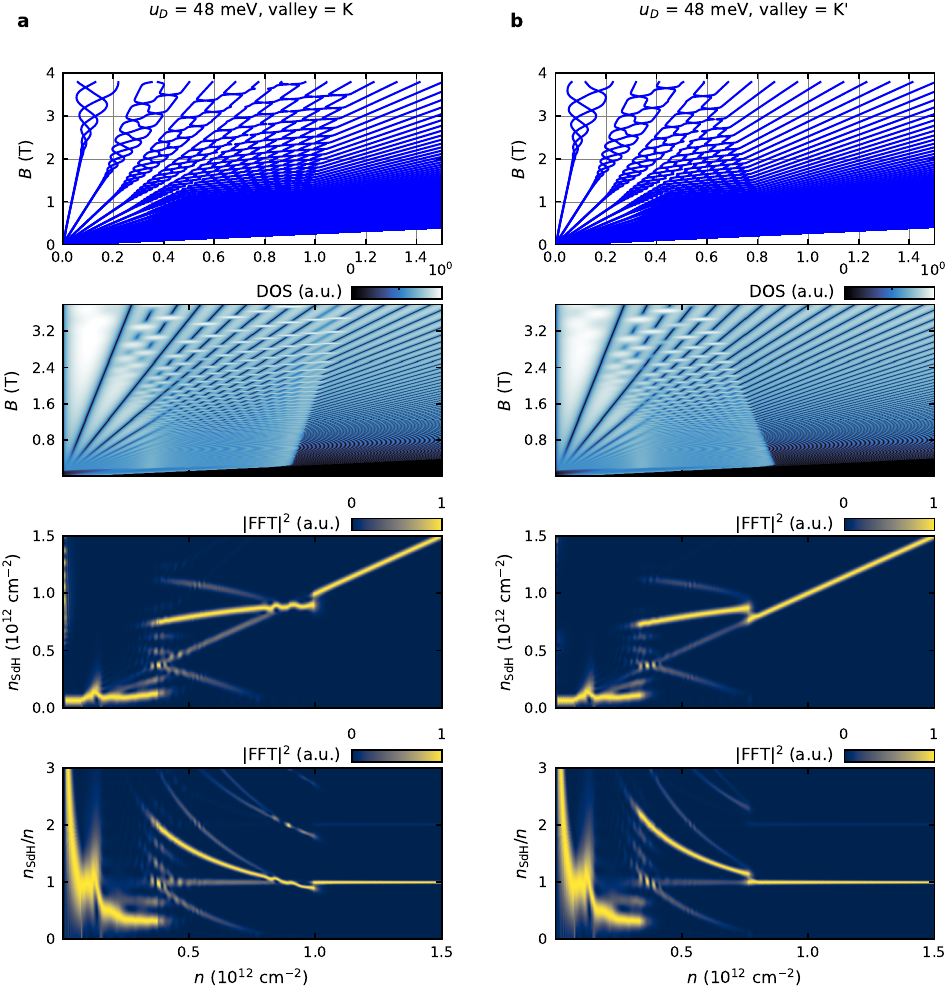}
   \caption{Theoretical Landau levels and corresponding FFTs as a function of density for $u_D = 48\,\si{\milli\electronvolt}$}
   \label{fig-theory-LL-48}
\end{figure*}

Figs.~\ref{fig-theory-LL-39},~\ref{fig-theory-LL-42},~\ref{fig-theory-LL-43},~\ref{fig-theory-LL-45}, and \ref{fig-theory-LL-48} show the resulting Landau level spectra for both valleys. All Landau level calculations use the Set 1 parameters (see Sec.~\ref{secapp:continuum_model}).

Using the same Fourier transform procedure as Fig.~3 of the main text, we then compute the frequencies of quantum oscillations in the density of states at the Fermi level. We use thermal broadening of $T = \SI{0.29}{K}$ when computing the density of states.  In all cases, the high density regime has a single frequency $n_{\mathrm{SdH}} = n$. Below the Lifshitz transition to an annular region (depending on $u_D$), another branch of Landau levels becomes visible that disperses starting not from zero but from finite density corresponding to the potential maximum at $K$. Part of this branch disperses towards lower density, but the highest slope is set by the valley, with a positive slope for $K$ and a negative slope for $K'$. This is consistent with the sign of the orbital magnetization. The intersections between the two branches of Landau levels create horizontal-in-density features resembling the experimental features. In the Fourier transform, these create roughly constant in density oscillation peaks in $n_{\mathrm{SdH}}$. However, only a single primary oscillation frequency is observed in either the three-pocket or annular regime. Secondary peaks arise from period-doubling and are much fainter. This is qualitatively inconsistent with the experimental SdH oscillations.

Severe magnetic breakdown can occur in the regime near the Lifshitz transitions. To visualize this, we resolve the Landau level wavefunctions in a coherent state basis in momentum space. As our model has continuous translation symmetry, we may resolve the wavefunction in the microscopic basis (restricted to $K$ for simplicity) $\ket{\mathsf{k},\sigma,\ell,K} = \sum_{n} U_{\mathsf{k},n} \ket{n,\sigma,\ell,K}$ where $\mathsf{k} = i \ell_B (k_x + i k_y)/\sqrt{2} \in \mathbb{C}$ and
\begin{equation}
\label{eq:LL_momentum_coherent_states}
	U_{\mathsf{k},n} = 	\braket{\mathsf{k}}{n} = \ell_B \frac{\mathsf{k}^n}{\sqrt{n!}} e^{- \mathsf{k} \overline{\mathsf{k}}/2}.
\end{equation}
(Guiding centers not involved due to continuous translations.) Care is needed to evaluate Eq.~\eqref{eq:LL_momentum_coherent_states} numerically due to catastrophic cancellations resulting in underflows and overflows. To avoid this, we use both direct evaluation and the asymptotic formula
\begin{equation}
	\braket{\mathsf{k} = r e^{i\theta}}{n}  
	= \frac{e^{i n \theta}}{(2\pi n)^{1/4}} \left( \frac{e r^2}{n} \right)^{n/2} e^{-r^2 /2 } e^{-\mu(n)/2},
\end{equation}
where $\mu(n) = \frac{1}{12 n} - \frac{1}{360 n^3} + \frac{1}{1260 n^5} - \cdots$ is the remainder term in Stirling's formula. For $\mathsf{k}$ of roughly order unity, the asymptotic formula is accurate to machine precision after about $n=15-20$. 

\begin{figure*}
   \centering 
   \includegraphics[width=\textwidth]{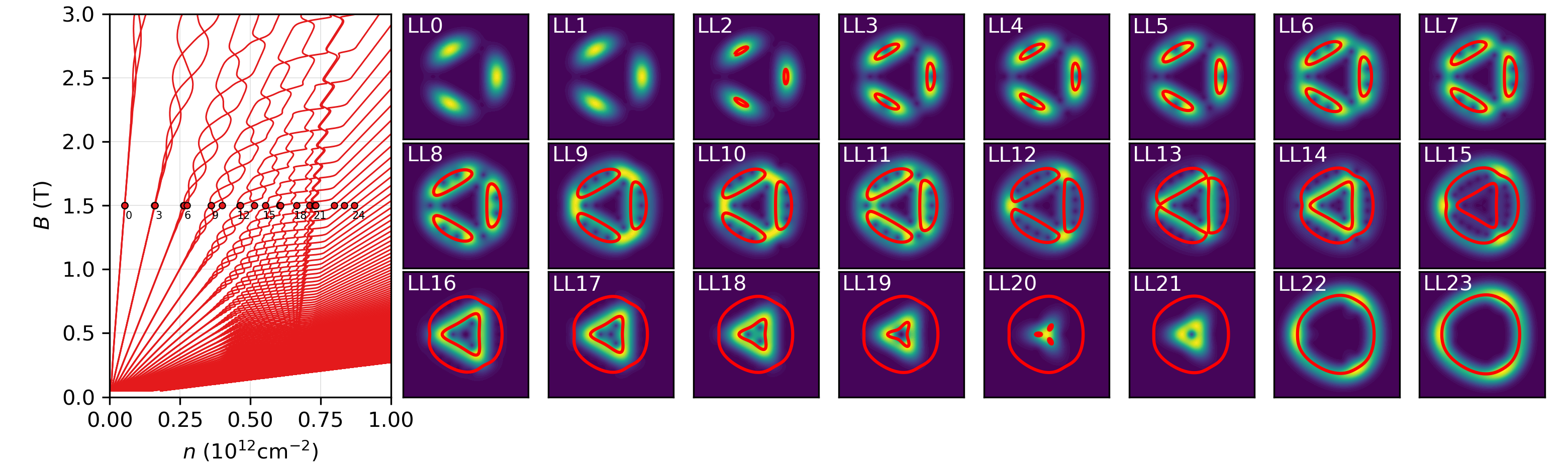}
   \caption{Landau level coherent states for $u_D =\SI{43}{meV}$ at $B = \SI{1.5}{T}$. The individual Landau levels shown are marked with red dots in the Landau fan on the left. Red contours show the $B=0$ Fermi surface at the same densities. Note that due to Streda slopes, the $B=0$ contours are offset to higher or lower density than the corresponding Landau levels. Computed with $N_{LL}=150$ in the $K$ valley.}
   \label{fig-theory-coherent_states}
\end{figure*}

Fig.~\ref{fig-theory-coherent_states} shows the $k$-resolved wavefunctions for the 24 lowest conduction Landau levels in the coherent state basis.  The uncertainty or characteristic spread for the wavefunctions are on the expected inverse magnetic length level. Strong breakdown effects are present in the interference region, such as Landau levels localized not along Fermi surfaces but in the regions between different pockets (see e.g.~LL10 and LL11).

\clearpage
\newpage
% =============================================================================
\section{Self-consistent Hartree-Fock calculations}\label{supp_sec:SCHF}
% =============================================================================

% =============================================================================
\subsection{Formalism}\label{app:HF_formalism}
% =============================================================================

Since we are mainly interested in moderate/large displacement fields and dilute electron doping $n>0$, we consider the interacting continuum model projected to the lowest conduction band
\begin{equation}
    \hat{H}_{\text{con}.}=\sum_{\bm{k} \tau s}\varepsilon_\tau(\bm{k})d^\dagger_{\bm{k},\tau,s}d_{\bm{k},\tau,s}+\frac{1}{2 A}\sum_{\bm{k}\bm{k}'\bm{q}}\sum_{\tau\tau'ss'}V(q)\lambda_{\tau}(\bm{k},\bm{q})\lambda_{\tau'}(\bm{k}',-\bm{q})d^\dagger_{\bm{k},\tau,s}d^\dagger_{\bm{k}',\tau',s'}d_{\bm{k}'-\bm{q},\tau',s'}d_{\bm{k}+\bm{q},\tau,s},
\end{equation}
where $A$ is the total system area, and $d^\dagger_{\bm{k},\tau,s}$ is the creation operator for the lowest conduction band in valley $\tau$ and spin $s$.
$\varepsilon_\tau(\bm{k})$ is the corresponding kinetic energy obtained from diagonalizing $h_0^\tau(\bm{k})$ in App.~\ref{secapp:continuum_model} (which may include the internal screening effects discussed later in App.~\ref{secapp:interlayer_screening}), and the form factor $\lambda_\tau(\bm{k},\bm{q})=\langle{u_\tau(\bm{k})}|{u_\tau(\bm{k}+\bm{q})}\rangle$ encodes the overlaps of the Bloch vectors.
We neglect spin-orbit coupling and intervalley Hund's couplings that are needed to break the degeneracy between various partially-polarized phases.
However, such couplings are not expected to significantly affect the boundaries between phases with different levels of flavor polarization, or the Fermiology within each phase.
Hence we only include the (dual-gate-screened) long-range density-density interaction with interaction potential
\begin{equation}
    V(q)=\frac{\frac{e^2}{2\epsilon_0\epsilon q}\tanh(qd)}{1+\frac{q_\text{TF}}{q}\tanh(qd)},
\end{equation}
where $d=20\,$nm is the gate-screening parameter, and $\epsilon$ is the relative permittivity which captures screening from the hBN dielectric as well as remote bands.
For simplicity, we have used a layer-independent $V(q)$, which is expected to be a good approximation for large $D$ where the low-energy conduction states are mainly localized to the top layer.
(Note that the interlayer screening of App.~\ref{secapp:interlayer_screening} incorporates $\bm{q}=0$ layer-dependent effects).
Following Ref.~\cite{vituri2025incommensurate}, we have also included phenomenological Thomas-Fermi screening with wavevector $q_\text{TF}=0.04/a$. Since there is no consensus on the most appropriate value of $\epsilon$ in HF calculations of rhombohedral graphene, we will treat $\epsilon $ as a theoretical tuning parameter.

The self-consistent HF procedure involves finding the HF density matrix $P_{\tau,\tau',s}(\bm{k})=\langle d^\dagger_{\bm{k},\tau,s}d_{\bm{k},\tau',s} \rangle$ that minimizes the total energy.
Note that our parameterization of $P_{\tau,\tau',s}(\bm{k})$ is limited to states that preserve spin-$U(1)$ symmetry, and permits intervalley coherence (IVC) but only at intervalley wavevector $\bm{q}_\text{IVC}=0$, which connects the $K$ and $K'$ points. $P_{\tau,\tau',s}(\bm{k})$ further preserves continuous translation symmetry, i.e.~our calculations do not include Wigner crystals or stripe states.
For the majority of the calculations (main text Fig.~\ref{fig1}e, and App.~\ref{subsubsection:HF_flavor_phase} and \ref{subsubsection:quarter_metal_phase}), we do not allow any IVC, i.e.~$P_{\tau,\tau',s}(\bm{k})$ is constrained to vanish for $\tau\neq \tau'$. In App.~\ref{subsection:intervalley_coherence}, we consider $\bm{q}_\text{IVC}=0$ IVC phases.

% =============================================================================
\subsection{Interlayer Hartree screening}\label{secapp:interlayer_screening}
% =============================================================================

The most straightforward way to incorporate the external displacement field $D$ is to model it as a linearly varying layer potential that satisfies $u_D\equiv V_{l+1}-V_l=edD/\epsilon_0\epsilon_\perp$, where $d=3.33\,$\r{A} is the interlayer graphene spacing.
For the perpendicular dielectric constant, we use $\epsilon_\perp\simeq 3.5$ appropriate for the hBN substrate.
This simple approach has two limitations.
First, the relation $u_D\equiv V_{l+1}-V_l=edD/\epsilon_0\epsilon_\perp$ significantly overestimates the effective value of $u_D$ that is present in R4G for a given experimentally-applied $D$.
This is because the displacement field drives a charge imbalance localized near the outermost graphene layers, which will tend to screen $D$.
In the conventions used here, a positive $D$ pointing upwards generates a net excess of electrons (holes) on the bottom (top) layer, which will induce an electric field pointing downwards.
(Note that the doped electrons for $n>0$ are localized towards the top layer.)
Second, the screened layer potentials generally vary non-linearly in the layer index, since the charge imbalance is not strictly localized on the outermost graphene layers.
This effect is expected to be more important for smaller values of displacement field.

To account for these screening effects, in some of the HF calculations, we first implement a self-consistent interlayer Hartree calculation to estimate the screened layer potentials $V_l(D)$, following the formalism of Ref.~\cite{Kolar2026singlegatetracking}.
The experimentally relevant parameters are the areal carrier density $n=\sum_l n_l$ and the displacement field $D=e\frac{n_t-n_b}{2}$, where we have introduced the layer-resolved graphene densities $n_l$, and the top and bottom gate densities $n_t$ and $n_b$.
These are all measured relative to a charge neutral reference point.
Overall charge neutrality requires $n_t+n_b+\sum_l n_l=0$.
By applying Gauss's law around layer $l$, we can derive that the vertical electric field $E_l$ between layers $l$ and $l+1$ is
\begin{equation}
    E_l=-e\frac{n_b+\sum_{j\leq l}n_l}{\epsilon_0\epsilon_\perp}.
\end{equation}
The interlayer potential differences are $V_{l+1}-V_{l}=eE_ld$.
By substituting $n_e,D$ for $n_b,n_t$, we obtain the recursion relation
\begin{equation}\label{SIeq:Vl_recursion}
    V_{l+1}-V_l=\frac{edD}{\epsilon_0\epsilon_{\perp}}+\frac{e^2 d}{2\epsilon_0\epsilon_{\perp}}\left(n-2\sum_{j\leq l}n_l\right).
\end{equation}
These equations can be solved iteratively given a prescription for determining the layer densities $n_l$ from a set of layer potentials $V_l$.
In this work, we first diagonalize the R4G continuum model with layer potentials $V_l$.
Since we are interested in $n>0$, we occupy all valence band states in all flavors.
We further occupy the lowest conduction band states up to the target density $n$.
The occupied conduction band states are either fully spin-valley polarized (corresponding to $N_{\text{scr.\,con.}}=1$), or symmetrically distributed across all flavors (corresponding to $N_{\text{scr.\,con.}}=4$).

The screened layer potentials $V_l$ are then used as inputs to the self-consistent HF calculations for the same values of $n$ and $D$.
For each HF calculation, we will indicate the flavor distribution of the conduction states within the screening procedure, though we find that the final results with either choice are qualitatively similar.
Note that our HF calculations do not enable `feedback' to the interlayer screening.
For example, the potentials $V_l$ remain fixed even if the HF procedure leads to non-trivial restructuring of the Fermi surfaces due to Fock exchange effects.
However, such considerations are not expected to affect qualitative aspects of the phase diagram.

For the screening calculations, we utilize a triangular momentum mesh with mesh spacing $0.04\,\text{nm}^{-1}$ and a circular cutoff of radius $1.5\,\text{nm}^{-1}$ (corresponding to 5101 momentum points).
Note that the momentum cutoff needs to be significantly larger than the typical size of Fermi surfaces for the experimentally relevant density range $n\lesssim 10^{12}\,\text{cm}^{-2}$ to obtain converged results.

% =============================================================================
\subsection{Results}\label{subsection:HF_results}
% =============================================================================

% =============================================================================
\subsubsection{Flavor phase diagram}\label{subsubsection:HF_flavor_phase}
% =============================================================================

\begin{figure*}
   \centering 
   \includegraphics[width=1\textwidth]{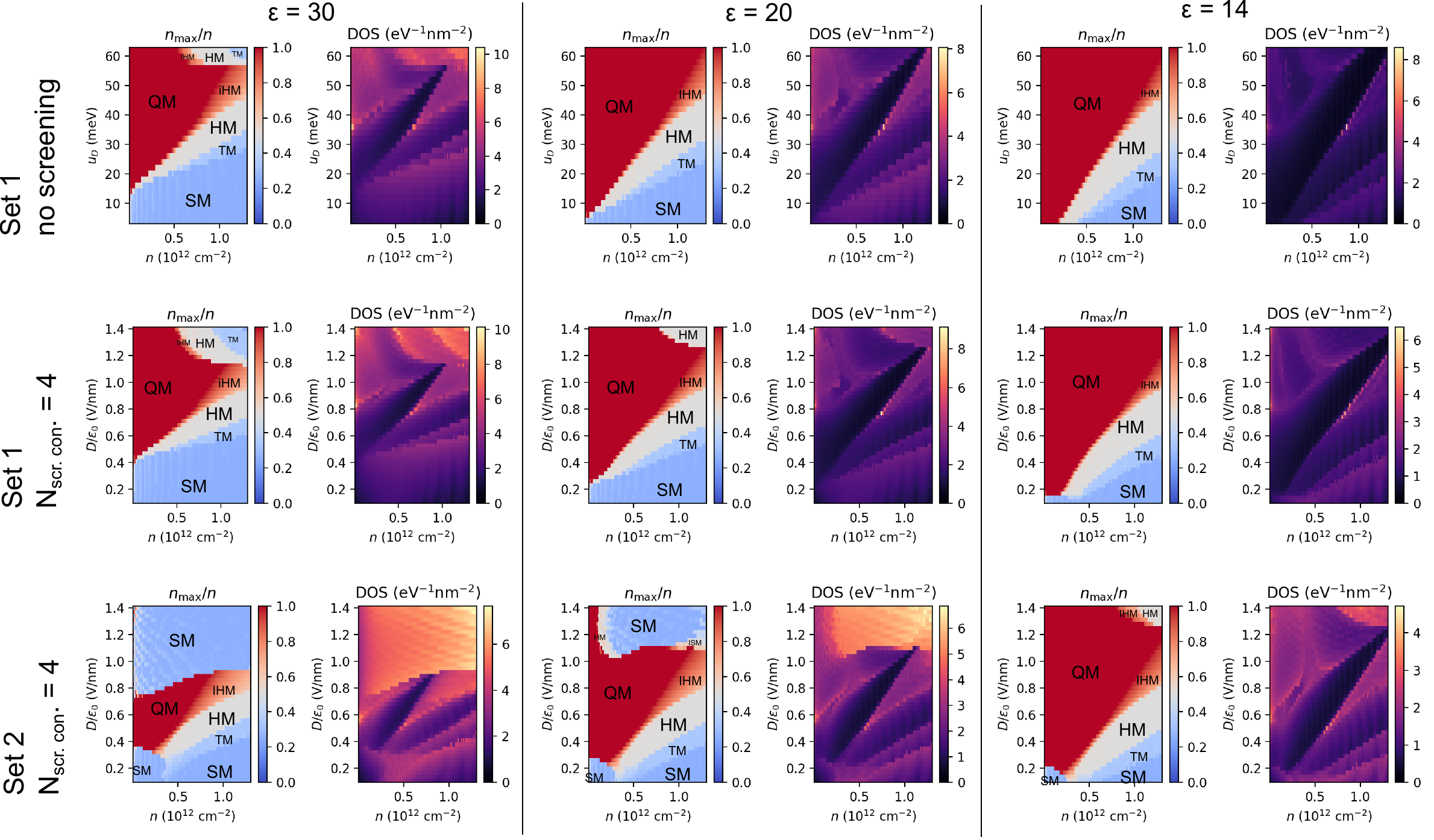}
   \caption{Flavor phase diagram computed using self-consistent HF calculations for dielectric constants $\epsilon=30,20,14$. The top and middle rows use the Set 1 parameters (see App.~\ref{secapp:continuum_model}). The top row uses a linear interlayer potential $u_D$, while the middle and bottom rows use  Hartree-screened layer potentials with conduction band degeneracy $N_\text{scr.\,con.}=4$ in the screening calculation. The bottom row uses the Set 2 parameters. }
   \label{HF_multi_combined}
\end{figure*}

We perform self-consistent HF calculations in order to determine the flavor polarization of the ground state.
Here, we restrict to states that preserve valley number and spin $S_z$ (IVC states will be discussed in App.~\ref{subsection:intervalley_coherence}).
A simple measure of the degree of flavor polarization is the ratio $n_\text{max}/n$, where $n_{\text{max}}$ is the density of the highest-occupied flavor.
A symmetric phase with all flavors populated equally corresponds to $n_\text{max}/n=1/4$, while a fully-polarized phase corresponds to $n_\text{max}/n=1$. We do not enforce $C_{3z}$ symmetry.
We also compute the DOS at the Fermi level of the HF band structure. In the DOS computation, we use Lorentzian broadening with linewidth $0.2\,$meV.
For the HF, we utilize a triangular momentum mesh with mesh spacing $0.015\,\text{nm}^{-1}$ and a circular cutoff of radius $0.5\,\text{nm}^{-1}$ (corresponding to 4033 momentum points). For Fig.~\ref{fig1}e in the main text, because we perform calculations to a larger maximum electron density $n=1.8\times 10^{12}\,\text{cm}^{-2}$, we use a larger circular cutoff of radius $0.65\,\text{nm}^{-1}$ (corresponding to 6817 momentum points) in order to accommodate all the candidate phases. Fig.~\ref{fig1}e uses the Set 1 parameters, a linear interlayer potential, and $\epsilon=30$.

In Fig.~\ref{HF_multi_combined}, we present flavor phase diagrams for three different sets of single-particle models (given by the three rows), and three dielectric constants $\epsilon=30,20,14$ (given by the three columns). The $n_\text{max}/n$ subplot contains labels for the different phases:
\begin{itemize}
    \item A quarter metal (QM) contains one occupied flavor.
    \item A half metal (HM) contains two equally occupied flavors.
    \item A triple metal (TM) contains three equally occupied flavors.
    \item A symmetric metal (SM) contains four equally occupied flavors.
\end{itemize}
A prefix `I' indicates that the populations in the occupied flavors are not equal.

In the first row of Fig.~\ref{HF_multi_combined}, we first consider the Set 1 parameters in App.~\ref{secapp:continuum_model} and use a linear interlayer potential $u_D$ to model the external displacement field. For all interaction strengths considered, the phase diagram is dominated by a sequence of phase transitions that slope upwards in the $n$-$u_D$ plane. At small $u_D$ or large $n$, the ground state is Sym, i.e.~all flavors equally occupied. As the interlayer potential is increased, which flattens the bottom of the conduction band and increases the non-interacting DOS, we find the sequence $\text{SM}\rightarrow\text{TM}\rightarrow\text{HM}\rightarrow\text{QM}$. This generates an associated `resetting' of the interacting DOS, which may be connected to the resistance features in the experiment (see main text Fig.~\ref{fig1}d). Note that TM occupies a narrow region compared to the other phases, and there is no clear evidence for TM in the experiment. Between QM and HM, we also find IHM, which corresponds to an imbalanced half metal. The transition between QM and IHM appears to be continuous within our mean-field calculations. For increasing interaction strength (smaller $\epsilon$), QM occupies an increasingly larger region of the phase diagram. For $\epsilon=30$, we also observe partially isospin-polarized phases re-emerging at the largest layer potentials $u_D\gtrsim 56\,\text{meV}$.

In the second row of Fig.~\ref{HF_multi_combined}, we continue to use the Set 1 parameters, but incorporate interlayer screening effects according to the prescription of App.~\ref{secapp:interlayer_screening}. Inclusion of interlayer screening enables a more direct comparison of the vertical axis with the experimental displacement field $D$. The displacement field scales of the major flavor transitions in HF are broadly in agreement with experiment. However, we caution that the precise positions of the phases in HF are sensitive to both the value of the effective in-plane dielectric constant $\epsilon$, as well as the perpendicular dielectric constant $\epsilon_\perp$ used in the screening calculation. Furthermore, quantum fluctuations beyond HF are expected to further renormalize the phase boundaries. Hence, we refrain from making a more quantitative comparison of the positions of the flavor transitions. The shape of the phase diagram is similar to that obtained using a linear interlayer potential (first row), implying that non-linear interlayer screening effects are not important for the regime of interest. 

In the bottom row of Fig.~\ref{HF_multi_combined}, we use Set 2 parameters with interlayer screening. Compared to Set 1 (middle row), the flavor transitions move to lower $D$. Notably, a pocket of SM emerges at small $D$ and small $n$.

% =============================================================================
\clearpage
\newpage
\subsubsection{Quarter metal phase diagram}\label{subsubsection:quarter_metal_phase}
% =============================================================================

\begin{figure*}
   \centering 
   \includegraphics[width=0.6\textwidth]{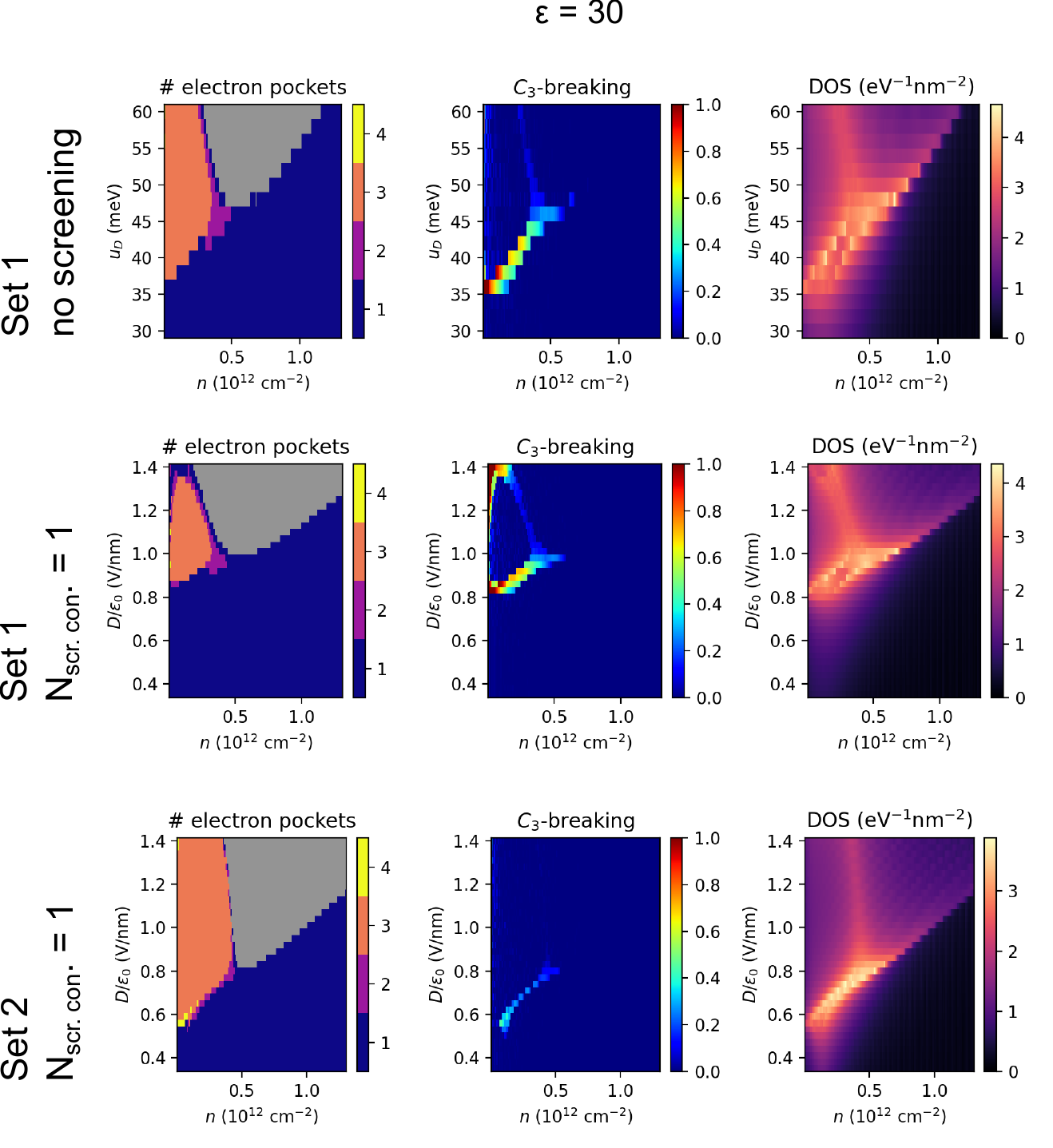}
   \caption{Phase diagram assuming full spin- and valley-polarization (i.e.~a QM), computed using self-consistent HF calculations for dielectric constant $\epsilon=30$. The top and middle rows use the Set 1 parameters (see App.~\ref{secapp:continuum_model}). The top row uses a linear interlayer potential $u_D$, while the middle row uses  Hartree-screened layer potentials with conduction band degeneracy $N_\text{scr.\,con.}=1$ in the screening calculation. The bottom row uses the Set 2 parameters. In the subplots labelled `\# electron pockets', the grey region indicates a single annular Fermi surface. The $C_3$-breaking order parameter is defined in Eq.~\ref{appeq:O_C3}.}
   \label{HF_single_epsr30}
\end{figure*}

\begin{figure*}
   \centering 
   \includegraphics[width=0.6\textwidth]{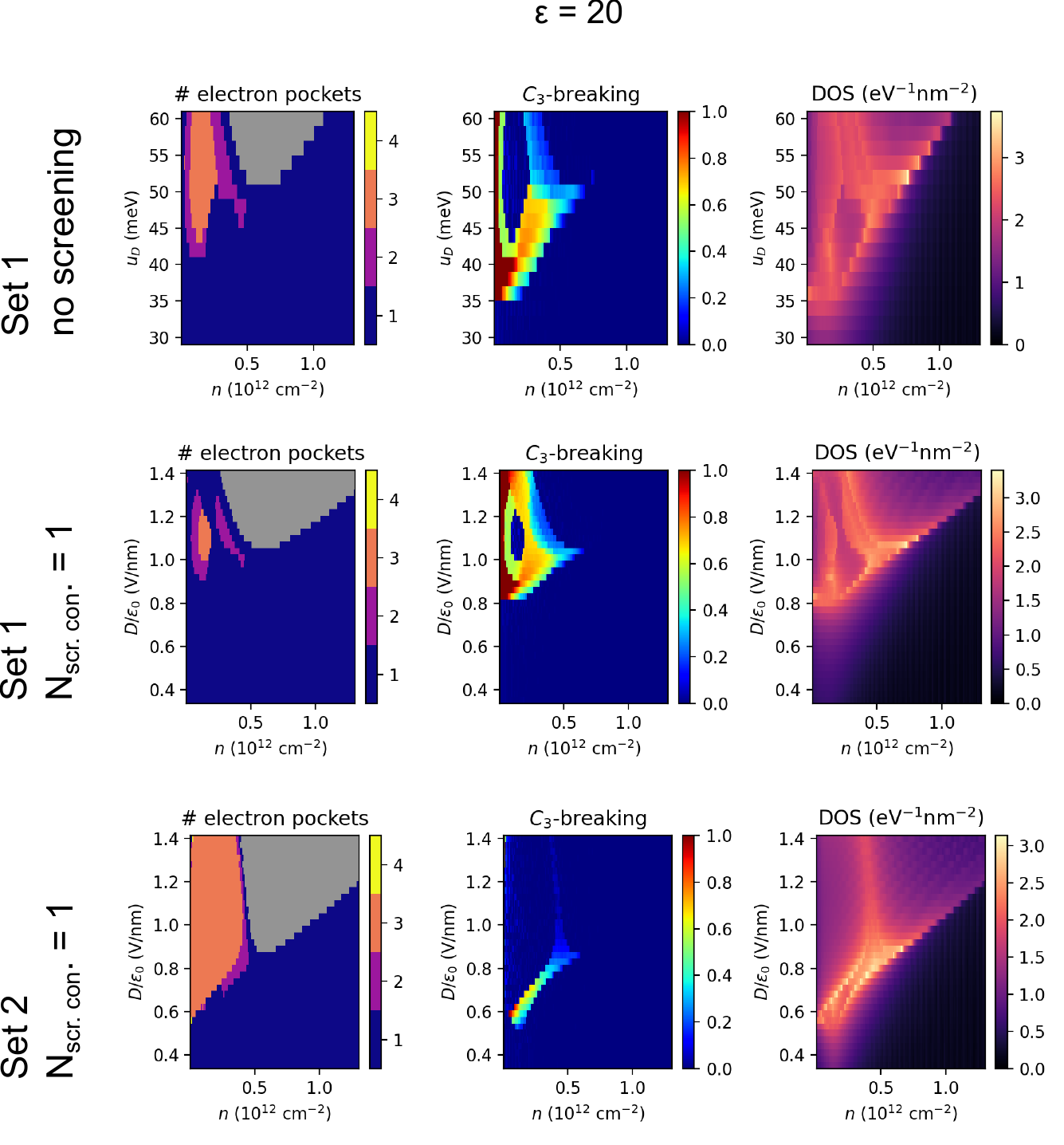}
   \caption{Phase diagram assuming full spin- and valley-polarization (i.e.~a QM), computed using self-consistent HF calculations for dielectric constant $\epsilon=20$. The top and middle rows use the Set 1 parameters (see App.~\ref{secapp:continuum_model}). The top row uses a linear interlayer potential $u_D$, while the middle row uses  Hartree-screened layer potentials with conduction band degeneracy $N_\text{scr.\,con.}=1$ in the screening calculation. The bottom row uses the Set 2 parameters. In the subplots labelled `\# electron pockets', the grey region indicates a single annular Fermi surface. The $C_3$-breaking order parameter is defined in Eq.~\ref{appeq:O_C3}.}
   \label{HF_single_epsr20}
\end{figure*}

\begin{figure*}
   \centering 
   \includegraphics[width=0.6\textwidth]{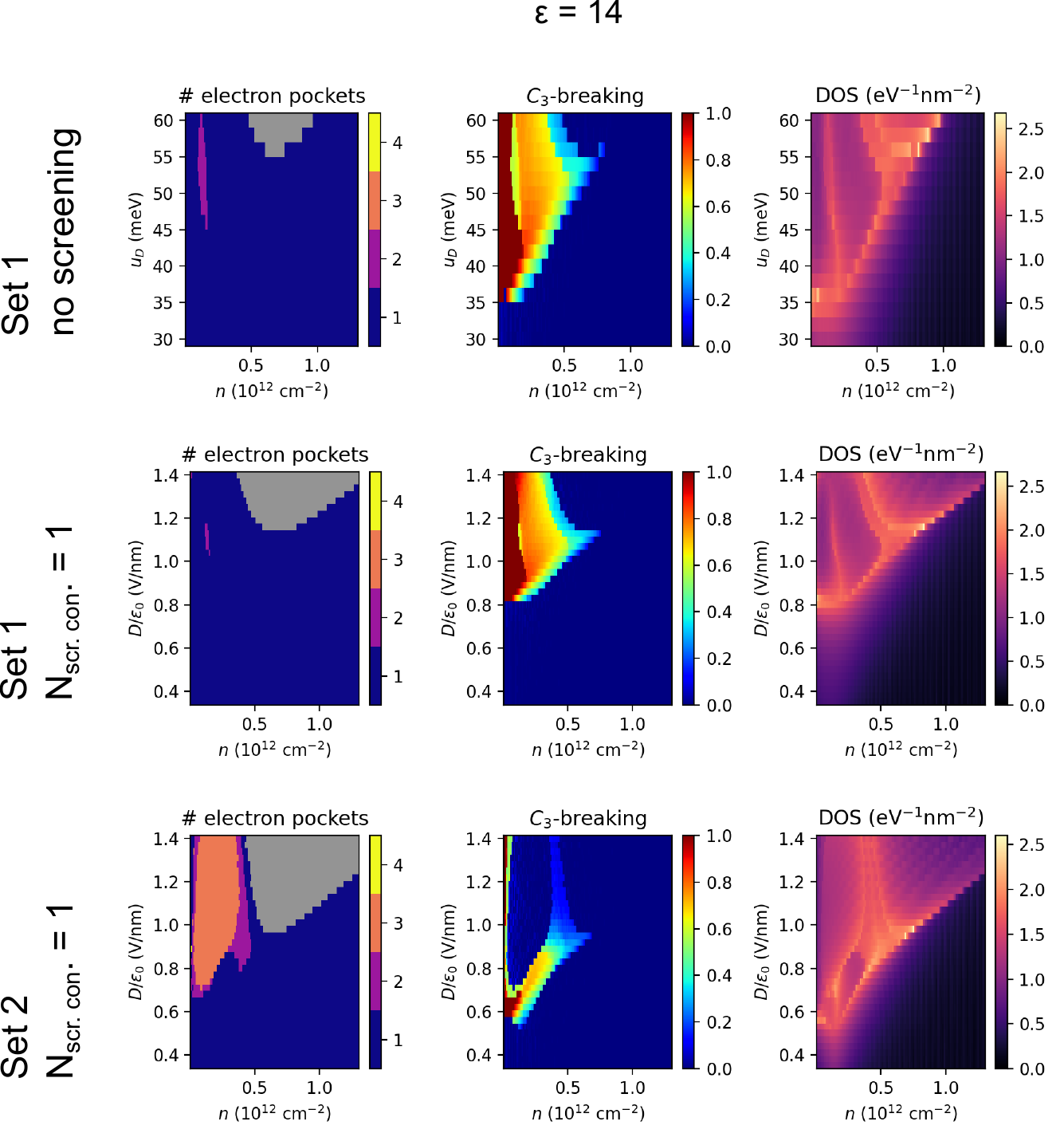}
   \caption{Phase diagram assuming full flavor polarization (i.e.~QM), computed using self-consistent HF calculations for dielectric constant $\epsilon=14$. The top and middle rows use the Set 1 parameters (see App.~\ref{secapp:continuum_model}). The top row uses a linear interlayer potential $u_D$, while the middle row uses  Hartree-screened layer potentials with conduction band degeneracy $N_\text{scr.\,con.}=1$ in the screening calculation. The bottom row uses the Set 2 parameters. In the subplots labelled `\# electron pockets', the grey region indicates a single annular Fermi surface. The $C_3$-breaking order parameter is defined in Eq.~\ref{appeq:O_C3}.}
   \label{HF_single_epsr14}
\end{figure*}

To better understand the Fermiology in the fully-polarized quarter metal (QM), we zoom in on the region at high displacement fields and restrict to HF states polarized in valley $K$ and spin $\uparrow$. 
The HF solution can be characterized by the electron occupation function $n(\bm{k})=0,1$.
We diagnose the Fermiology by computing the number of connected electron pockets.
To detect annular Fermi surfaces, we also compute the number of hole pockets.
In our calculations, the only solutions that contain hole pockets correspond to a single annular Fermi surface.
To quantify nematicity, we define a $C_3$-breaking order parameter according to
\begin{equation}\label{appeq:O_C3}
    O_{C_3}=1-\frac{\sum_{\bm{k}}n(\bm{k})n(C_3\bm{k})}
    {\sum_{\bm{k}}n(\bm{k})}.
\end{equation}
A $C_3$-symmetric Fermi surface corresponds to $O_{C_3}=0$, while a maximally $C_3$-breaking Fermi surface (e.g.~one where if $\bm{k}$ is occupied, then neither $C_3\bm{k}$ nor $C_3^2\bm{k}$ are occupied) corresponds to $O_{C_3}=1$.
As in App.~\ref{subsubsection:HF_flavor_phase}, we compute the DOS at the Fermi level using the HF band structure.
We utilize a triangular momentum mesh with mesh spacing $0.01\,\text{nm}^{-1}$ and a circular cutoff of radius $0.5\,\text{nm}^{-1}$ (corresponding to 9055 momentum points).

In Fig.~\ref{HF_single_epsr30}, we present HF results for three different sets of single-particle parameters (given by the three rows), and dielectric constant $\epsilon=30$. Fig.~\ref{HF_single_epsr20} and \ref{HF_single_epsr14} provide analogous results for $\epsilon=20,14$ respectively. In the `\# electron pockets' subplot, we show the number of electron pockets. The grey region indicates the annular regime. 

The first row of Figs.~\ref{HF_single_epsr30}-\ref{HF_single_epsr14} corresponds to the Set 1 parameters with linear interlayer potential. At the smallest $u_D$, the ground state consists of a single connected pocket. At larger $u_D$, two new dominant phases occur that descend directly from the non-interacting Fermiology: a three-pocket regime, and an annular regime. However, interactions lead to several notable effects. First, $C_3$-breaking nematic states appear with qualitatively distinct Fermiology, especially at the boundary of the three-pocket regime. These mainly consist of either one or two connected Fermi surfaces, which will be described in more detail below. Second, the arcs of vHSs in the non-interacting model are split, and the peak DOS is suppressed. Finally, the annular region shifts to higher $u_D$ with increasing interaction strength. This occurs due to Fock renormalization near the Dirac momentum in the occupied valley, which tends to lower the self-energy and partially counteract the effect of the interlayer potential. The inclusion of interlayer screening effects in the second row  does not qualitatively impact the phase diagram. In the third row, which uses the Set 2 parameters, the main difference is that nematic states take up a significantly smaller region of the $n$-$D$ plane. This occurs because compared with Set 1, the non-interacting dispersion with Set 2 is not as flat over an extended region of momentum space.

\begin{figure*}
   \centering 
   \includegraphics[width=0.9\textwidth]{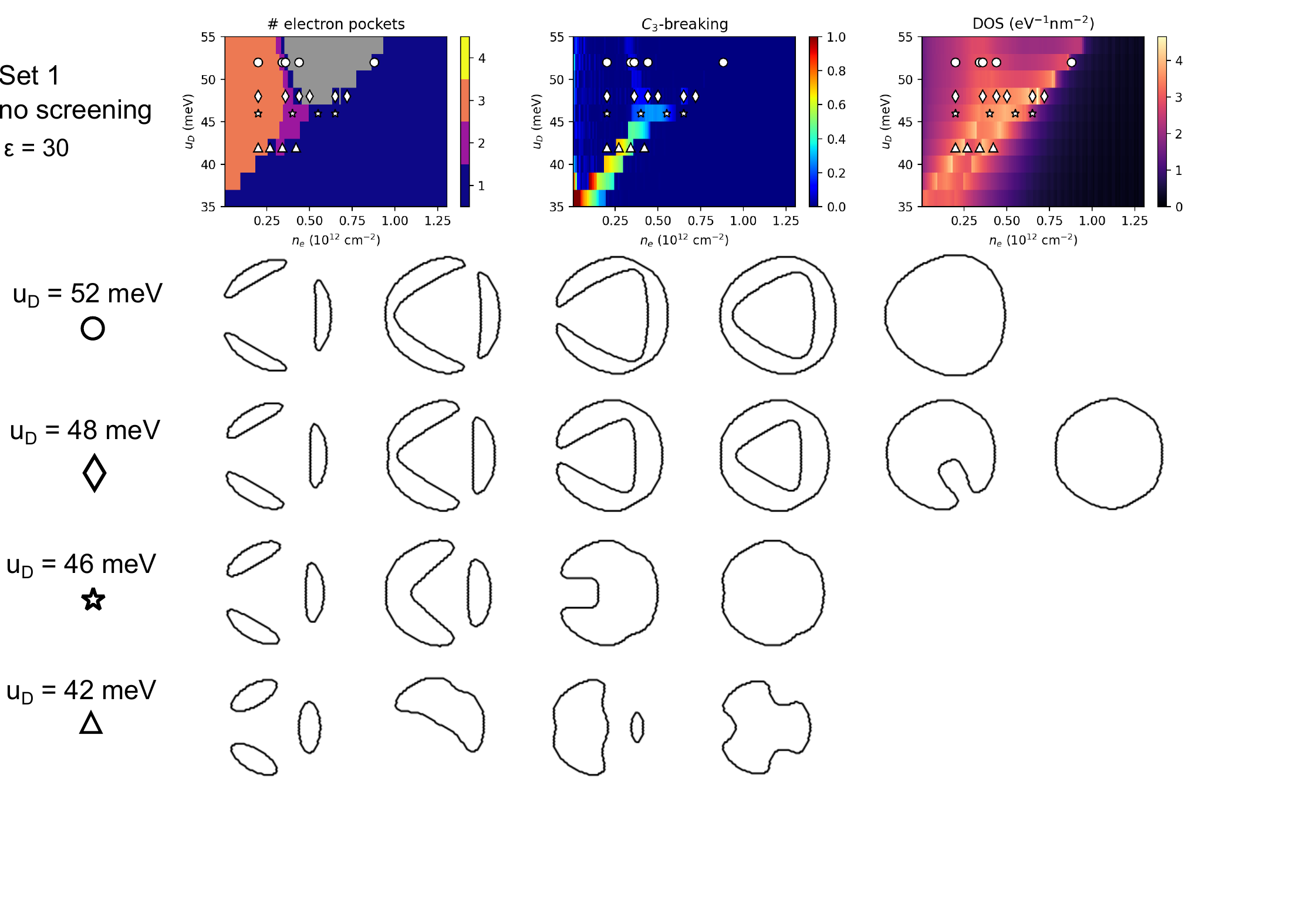}
   \caption{Fermi surfaces computed using self-consistent HF calculations that assume full spin-valley polarization. The calculations use the Set 1 parameters (see App.~\ref{secapp:continuum_model}), a dielectric constant $\epsilon=30$, and a linear interlayer potential $u_D$. For each interlayer potential listed, the Fermi surfaces are ordered from left to right according to the positions of the symbols on the phase diagrams. }
   \label{HF_single_unscreened_set1_eps30_FS}
\end{figure*}

\begin{figure*}
   \centering 
   \includegraphics[width=0.9\textwidth]{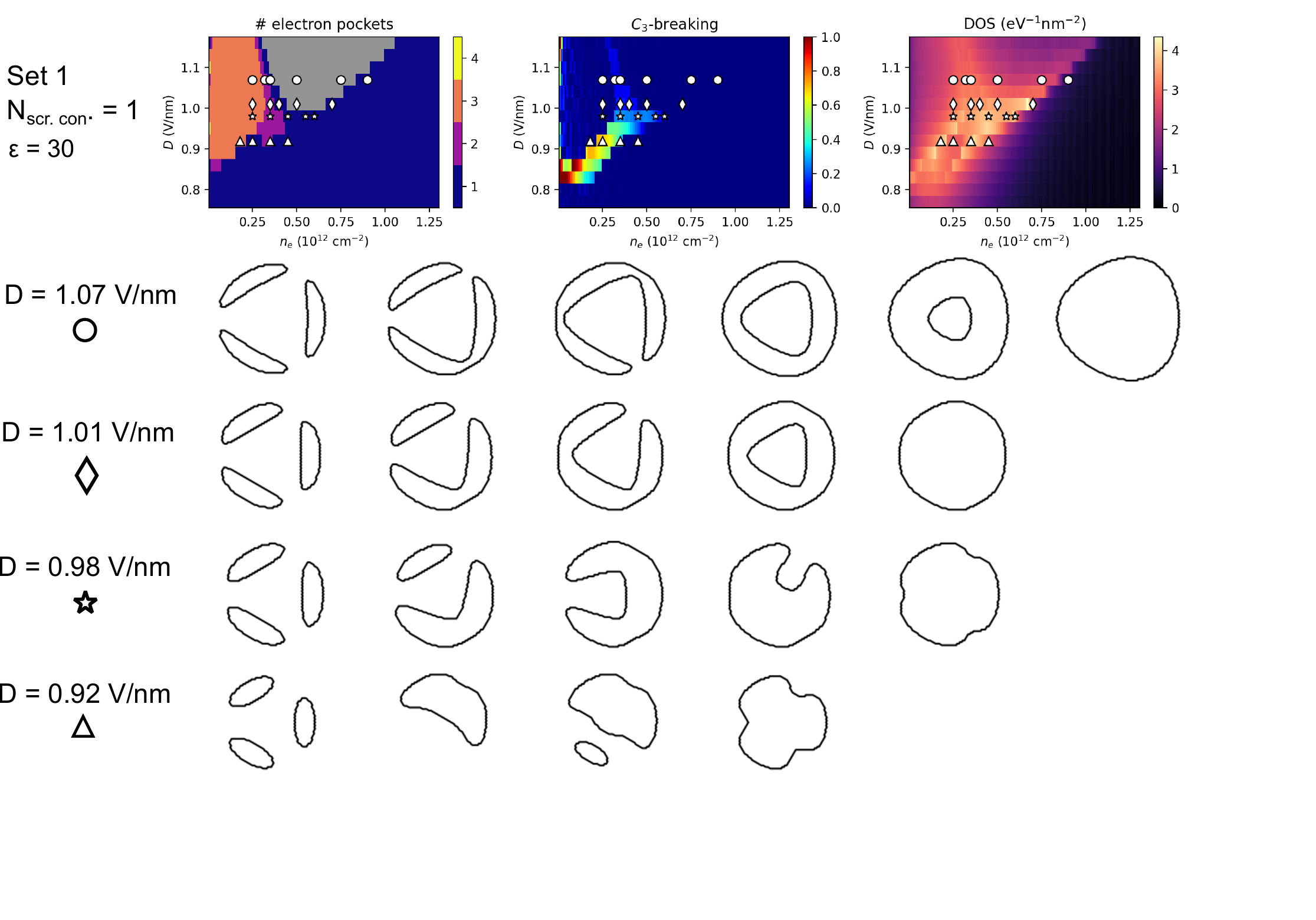}
   \caption{Fermi surfaces computed using self-consistent HF calculations that assume full spin-valley polarization. The calculations use the Set 1 parameters (see App.~\ref{secapp:continuum_model}), a dielectric constant $\epsilon=30$, and Hartree-screened layer potentials with conduction band degeneracy $N_\text{scr.\,con.}=1$ in the screening calculation. For each displacement field listed, the Fermi surfaces are ordered from left to right according to the positions of the symbols on the phase diagrams. }
   \label{HF_single_screened_set1_eps30_FS}
\end{figure*}

\begin{figure*}
   \centering 
   \includegraphics[width=0.9\textwidth]{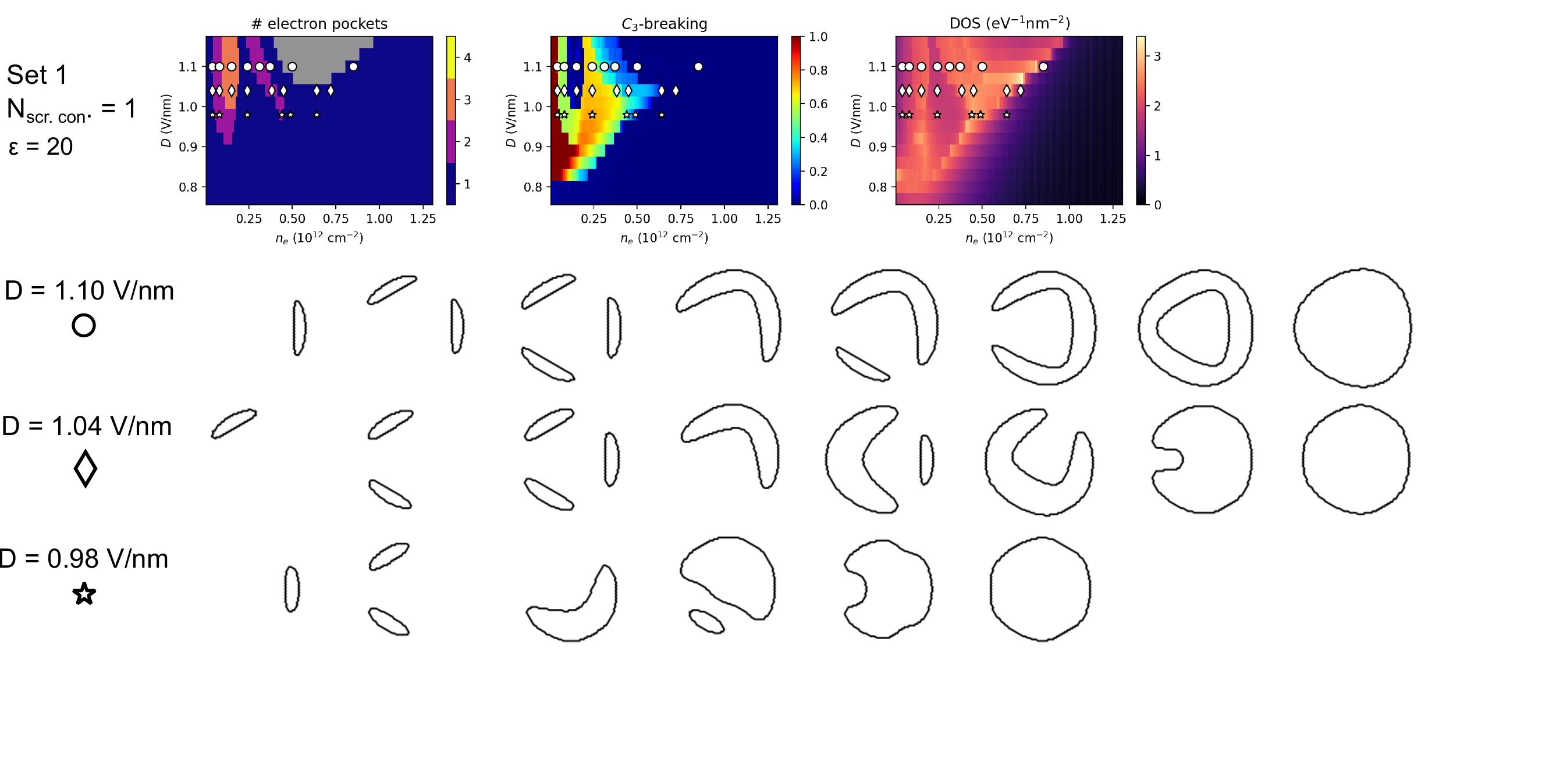}
   \caption{Fermi surfaces computed using self-consistent HF. The calculations use the Set 1 parameters (see App.~\ref{secapp:continuum_model}), a dielectric constant $\epsilon=20$, and Hartree-screened layer potentials with conduction band degeneracy $N_\text{scr.\,con.}=1$ in the screening calculation. For each displacement field listed, the Fermi surfaces are ordered from left to right according to the positions of the symbols on the phase diagrams. }
   \label{HF_single_screened_set1_eps20_FS}
\end{figure*}

\begin{figure*}
   \centering 
   \includegraphics[width=0.9\textwidth]{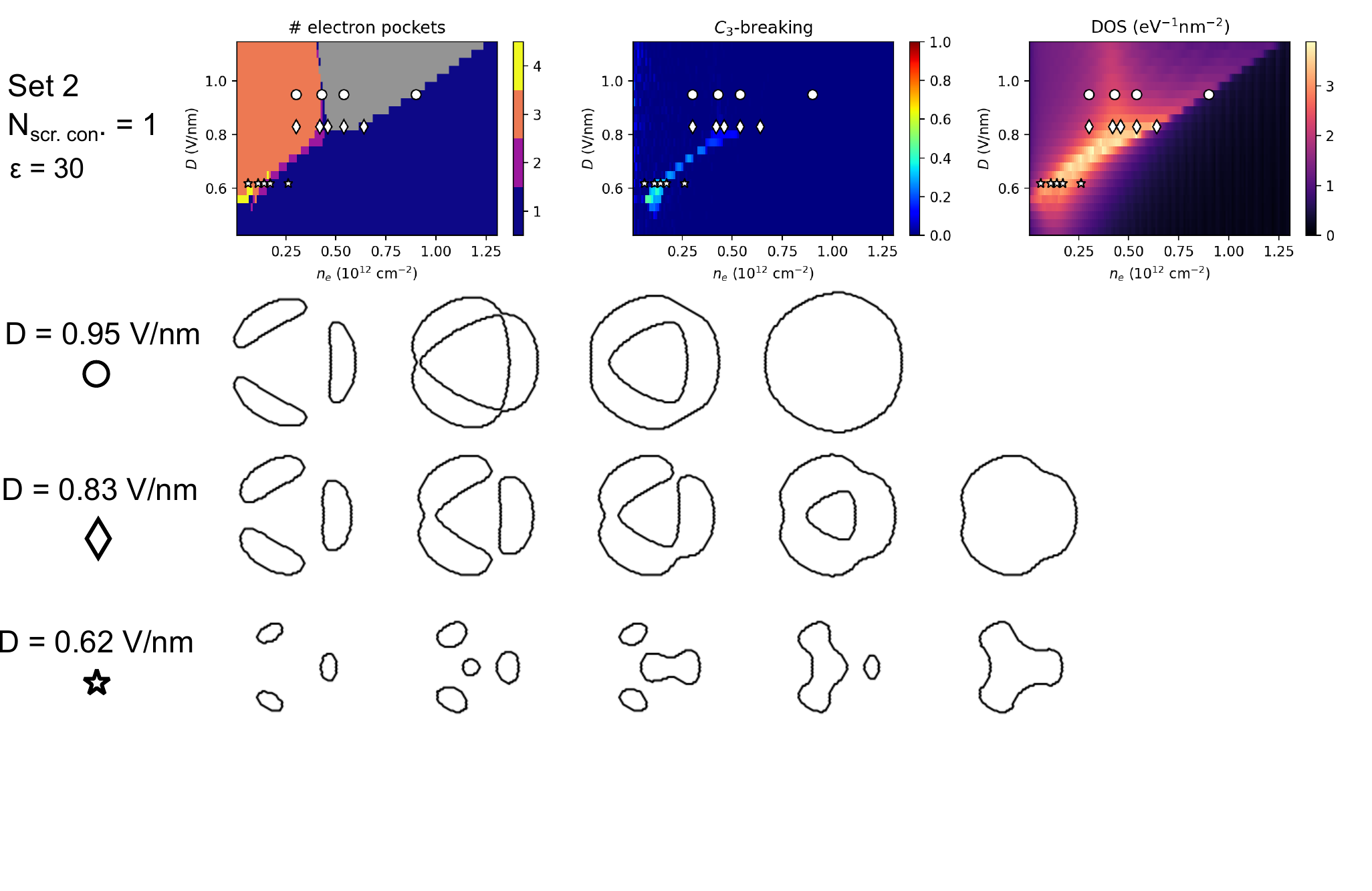}
   \caption{Fermi surfaces computed using self-consistent HF calculations that assume full spin-valley polarization. The calculations use the Set 2 parameters (see App.~\ref{secapp:continuum_model}), a dielectric constant $\epsilon=30$, and Hartree-screened layer potentials with conduction band degeneracy $N_\text{scr.\,con.}=1$ in the screening calculation. For each displacement field listed, the Fermi surfaces are ordered from left to right according to the positions of the symbols on the phase diagrams. }
   \label{HF_single_screened_set2_eps30_FS}
\end{figure*}

In Fig.~\ref{HF_single_unscreened_set1_eps30_FS}, \ref{HF_single_screened_set1_eps30_FS}, \ref{HF_single_screened_set1_eps20_FS} and \ref{HF_single_screened_set2_eps30_FS}, we plot representative HF Fermi surfaces. We highlight the different classes of nematic states:
\begin{itemize}
    \item For very low densities and sufficiently strong interactions, such as in Fig.~\ref{HF_single_screened_set1_eps20_FS}, we find pocket polarized phases where electrons only occupy one or two of the trigonal pockets.
    \item Two of these pockets may merge into a single connected `bean' (e.g.~second panel for $D=0.92\,$V/nm in Fig.~\ref{HF_single_screened_set1_eps30_FS}) or `boomerang' (e.g.~fourth panel for $D=1.10\,$V/nm in Fig.~\ref{HF_single_screened_set1_eps20_FS}) shape. These can be accompanied by a separate filled trigonal pocket for both the `bean' (e.g.~third panel for $D=0.92\,$V/nm in Fig.~\ref{HF_single_screened_set1_eps30_FS}) and `boomerang' (e.g.~fifth panel for $D=1.10\,$V/nm in Fig.~\ref{HF_single_screened_set1_eps20_FS}) cases.
    \item All three trigonal pockets can merge to form a single connected `horseshoe' Fermi surface (e.g.~sixth panel for $D=1.10\,$V/nm in Fig.~\ref{HF_single_screened_set1_eps20_FS}).
    \item Just below the bottom tip of the annular region, HF can find a singly-connected nematic Fermi surface that resembles a disk centered at the Dirac momentum but with a small `notch' cut out (e.g.~fifth panel for $u_D=48\,$meV in Fig.~\ref{HF_single_unscreened_set1_eps30_FS}).
    \item For small $D$ and $n_e$ for Set 2 (e.g.~$D=0.62\,$V/nm in Fig.~\ref{HF_single_screened_set2_eps30_FS}), we also find states that involve an electron pocket at the Dirac momentum. These include $C_3$-symmetric states with four connected components, and nematic states with two or three components.
\end{itemize}

\begin{figure*}
   \centering 
   \includegraphics[width=0.8\textwidth]{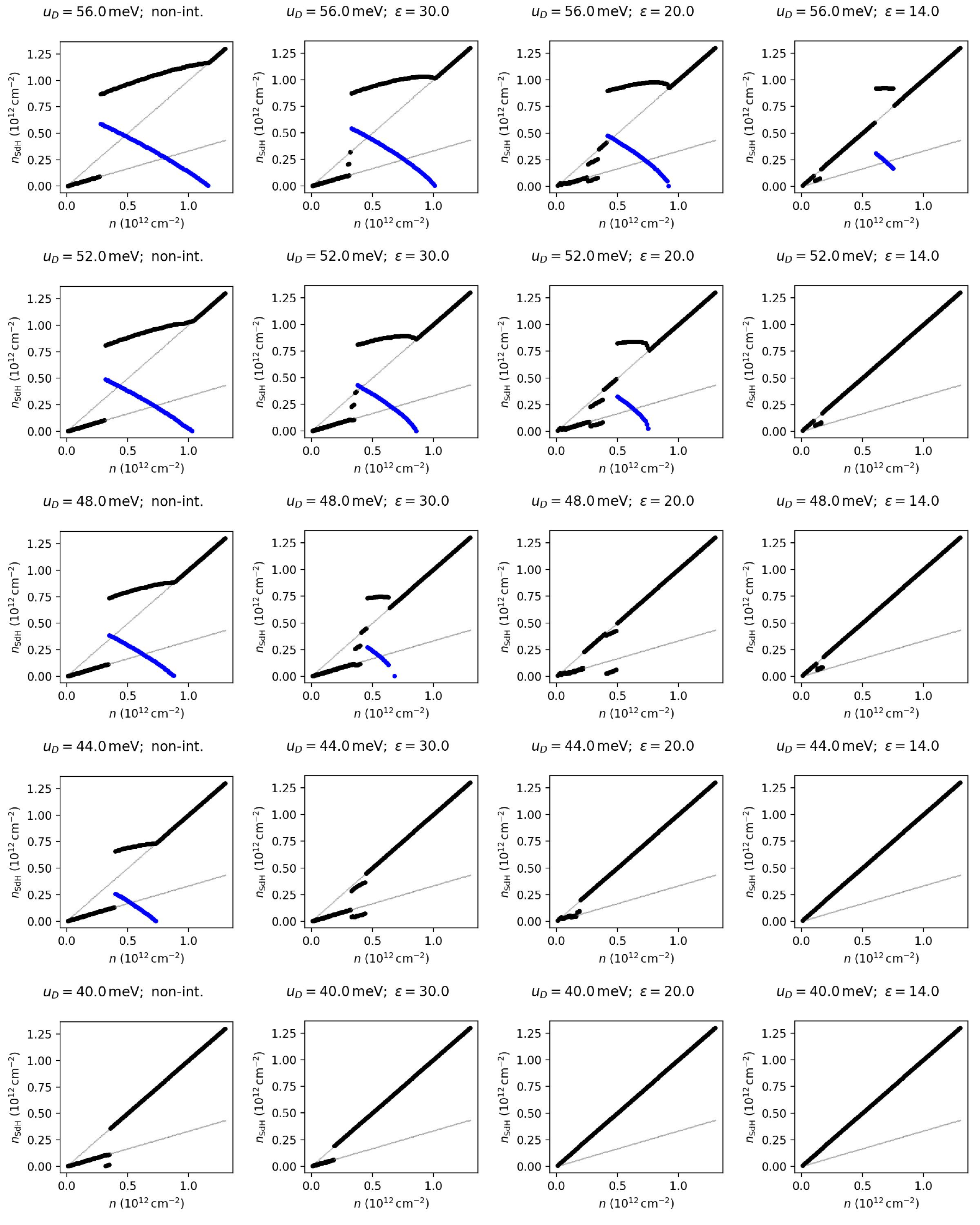}
   \caption{Fermi surface orbit areas, expressed as an effective density $n_\text{SdH}$ using the semiclassical Onsager relation Eq.~\ref{eq:onsager}, as a function of electronic density $n$ within self-consistent HF calculations that assume full spin-valley polarization. Black (blue) dots correspond to electron-like (hole-like) orbits. Grey lines indicate $n_\text{SdH}=n$ and $n_\text{SdH}=n/3$. The calculations use the Set 1 parameters (see App.~\ref{secapp:continuum_model}) and a linear interlayer potential. We consider different interlayer potentials (rows) and dielectric constants (columns). `Non-int.' corresponds to the non-interacting case where $\epsilon\rightarrow\infty$.}
   \label{HF_single_onsager}
\end{figure*}

In Fig.~\ref{HF_single_onsager}, we compute the areas $A_k$ of the Fermi surface orbits in the HF state, and plot the corresponding effective densities $n_\text{SdH}=\frac{A_k}{(2\pi)^2}$ vs the electron density $n$. Within semiclassical Onsager theory, these quantities are related to the quantum oscillation frequencies according to Eq.~\ref{eq:onsager} in the main text. We consider the Set 1 parameters with a linear interlayer potential. For small $u_D$, the calculations mainly show a single `tone' $n_\text{SdH}=n$ at high $n$ (corresponding to a simply-connected Fermi pocket) and a three-fold degenerate $n_\text{SdH}=n/3$ at low $n$ (corresponding to the three-pocket regime). Interactions can drive nematic states (such as polarization into a subset of the trigonal pockets) which generate more complex behavior for small $n$. For larger $u_D$, an annular regime is present, signalled by the emergence of a hole-like orbit (blue) and a deviation of the electron-like orbit from $n_\text{SdH}=n$. The transition from the simply-connected Fermi pocket at higher density can be either (nearly) continuous or first-order depending on the value of $u_D$ and the interaction strength. In all cases though, when entering the annular regime from high $n$, the hole-like orbit starts off with an area that is significantly smaller than that of the simply-connected Fermi pocket.

\clearpage
\newpage
% =============================================================================
\subsubsection{Intervalley coherence}\label{subsection:intervalley_coherence}
% =============================================================================

\begin{figure*}
   \centering 
   \includegraphics[width=0.9\textwidth]{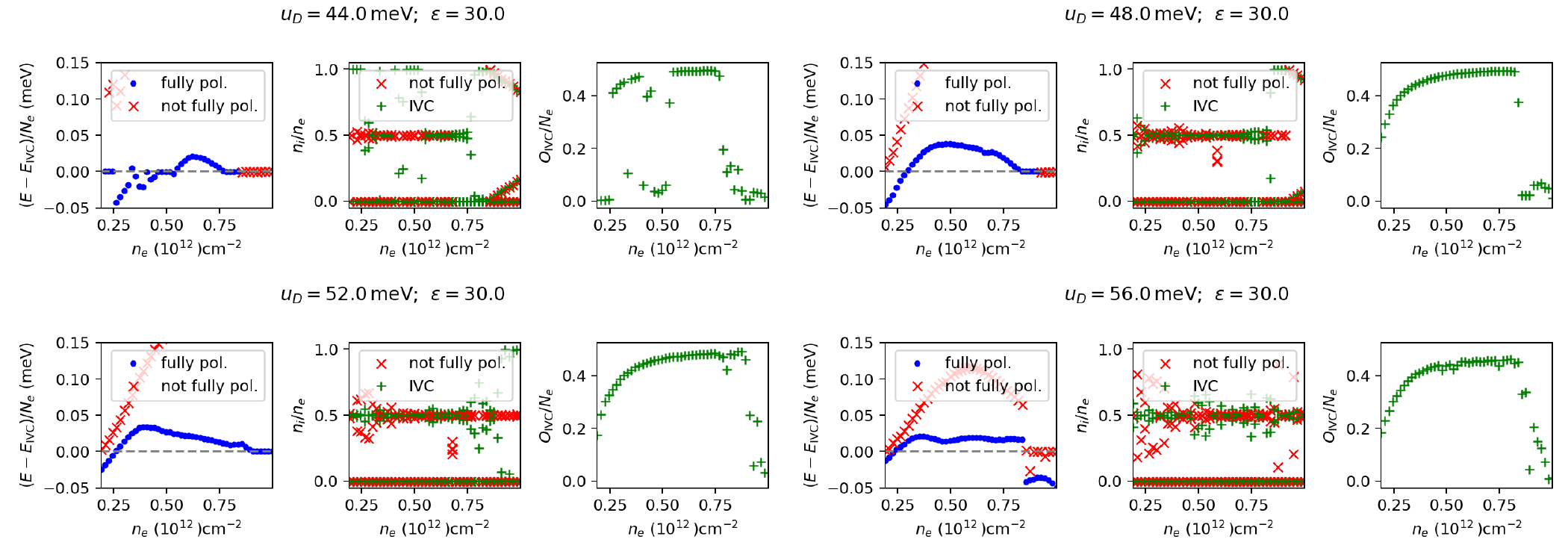}
   \caption{Competition between fully-polarized, not-fully polarized, and IVC states in HF. The calculations use the Set 1 parameters (see App.~\ref{secapp:continuum_model}), dielectric constant $\epsilon=30$, and a linear interlayer potential $u_D$.}
   \label{HF_IVC_unscreened_set1_epsr30}
\end{figure*}
\begin{figure*}
   \centering 
   \includegraphics[width=0.9\textwidth]{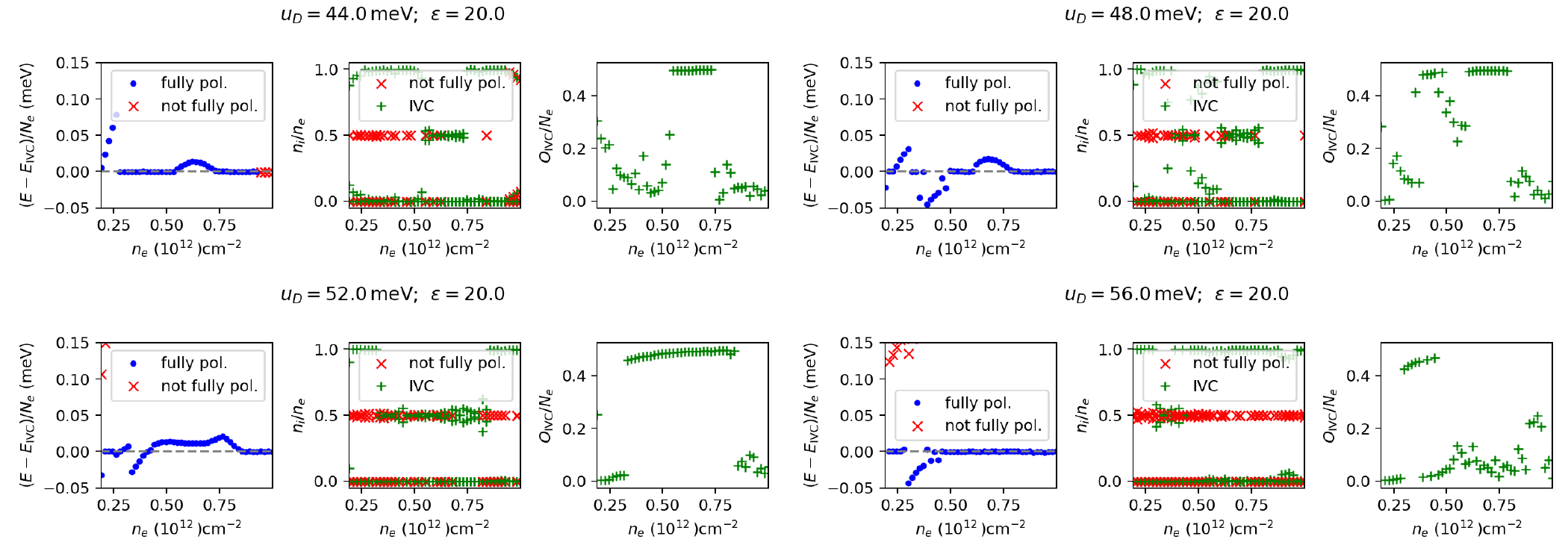}
   \caption{Competition between fully-polarized, not-fully polarized, and IVC states in HF. The calculations use the Set 1 parameters (see App.~\ref{secapp:continuum_model}), dielectric constant $\epsilon=20$, and a linear interlayer potential $u_D$.}
   \label{HF_IVC_unscreened_set1_epsr20}
\end{figure*}
\begin{figure*}
   \centering 
   \includegraphics[width=0.9\textwidth]{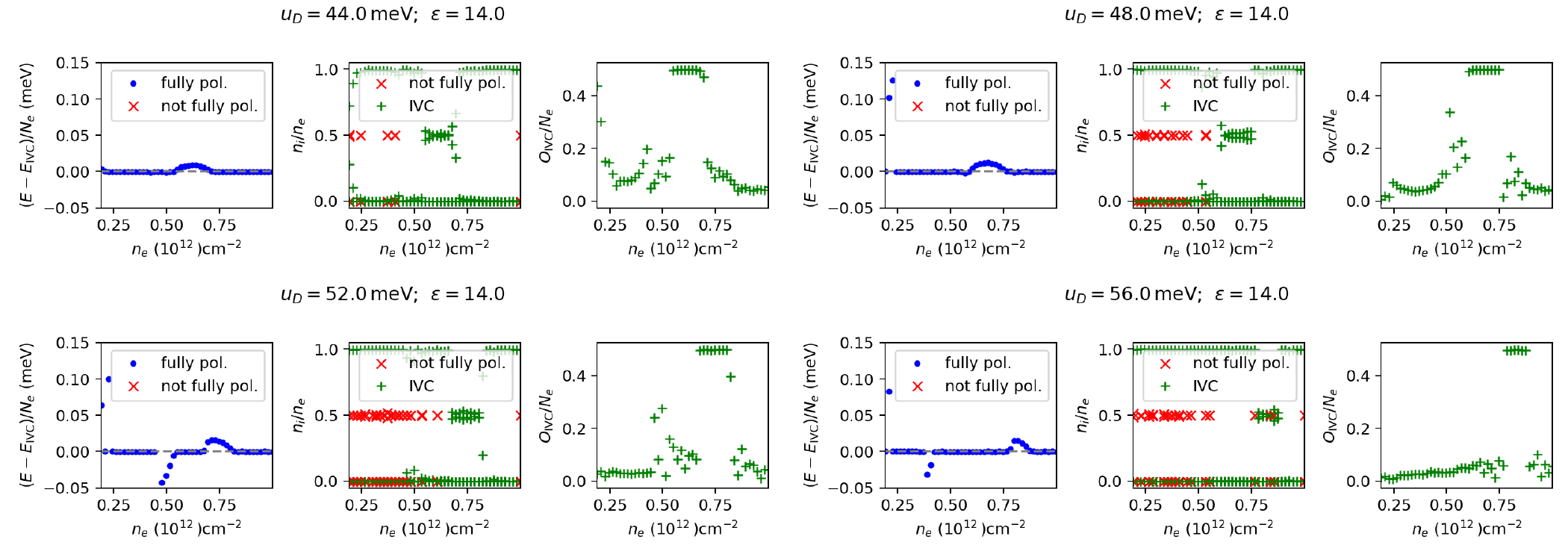}
   \caption{Competition between fully-polarized, not-fully polarized, and IVC states in HF. The calculations use the Set 1 parameters (see App.~\ref{secapp:continuum_model}), dielectric constant $\epsilon=14$, and a linear interlayer potential $u_D$.}
   \label{HF_IVC_unscreened_set1_epsr14}
\end{figure*}

\begin{figure*}
   \centering 
   \includegraphics[width=0.9\textwidth]{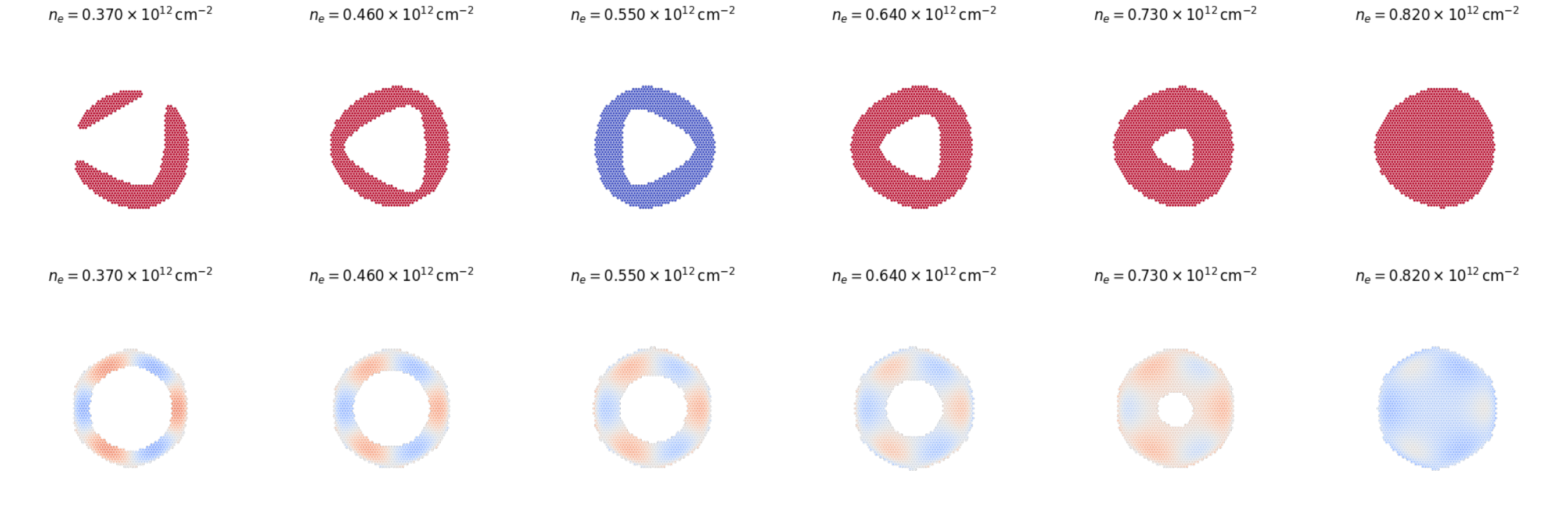}
   \caption{Fermi surfaces for the lowest energy fully-polarized (top row) and IVC (bottom) states. Blue vs red shading indicates the weight of the occupied HF Bloch states in valley $K$ vs valley $K'$. The calculations use the Set 1 parameters (see App.~\ref{secapp:continuum_model}), dielectric constant $\epsilon=20$, and linear interlayer potential $u_D=52\,$meV.}
   \label{IVC_FS_uD52_eps20}
\end{figure*}

\begin{figure*}
   \centering 
   \includegraphics[width=0.45\textwidth]{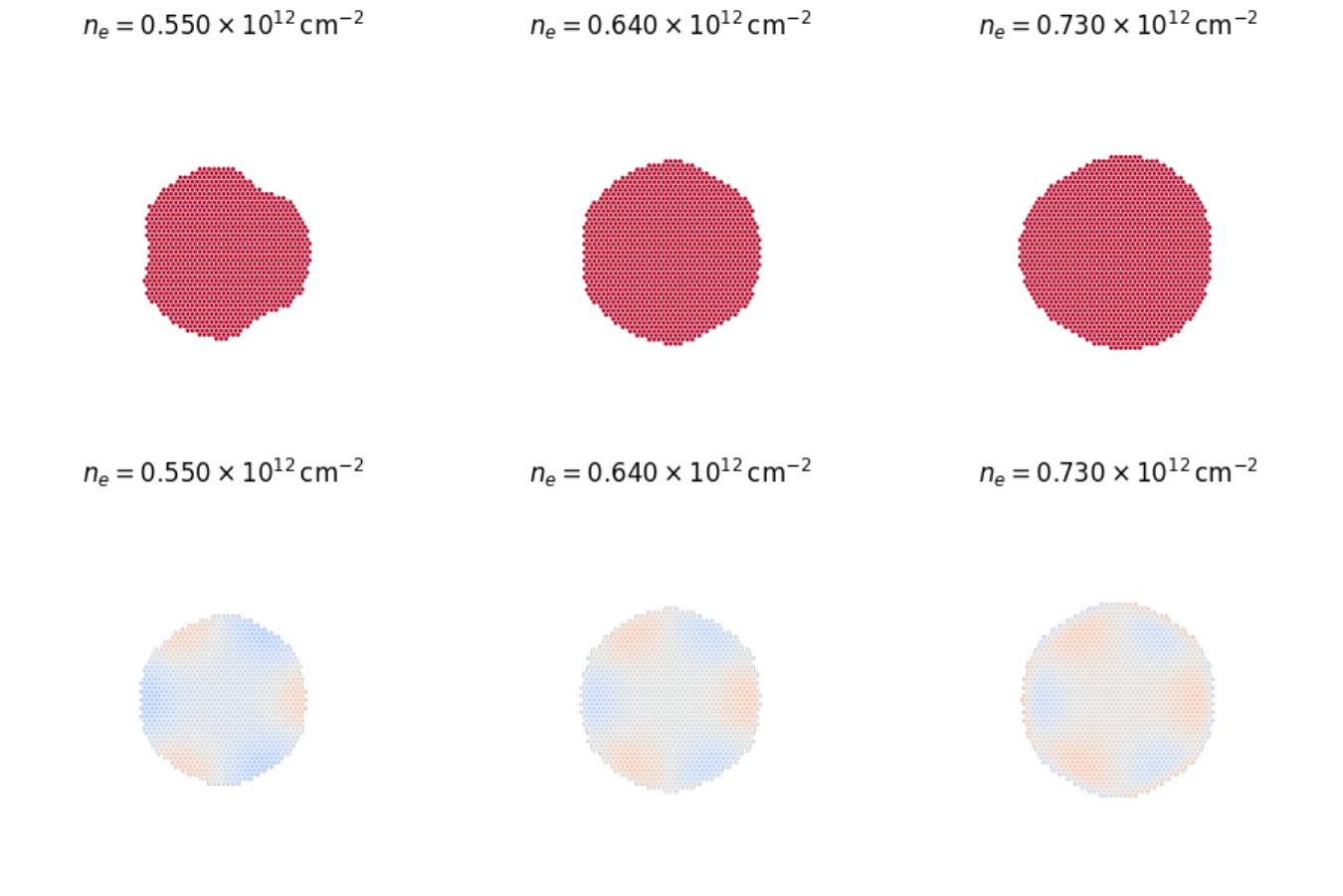}
   \caption{Fermi surfaces for the lowest energy fully-polarized (top row) and IVC (bottom) states. Blue vs red shading indicates the weight of the occupied HF Bloch states in valley $K$ vs valley $K'$. The calculations use the Set 1 parameters (see App.~\ref{secapp:continuum_model}), dielectric constant $\epsilon=20$, and linear interlayer potential $u_D=44\,$meV.}
   \label{IVC_FS_uD44_eps20}
\end{figure*}

In this section, we investigate IVC states with $\bm{q}_\text{IVC}$ in HF. We quantify the magnitude of IVC via
\begin{equation}
    O_\text{IVC}=\sum_{  \bm {k },s}| P_{+,-,s}(\bm{k}) |.
\end{equation}
The maximum possible value of $O_\text{IVC}=0$ per particle is $1/2$. States that preserve valley-$U(1)$ symmetry have $O_\text{IVC}=0$. 

In Figs.~\ref{HF_IVC_unscreened_set1_epsr30}, \ref{HF_IVC_unscreened_set1_epsr20}, \ref{HF_IVC_unscreened_set1_epsr14} (corresponding to $\epsilon=30,20,14$ respectively), we perform HF calculations as a function of $n$ using the Set 1 for large interlayer potentials $u_D$. The momentum mesh is the same as in App.~\ref{subsubsection:quarter_metal_phase}. We compare the properties of the following self-consistent HF solutions:
\begin{enumerate}
    \item \underline{Blue dots:} Fully-polarized quarter metals that preserve valley-$U(1)$. This corresponds to a spin- and valley-polarized QM.
    \item \underline{Red crosses:} Not-fully-polarized metals that preserve valley-$U(1)$.
    \item \underline{Green pluses:} IVC states that break valley-$U(1)$ symmetry. In the calculations, we denote a solution as containing IVC if $O_\text{IVC}/N_e>5\times10^{-3}$, where $N_e$ is the total particle number. These solutions also count as quarter metals, since we find that only one spin and IVC sector is occupied.
\end{enumerate}
In the first subplot for each parameter, we plot the total energies of the fully-polarized and not-fully-polarized solutions relative to the IVC solution. In the second subplot, we plot $n_i/n$ for across all flavors for the not-fully-polarized and IVC solutions, where $n_i$ is the average density of flavor $i$. In the third subplot, we plot $O_{\text{IVC}}/N_e$ for the IVC solution.

In our calculations, the not-fully-polarized solution, where it exists, is HM, except for larger $n$ where it is IHM. For the parameters studied, this (metastable) HM is typically has energy at least $\gtrsim 0.2\,$meV above the fully spin- and valley-polarized QM. This energy difference increases for stronger interactions strengths (i.e.~smaller $\epsilon$), such that the energy curve for HM is above the vertical scale in our energy difference subplots. Hence, we believe that the HM is not relevant for this parameter region.

On the other hand, we find that the spin-polarized IVC solution (SIVC), which mainly exhibits a balanced occupation $n_i/n_e\approx 0.5$ across the two valleys in the polarized spin sector, closely competes with QM over a large density range. (Note that in our framework, the SIVC is degenerate with e.g.~a spin-valley locked IVC solution due to our neglect of spin-orbit coupling and other anisotropies.) In fact, SIVC can have a lower energy than QM even for relatively strong interaction strength $\epsilon=14$. We caution though that the energy difference per electron is very small $\approx 0.02\,$meV (note that we terminate the self-consistent HF algorithm when the energy difference between iterations is less than $10^{-5}\,$meV per electron), such that the energetic ordering may be flipped by beyond mean-field correlation effects or changes in model parameters. Regardless, our results demonstrate that SIVC is a competitive phase that should be carefully considered. We have performed a small number of calculations at a non-zero intervalley spiral wavevector $\bm{q}_\text{IVC}\neq 0$~\cite{vituri2025incommensurate}, but have not found any solutions that are significantly lower in energy. We defer a more detailed study of SIVC to future theoretical work.  

In Fig.~\ref{IVC_FS_uD52_eps20}, we show representative Fermi surfaces for QM (top row) and SIVC (bottom) states as a function of density. For the dielectric constant $\epsilon=20$, we have chosen an interlayer potential $u_D=52\,$meV such that the QM phase goes from a simply-connected disk, to an annular Fermi surface, to a nematic solution with decreasing density. The SIVC Fermi surface maintains an annular shape to a relatively lower $n$ compared with QM. We note that the occupied HF Bloch functions in the SIVC smoothly modulate between having a net polarization in valley $K$ (blue) and valley $K'$ (red) around the annulus, which hints at the mechanism driving SIVC: The HF Bloch state polarizes towards the trigonal pocket of each valley to reduce the kinetic energy, and smoothly interpolates in between to maximize the exchange energy. This process is aided by the fact that the trigonal pockets from each valley approximately `nest' into each other.

In Fig.~\ref{IVC_FS_uD44_eps20}, we choose a lower interlayer potential $u_D=44\,$meV where the annular regime does not appear in HF for either the QM or SIVC solutions for $\epsilon=20$. The SIVC Bloch functions exhibit a smooth modulation of the valley polarization around the outer portion of the simply-connected Fermi pocket. The SIVC Fermi surface is typically more circular than the QM Fermi surface. 

We briefly discuss the quantum oscillation signatures of an SIVC state, considering for instance the annular Fermi surfaces of Ref.~\ref{IVC_FS_uD52_eps20}. Straightforward application of the Onsager relation in Eq.~\ref{eq:onsager} does not explain the qualitative signatures of the multitone state in experiment. However, there are additional considerations worth commenting on. The annular Fermi sea `stitches' together Fermi pockets with finite valley polarization via regions of IVC. Each such pocket contains a charge density $\lesssim n/6$ if there is no net imbalance between the valleys. If the semiclassical orbits prefer to maintain their valley character, then there could be breakdown frequencies with $n_\text{SdH}\lesssim n/6$, which may be relevant for the low tones of the multitone state. The opposite orbital magnetizations of the two valleys will also complicate a semiclassical analysis of the oscillations. As $B$ increases, the kinetic energy of one valley will change relative to the other, thereby changing the sizes of the pockets above. For finite magnetic fields, the nature of the ground state may change more dramatically~\cite{wang2024account}, such as a substantial re-orientation of the valley order parameter. Future theoretical work will be required to establish the quantum oscillation signatures of the IVC solutions.

\section{Magnetic field induced annular Fermi surface}\label{secapp:Ann_from_Mag}
\begin{figure*}
   \centering 
   \includegraphics[width=0.95\textwidth]{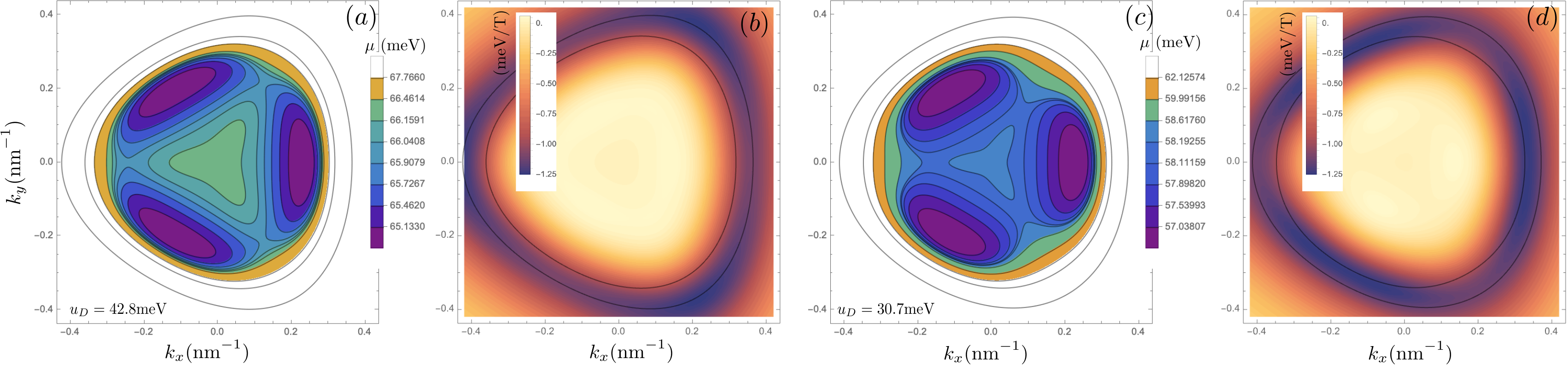}
   \caption{(a) Non-interacting Fermi surfaces for the Set 1 parameters, computed for equally spaced densities of $0.1\times10^{12}$\,cm$^{-2}$. We assume polarization into a single spin and valley $K$. The value of $u_D$ corresponds to the location of the higher-order vHS. (b) plots the orbital magnetization of the Bloch states in valley $K$. (c) and (d) show the corresponding results for the Set 2 parameters. In all panels, the Fermi surfaces corresponding to densities $n=0.9\times10^{12}$\,cm$^{-2}$ and $n=1.2\times10^{12}$\,cm$^{-2}$ are indicated with black lines. For Set 1, the corresponding chemical potentials are $69.7$\,meV and $79.7$\,meV compared to the van Hove at $\sim 66.2$meV and for Set 2 they are $65.1$\,meV, and  $77.9$\,meV compared to the van Hove at $\sim 58.6$meV.}
   \label{fig-sup-theory-nonint_DOS_sets}
\end{figure*}

It is worth remarking on the possibility that the two flat tones in the multi-tone state arise from a magnetically induced annular Fermi surface, and why this is not energetically favorable. 
In Fig.~\ref{fig-sup-theory-nonint_DOS_sets}(b) and (d), we compute the self-rotation orbital magnetization of the Bloch states~\cite{thonhauser2005orbital} in valley $K$ using the parameters from Set 1 and Set 2, respectively. We observe that the equal-energy contours corresponding to the two nearly flat tones enclose regions of high orbital magnetization. This is particularly apparent when using the parameters from Set 2.  {This selfsame region is the locus of the ring of Berry curvature~\cite{patri2025family} that dominates the quantum geometry of this model.} If one includes the external magnetic field via an orbital Zeeman coupling,
\begin{equation}
\label{eq:orbital_Zeeman}
    \varepsilon(\bm{k}) \rightarrow \varepsilon(\bm{k}) - B m_{\mathrm{SR}}(\bm{k}),
\end{equation}
where $B m_{\mathrm{SR}}$ is the orbital magnetization correction to the electronic dispersion (the so-called ``self-rotation term") \cite{thonhauser2005orbital,2015PhRvB..91u4405G,2014arXiv1411.5940R}. Explicitly,
\begin{equation}
    m_{\mathrm{SR}}(\bm{k}) = - \frac{i e}{2 \hbar} \epsilon_{ij} \bra{\partial_i u(\bm{k})}\big(h_0(\bm{k}) - \varepsilon(\bm{k})\big)\ket{\partial_j u(\bm{k})}
\end{equation}
where the cartesian indices $ij$ are summed implicitly, $h_0(\bm{k})$ is the single-particle Hamiltonian, and $\varepsilon(\bm{k}),u(\bm{k})$ are the eigen-energy and eigen-state of the active conduction band (considered here in the $K$ valley). The Berry curvature is $f(\bm{k}) = i \epsilon_{ij} \bra{\partial_i u(\bm{k})}\ket{\partial_j u(\bm{k})}$. An external magnetic field will lower the energy of the high magnetization regions, and, for sufficiently large fields, will pull the annular region down low enough to be the band minimum. However, pushing the low Fermi contour out to the marked large radius would require a magnetic field on the order of \SI{10}{T}.

\begin{figure}
     \centering
     \includegraphics[width=\linewidth]{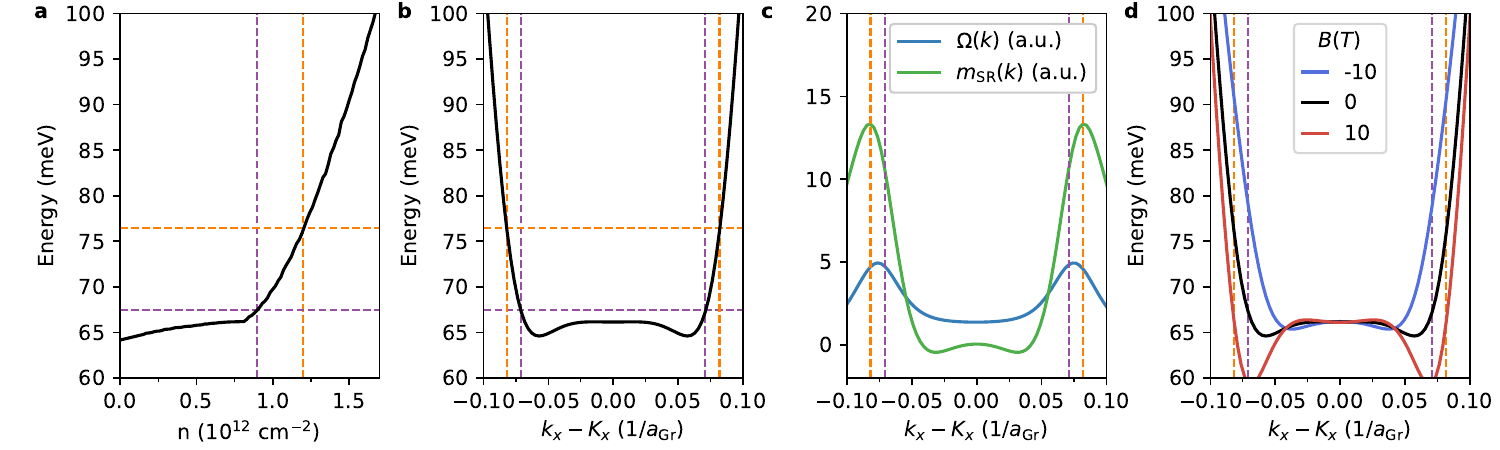}
     \caption{
     RMG with Set 1 parameters at $u_D = \SI{42.8}{meV}$.
a. Energy versus density. The dispersion increases rapidly after about \SI{0.8e12}{\per\centi\meter\squared}, once the flat minima is filled. Purple and orange lines correspond to two high tones at $n_{\mathrm{SdH}} = \SI{0.9e12}{\per\centi\meter\squared}$ and $n_{\mathrm{SdH}} = \SI{12e12}{\per\centi\meter\squared}$.
b. Bandstructure of RMG. Purple and orange lines again correspond to the high tones.
c. Berry curvature and the self-rotation part of the orbital magnetization in the $K$ valley.
d. Dispersion under an orbital Zeeman field. At \SI{10}{T} the orbital field is insufficient to push the band minima as far out as the annulus between the orange and purple lines.
     }
     \label{figS:R4G_dispersion_density_orbital_zeeman}
 \end{figure}

The need for such a large magnetic field is well illustrated by the Fermi surface contours shown in Fig.~\ref{fig-sup-theory-nonint_DOS_sets}(a) and (c). In both, the dispersion is very flat (within $1-2\,$meV) up to a density of $n\approx 0.8\times 10^{12}\,\text{cm}^{-2}$, beyond which the dispersion increases rapidly (Fig.~\ref{figS:R4G_dispersion_density_orbital_zeeman})~\cite{bernevig2025berrytrashcanmodelinteracting}. Since the flat tones correspond to fillings $n = 0.9$ and \SI{1.2e12}{\per\centi\meter\squared}, the relevant contours are high in energy ${\sim}5-15\si{meV}$ compared to the flat part of the band, and as such a large magnetic field is necessary to bring them down below the Fermi level at the relevant fillings.

In addition to requiring an excessively large orbital Zeeman field, the proposed magnetically induced annular Fermi pocket is also unstable to interactions. Hartree-Fock calculations, {performed with the orbital Zeeman coupling in  Eq.\eqref{eq:orbital_Zeeman}}, produced a state with a single simply-connected Fermi pocket (i.e., a CQM), even when the non-interacting band structure had an annular structure. {While such a dramatic consequence in the Fermiology may seem surprising, it is a natural consequence of the fact that the band minimum is quite flat with small variations compared to the scale of the exchange (Fock) interaction.}
 Since the non-interacting annular Fermi sea encloses large Berry curvature, the Bloch spinors on the inner and outer Fermi surfaces are rotated relative to one another. This decreases the Bloch wavefunction overlaps between the inner and outer surface, which, in turn, reduces the interaction energy. In contrast, the CQM  Fermi-surface lies inside the Berry curvature ring. Because of this, the relevant Bloch wavefunction overlaps are closer to unity, allowing the CQM to gain interaction energy and outcompete the annular state. The competition is not close, as the exchange scale of order \SI{10}{meV} is much larger than the scale of orbital magnetization.

% =============================================================================
\section{Quantum oscillations in a simplified single-band model}
% =============================================================================

\begin{figure*}[h]
    \centering
    \includegraphics[width=0.90\textwidth]{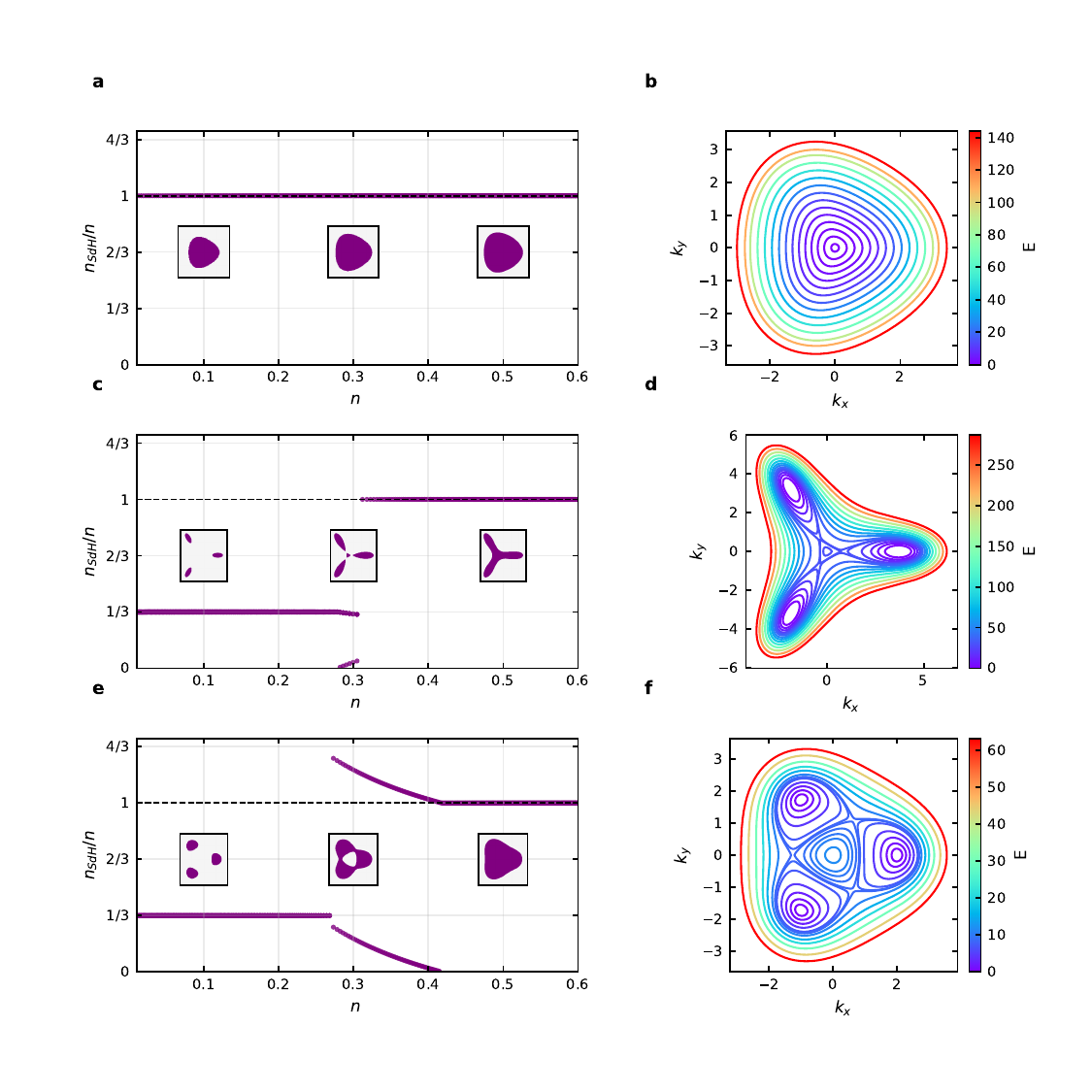}
    \caption{
    \textbf{Onsager analysis of a simplified single-band model.} 
    Normalized Shubnikov-de Haas frequency $n_{\mathrm{SdH}}/n$ versus dimensionless filling $n$ (left column) and the corresponding band dispersion $E(k_x,k_y)$ (right column) for the simplified single-band model with parameters (a,b) $(a_2,a_3,a_4) =  (4,1,1)$; ({c},{d}) $(a_2,a_3,a_4) =(5.5,6,1)$; and ({e},{f}) $(a_2,a_3,a_4) =(-5,1,1)$. 
    Insets in (a), (c), (e) show representative  Fermi pockets at $n = 0.1$, $0.3$, and $0.5$. 
    (a,b) A single pocket centered at the origin yields $n_{\mathrm{SdH}}/n = 1$  across the full filling range. 
    (c,d) Three off center minima in the dispersion produce three disconnected pockets at low $n$, giving $n_{\mathrm{SdH}}/n = 1/3$; a fourth pocket at the origin emerges in a small density regime giving three tones that dip below $n_{\mathrm{SdH}}/n = 1/3$ and a new tone that emerges above $n_{\mathrm{SdH}}/n = 0$, after which all four pockets merge into one giving $n_{\mathrm{SdH}}/n = 1$. 
    (e,f) Three off center minima in the dispersion produce three disconnected pockets at low $n$; successive Lifshitz transitions absorb the outer pockets into a single annular and then simply-connected Fermi pocket with $n_{\mathrm{SdH}}/n = 1$.}
    \label{fig-onsager-model}
\end{figure*}

The essential Fermiology of the rhombohedral tetralayer is well reproduced by a one-band model with dimensionless Hamiltonian
\begin{equation}
\label{eq:1band_model}
    h(\bm{k}) = a_2 |\bm{k}|^2 + a_3 (3 k_xk_y^2 - k_x^3) + a_4 |\bm{k}|^4.
\end{equation}
To model a conduction band, we fix $a_4 > 0$. For $a_2 > 9a^2_3/32 a_4$ there is a single simply-connected Fermi pocket for all densities. For  $ 9a^2_3/32 a_4 > a_2 > 0$, there is a single Fermi surface at high density, a $C_{3z}$ symmetric 4-pocket state at intermediate densities, and a 3-pocket state at low densities. For $0>a_2 $ there is a single Fermi surface at high density, an annular Fermi surface at intermediate densities, and a 3-pocket state at low densities. In Fig~\ref{fig-onsager-model} we show the dispersion of the one-band model in each of these three regimes. We also include plots of the corresponding normalized Shubnikov-de Haas frequency, $n_{\mathrm{SdH}}/n$, as a function of the total dimensionless filling $n$. Each point corresponds to a Fermi surface that encloses normalized density $n_i /n$. 

\stopcontents[SI]

%% file: refs.bib
@ARTICLE{2015PhRvB..91u4405G,
       author = {{Gao}, Yang and {Yang}, Shengyuan A. and {Niu}, Qian},
        title = "{Geometrical effects in orbital magnetic susceptibility}",
      journal = {\prb},
         year = 2015,
        month = jun,
       volume = {91},
       number = {21},
          eid = {214405},
        pages = {214405},
          doi = {10.1103/PhysRevB.91.214405},
archivePrefix = {arXiv},
       eprint = {1411.0324},
 primaryClass = {cond-mat.mes-hall},
       adsurl = {https://ui.adsabs.harvard.edu/abs/2015PhRvB..91u4405G}
}

@ARTICLE{2014arXiv1411.5940R,
       author = {{Raoux}, A. and {Pi{\'e}chon}, F. and {Fuchs}, J.~N. and {Montambaux}, G.},
        title = "{Orbital magnetism of coupled bands models}",
      journal = {arXiv e-prints},
         year = 2014,
        month = nov,
          eid = {arXiv:1411.5940},
        pages = {arXiv:1411.5940},
          doi = {10.48550/arXiv.1411.5940},
archivePrefix = {arXiv},
       eprint = {1411.5940},
 primaryClass = {cond-mat.mes-hall},
       adsurl = {https://ui.adsabs.harvard.edu/abs/2014arXiv1411.5940R}
}

@ARTICLE{2019NatCo..10.5769Y,
       author = {{Yuan}, Noah F.~Q. and {Isobe}, Hiroki and {Fu}, Liang},
        title = "{Magic of High-Order van Hove Singularity}",
      journal = {Nature Communications},
         year = 2019,
        month = dec,
       volume = {10},
          eid = {5769},
        pages = {5769},
          doi = {10.1038/s41467-019-13670-9},
archivePrefix = {arXiv},
       eprint = {1901.05432},
 primaryClass = {cond-mat.str-el},
       adsurl = {https://ui.adsabs.harvard.edu/abs/2019NatCo..10.5769Y}
}

@misc{data_repo,
author = {Kalantre, S. S. and Alexander, B. H. and May-Mann, J. and Herzog-Arbeitman, J. and Hocking, M. and Cao, Q. and
  Watanabe, K. and Taniguchi, T. and Goldhaber-Gordon, D. and Mannix, A. J. and Devakul, T. and Kwan, Y. H. and Parker, D. E. and Sharpe, A.},
journal = {Stanford Digital Repository},
title = {{Data for: Fermiology and the Candidate Chiral Superconductor in Rhombohedral Tetralayer Graphene, \text{Version} 1.0, \text{Stanford Digital Repository} (2026); https://doi.org/10.25740/pp266vn3552}}
}

@ARTICLE{2023PhRvB.108o5406P,
       author = {{Park}, Youngju and {Kim}, Yeonju and {Chittari}, Bheema Lingam and {Jung}, Jeil},
        title = "{Topological Flat Bands in Rhombohedral Tetralayer and Multilayer Graphene on Hexagonal Boron Nitride Moir{\'e} Superlattices}",
      journal = {\prb},
         year = 2023,
        month = oct,
       volume = {108},
       number = {15},
          eid = {155406},
        pages = {155406},
          doi = {10.1103/PhysRevB.108.155406},
archivePrefix = {arXiv},
       eprint = {2304.12874},
 primaryClass = {cond-mat.mes-hall},
       adsurl = {https://ui.adsabs.harvard.edu/abs/2023PhRvB.108o5406P}
}

@article{han2025signatures,
  title = {Signatures of Chiral Superconductivity in Rhombohedral Graphene},
  author = {Han, Tonghang and Lu, Zhengguang and Hadjri, Zach and Shi, Lihan and Wu, Zhenghan and Xu, Wei and Yao, Yuxuan and Cotten, Armel A. and Sharifi Sedeh, Omid and Weldeyesus, Henok and others},
  journal = {Nature},
  volume = {643},
  number = {8072},
  pages = {654--661},
  year = {2025},
  publisher = {Nature Publishing Group UK London}
}

@article{kallin2016chiral,
  title = {Chiral Superconductors},
  author = {Kallin, Catherine and Berlinsky, John},
  year = {2016},
  month = may,
  journal = {Reports on Progress in Physics},
  volume = {79},
  number = {5},
  pages = {054502},
  doi = {10.1088/0034-4885/79/5/054502}
}

@article{aokiCoexistenceSuperconductivityFerromagnetism2001,
  title = {Coexistence of Superconductivity and Ferromagnetism in {{URhGe}}},
  author = {Aoki, Dai and Huxley, Andrew and Ressouche, Eric and Braithwaite, Daniel and Flouquet, Jacques and Brison, Jean-Pascal and Lhotel, Elsa and Paulsen, Carley},
  year = {2001},
  month = oct,
  journal = {Nature},
  volume = {413},
  number = {6856},
  pages = {613--616},
  publisher = {Nature Publishing Group},
  doi = {10.1038/35098048}
}

@article{saxenaSuperconductivityBorderItinerantelectron2000,
  title = {Superconductivity on the Border of Itinerant-Electron Ferromagnetism in {{UGe2}}},
  author = {Saxena, S. S. and Agarwal, P. and Ahilan, K. and Grosche, F. M. and Haselwimmer, R. K. W. and Steiner, M. J. and Pugh, E. and Walker, I. R. and Julian, S. R. and Monthoux, P. and Lonzarich, G. G. and Huxley, A. and Sheikin, I. and Braithwaite, D. and Flouquet, J.},
  year = {2000},
  month = aug,
  journal = {Nature},
  volume = {406},
  number = {6796},
  pages = {587--592},
  publisher = {Nature Publishing Group},
  doi = {10.1038/35020500}
}

@article{chouIntravalleySpinpolarizedSuperconductivity2025,
  title = {Intravalley Spin-Polarized Superconductivity in Rhombohedral Tetralayer Graphene},
  author = {Chou, Yang-Zhi and Zhu, Jihang and Das Sarma, Sankar},
  year = {2025},
  month = may,
  journal = {Physical Review B},
  volume = {111},
  number = {17},
  pages = {174523},
  doi = {10.1103/PhysRevB.111.174523}
}

@Article{geier2025chiral,
author={Geier, Max
and Davydova, Margarita
and Fu, Liang},
title={Chiral and Topological Superconductivity in Isospin Polarized Multilayer Graphene},
journal={Nature Communications},
year={2025},
month={Dec},
day={01},
volume={17},
number={1},
pages={232},
issn={2041-1723},
doi={10.1038/s41467-025-66902-6},
url={https://doi.org/10.1038/s41467-025-66902-6}
}

@article{parra-martinezBandRenormalizationQuarter2025,
  title = {Band {{Renormalization}}, {{Quarter Metals}}, and {{Chiral Superconductivity}} in {{Rhombohedral Tetralayer Graphene}}},
  author = {Parra-Mart{\'i}nez, Guillermo and Jimeno-Pozo, Alejandro and Phong, V{\~o} Ti{\'e}n and Sainz-Cruz, H{\'e}ctor and Kaplan, Daniel and Emanuel, Peleg and Oreg, Yuval and Pantale{\'o}n, Pierre A. and Silva-Guill{\'e}n, Jos{\'e} {\'A}ngel and Guinea, Francisco},
  year = {2025},
  month = sep,
  journal = {Physical Review Letters},
  volume = {135},
  number = {13},
  pages = {136503},
  doi = {10.1103/zfmh-rjzc}
}

@article{wangChiralSuperconductivityParent2024,
  title = {Chiral Superconductivity from a Parent {{Chern band}} and its {{Non-Abelian}} Generalization},
  author = {Wang, Yan-Qi and Gao, Zhi-Qiang and Yang, Hui},
  journal = {Phys. Rev. B},
  volume = {113},
  issue = {17},
  pages = {174507},
  numpages = {9},
  year = {2026},
  month = {May},
  publisher = {American Physical Society},
  doi = {10.1103/fdz1-dbf6},
  url = {https://link.aps.org/doi/10.1103/fdz1-dbf6}
}

@article{yoonQuarterMetalSuperconductivityRhombohedral2026,
  title = {Quarter-{{Metal Superconductivity}} in {{Rhombohedral Graphene}}},
  author = {Yoon, Chiho and Xu, Tianyi and Barlas, Yafis and Zhang, Fan},
  year = {2026},
  month = jan,
  journal = {Physical Review Letters},
  volume = {136},
  number = {2},
  pages = {026603},
  doi = {10.1103/fcdc-9lm3}
}

@book{shoenberg1984magnetic,
  title = {Magnetic Oscillations in Metals},
  author = {Shoenberg, David},
  year = {1984},
  publisher = {Cambridge University Press}
}

@article{auerbachIsospinMagneticTexture2025,
  title = {Isospin Magnetic Texture and Intervalley Exchange Interaction in Rhombohedral Tetralayer Graphene},
  author = {Auerbach, Nadav and Dutta, Surajit and Uzan, Matan and Vituri, Yaar and Zhou, Yaozhang and Meltzer, Alexander Y. and Grover, Sameer and Holder, Tobias and Emanuel, Peleg and Huber, Martin E. and Myasoedov, Yuri and Watanabe, Kenji and Taniguchi, Takashi and Oreg, Yuval and Berg, Erez and Zeldov, Eli},
  year = {2025},
  month = nov,
  journal = {Nature Physics},
  volume = {21},
  number = {11},
  pages = {1765--1772},
  doi = {10.1038/s41567-025-03035-z}
}

@article{zhang2011magnetoelectric,
  title={{{Magnetoelectric Coupling, Berry phase, and Landau Level Dispersion in a Biased Bilayer Graphene}}},
  author={Zhang, Lingfeng M and Fogler, Michael M and Arovas, Daniel P},
  journal={Physical Review B—Condensed Matter and Materials Physics},
  volume={84},
  number={7},
  pages={075451},
  year={2011},
  publisher={APS}
}

@article{slizovskiy2019films,
  title={Films of Rhombohedral Graphite as Two-Dimensional Topological Semimetals},
  author={Slizovskiy, Sergey and McCann, Edward and Koshino, Mikito and Fal’ko, Vladimir I},
  journal={Communications Physics},
  volume={2},
  number={1},
  pages={164},
  year={2019},
  publisher={Nature Publishing Group UK London}
}

@article{alexandradinataRevealingTopologyFermiSurface2018,
  title = {Revealing the {{Topology}} of {{Fermi-Surface Wave Functions}} from {{Magnetic Quantum Oscillations}}},
  author = {Alexandradinata, A. and Wang, Chong and Duan, Wenhui and Glazman, Leonid},
  year = {2018},
  month = feb,
  journal = {Physical Review X},
  volume = {8},
  number = {1},
  pages = {011027},
  doi = {10.1103/PhysRevX.8.011027}
}

@article{alexandradinataGeometricPhaseOrbital2017,
  title = {Geometric {{Phase}} and {{Orbital Moment}} in {{Quantization Rules}} for {{Magnetic Breakdown}}},
  author = {Alexandradinata, A. and Glazman, Leonid},
  year = {2017},
  month = dec,
  journal = {Physical Review Letters},
  volume = {119},
  number = {25},
  pages = {256601},
  doi = {10.1103/PhysRevLett.119.256601}
}

@article{alexandradinataSemiclassicalTheoryLandau2018,
  title = {Semiclassical Theory of {{Landau}} Levels and Magnetic Breakdown in Topological Metals},
  author = {Alexandradinata, A. and Glazman, Leonid},
  year = {2018},
  month = apr,
  journal = {Physical Review B},
  volume = {97},
  number = {14},
  pages = {144422},
  doi = {10.1103/PhysRevB.97.144422}
}

@article{cohen_magnetic_1961,
  title = {Magnetic {Breakdown} in {Crystals}},
  author = {Cohen, Morrel H. and Falicov, L. M.},
  year = {1961},
  month = sep,
  journal = {Physical Review Letters},
  volume = {7},
  number = {6},
  pages = {231--233},
  doi = {10.1103/PhysRevLett.7.231}
}

@article{Kolar2026singlegatetracking,
  title = {Single-Gate Tracking Behavior in Flat-band Multilayer Graphene Devices},
  author = {Kol{\'a}{\v r}, Kry{\v s}tof and Waters, Dacen and Folk, Joshua and Yankowitz, Matthew and Lewandowski, Cyprian},
  journal = {Physical Review B},
  volume = {113},
  number = {7},
  pages = {075131},
  year = {2026},
  month = feb,
  publisher = {American Physical Society},
  doi = {10.1103/j1zf-v3j5}
}

@article{minChiralDecompositionElectronic2008,
  title = {Chiral Decomposition in the Electronic Structure of Graphene Multilayers},
  author = {Min, Hongki and MacDonald, Allan H.},
  year = {2008},
  month = apr,
  journal = {Physical Review B},
  volume = {77},
  number = {15},
  pages = {155416},
  doi = {10.1103/PhysRevB.77.155416}
}

@article{koshinoTrigonalWarpingBerry2009,
  title = {Trigonal Warping and {{Berry}}'s Phase {{N}}$\pi$ in {{ABC-stacked}} Multilayer Graphene},
  author = {Koshino, Mikito and McCann, Edward},
  year = {2009},
  month = sep,
  journal = {Physical Review B},
  volume = {80},
  number = {16},
  pages = {165409},
  doi = {10.1103/PhysRevB.80.165409}
}

@article{luttinger1960ground,
  title={Ground-State Energy of a Many-Fermion System. II},
  author={Luttinger, Joaquin Mazdak and Ward, John Clive},
  journal={Physical Review},
  volume={118},
  number={5},
  pages={1417},
  year={1960},
  publisher={APS}
}

@article{oshikawa2000topo,
  title={{{Topological Approach to Luttinger's Theorem and the Fermi Surface of a Kondo Lattice}}},
  author={Oshikawa, Masaki},
  journal={Phys. Rev. Lett.},
  volume={84},
  issue={15},
  pages={3370--3373},
  numpages={0},
  year={2000},
  month={Apr},
  publisher={American Physical Society},
  doi={10.1103/PhysRevLett.84.3370},
  url={https://link.aps.org/doi/10.1103/PhysRevLett.84.3370}
}

@article{vituri2025incommensurate,
  title = {Incommensurate Intervalley Coherent States in {{ABC}} Graphene: Collective Modes and Superconductivity},
  author = {Vituri, Yaar and Xiao, Jiewen and Pareek, Keshav and Holder, Tobias and Berg, Erez},
  journal = {Phys. Rev. B},
  volume = {111},
  issue = {7},
  pages = {075103},
  numpages = {13},
  year = {2025},
  month = {Feb},
  publisher = {American Physical Society},
  doi = {10.1103/PhysRevB.111.075103},
  url = {https://link.aps.org/doi/10.1103/PhysRevB.111.075103}
}

@article{patri2025family,
  title={Family of Multilayer Graphene Superconductors with Tunable Chirality: Momentum-Space Vortices Nucleated by a Ring of {{Berry}} Curvature},
  author={Patri, Adarsh S and Franz, Marcel},
  journal={Physical Review B},
  volume={112},
  number={21},
  pages={214505},
  year={2025},
  publisher={APS}
}

@article{yang2025topological,
  title={Topological Incommensurate {{Fulde-Ferrell-Larkin-Ovchinnikov}} Superconductor and {{Bogoliubov}} {{Fermi}} Surface in Rhombohedral Tetralayer Graphene},
  author={Yang, Hui and Zhang, Ya-Hui},
  journal={Physical Review B},
  volume={112},
  number={2},
  pages={L020506},
  year={2025},
  publisher={APS}
}

@article{qin2026chiral,
doi = {10.1088/0256-307X/43/3/030708},
url = {https://doi.org/10.1088/0256-307X/43/3/030708},
year = {2026},
month = {mar},
publisher = {Chinese Physical Society and IOP Publishing Ltd},
volume = {43},
number = {3},
pages = {030708},
author = {Qin, Qiong and Wu, Congjun},
title = {Chiral Finite-Momentum Superconductivity in the Tetralayer Graphene},
journal = {Chinese Physics Letters}}

@article{christos2025finite,
  title={Finite-Momentum Pairing and Superlattice Superconductivity in Valley-Imbalanced Rhombohedral Graphene},
  author={Christos, Maine and Bonetti, Pietro M and Scheurer, Mathias S},
  journal={arXiv preprint arXiv:2503.15471},
  year={2025}
}

@Article{may2026pairing,
author={May-Mann, Julian
and Helbig, Tobias
and Devakul, Trithep},
title={How Pairing Mechanism Dictates Topology in Valley-Polarized Superconductors with {{Berry}} Curvature},
journal={npj Quantum Materials},
year={2026},
month={Apr},
day={15},
issn={2397-4648},
doi={10.1038/s41535-026-00878-4},
url={https://doi.org/10.1038/s41535-026-00878-4}
}

@article{nadeem2023superconducting,
  title={The Superconducting Diode effect},
  author={Nadeem, Muhammad and Fuhrer, Michael S and Wang, Xiaolin},
  journal={Nature Reviews Physics},
  volume={5},
  number={10},
  pages={558--577},
  year={2023},
  publisher={Nature Publishing Group UK London}
}

@article{chen2025intrinsic,
  title={Intrinsic Superconducting Diode Effect and Nonreciprocal Superconductivity in Rhombohedral Graphene Multilayers},
  author={Chen, Yinqi and Scheurer, Mathias S and Schrade, Constantin},
  journal={Physical Review B},
  volume={112},
  number={6},
  pages={L060505},
  year={2025},
  publisher={APS}
}

@article{read2000paired,
  title={Paired States of Fermions in Two Dimensions with Breaking of Parity and Time-Reversal Symmetries and the Fractional Quantum {{Hall}} Effect},
  author={Read, Nicholas and Green, Dmitry},
  journal={Physical Review B},
  volume={61},
  number={15},
  pages={10267},
  year={2000},
  publisher={APS}
}

@article{pierre2003dephasing,
  title={Dephasing of Electrons in Mesoscopic Metal Wires},
  author={Pierre, F and Gougam, AB and Anthore, A and Pothier, H and Esteve, Daniel and Birge, Norman O},
  journal={Physical Review B},
  volume={68},
  number={8},
  pages={085413},
  year={2003},
  publisher={APS}
}

@article{kim2025topologicalchiral,
  title = {Topological Chiral Superconductivity Beyond Pairing in a {{Fermi Liquid}}},
  author = {Kim, Minho and Timmel, Abigail and Ju, Long and Wen, Xiao-Gang},
  journal = {Phys. Rev. B},
  volume = {111},
  issue = {1},
  pages = {014508},
  numpages = {20},
  year = {2025},
  month = {Jan},
  publisher = {American Physical Society},
  doi = {10.1103/PhysRevB.111.014508},
  url = {https://link.aps.org/doi/10.1103/PhysRevB.111.014508}
}

@article{jahin2026enhanced,
  title = {Enhanced {{Kohn-Luttinger}} Superconductivity in Geometric Bands},
  author = {Jahin, Ammar and Lin, Shi-Zeng},
  journal = {Phys. Rev. B},
  volume = {113},
  issue = {1},
  pages = {014504},
  numpages = {8},
  year = {2026},
  month = {Jan},
  publisher = {American Physical Society},
  doi = {10.1103/gt8h-czf3},
  url = {https://link.aps.org/doi/10.1103/gt8h-czf3}
}

@article{pendharkarTorsionalForceMicroscopy2024,
    title = {Torsional Force Microscopy of van der {Waals} Moirés and Atomic Lattices},
    volume = {121},
    url = {https://www.pnas.org/doi/abs/10.1073/pnas.2314083121},
    doi = {10.1073/pnas.2314083121},
    number = {10},
    urldate = {2025-04-20},
    journal = {Proceedings of the National Academy of Sciences},
    publisher = {Proceedings of the National Academy of Sciences},
    author = {Pendharkar, Mihir and Tran, Steven J. and Zaborski, Gregory and Finney, Joe and Sharpe, Aaron L. and Kamat, Rupini V. and Kalantre, Sandesh S. and Hocking, Marisa and Bittner, Nathan J. and Watanabe, Kenji and Taniguchi, Takashi and Pittenger, Bede and Newcomb, Christina J. and Kastner, Marc A. and Mannix, Andrew J. and Goldhaber-Gordon, David},
    month = mar,
    year = {2024},
    pages = {e2314083121},
}

@article{satoTopologicalSuperconductorsReview2017,
    title = {Topological Superconductors: a Review},
    volume = {80},
    issn = {0034-4885},
    shorttitle = {Topological superconductors},
    url = {https://doi.org/10.1088/1361-6633/aa6ac7},
    doi = {10.1088/1361-6633/aa6ac7},
    number = {7},
    urldate = {2026-05-24},
    journal = {Reports on Progress in Physics},
    publisher = {IOP Publishing},
    author = {Sato, Masatoshi and Ando, Yoichi},
    month = may,
    year = {2017},
    pages = {076501},
}

@misc{yoon2026majoranacrystal,
      title={{{Majorana Crystal in Rhombohedral Graphene}}}, 
      author={Chiho Yoon and Fan Zhang},
      year={2026},
      eprint={2603.16828},
      archivePrefix={arXiv},
      primaryClass={cond-mat.mes-hall},
      url={https://arxiv.org/abs/2603.16828}, 
}

@article{gaggioli2025vortex,
  title = {{{Spontaneous Vortex-Antivortex Lattice and Majorana Fermions in Rhombohedral Graphene}}},
  author = {Gaggioli, Filippo and Guerci, Daniele and Fu, Liang},
  journal = {Phys. Rev. Lett.},
  volume = {135},
  issue = {11},
  pages = {116001},
  numpages = {7},
  year = {2025},
  month = {Sep},
  publisher = {American Physical Society},
  doi = {10.1103/k8sb-rqxf},
  url = {https://link.aps.org/doi/10.1103/k8sb-rqxf}
}

@misc{han2025exactmodelschiralflatband,
      title={Exact Models of Chiral Flat-Band Superconductors}, 
      author={Zhaoyu Han and Jonah Herzog-Arbeitman and Qiang Gao and Eslam Khalaf},
      year={2025},
      eprint={2508.21127},
      archivePrefix={arXiv},
      primaryClass={cond-mat.str-el},
      url={https://arxiv.org/abs/2508.21127}, 
}

@article{li2025berry,
  title={{{Berry Trashcan With Short Range Attraction: Exact {$p_x+ip_y$} Superconductivity in Rhombohedral Graphene}}},
  author={Li, Ming-Rui and Kwan, Yves H and Yao, Hong and Bernevig, B Andrei},
  journal={arXiv preprint arXiv:2509.16312},
  year={2025}
}

@article{nuckolls2023quantum,
  title={Quantum Textures of the Many-body Wavefunctions in Magic-Angle Graphene},
  author={Nuckolls, Kevin P and Lee, Ryan L and Oh, Myungchul and Wong, Dillon and Soejima, Tomohiro and Hong, Jung Pyo and C{\u{a}}lug{\u{a}}ru, Dumitru and Herzog-Arbeitman, Jonah and Bernevig, B Andrei and Watanabe, Kenji and others},
  journal={Nature},
  volume={620},
  number={7974},
  pages={525--532},
  year={2023},
  publisher={Nature Publishing Group UK London}
}

@ARTICLE{2024PhRvB.109t5122H,
       author = {{Herzog-Arbeitman}, Jonah and {Wang}, Yuzhi and {Liu}, Jiaxuan and {Tam}, Pok Man and {Qi}, Ziyue and {Jia}, Yujin and {Efetov}, Dmitri K. and {Vafek}, Oskar and {Regnault}, Nicolas and {Weng}, Hongming and {Wu}, Quansheng and {Bernevig}, B. Andrei and {Yu}, Jiabin},
        title = "{Moir{\'e} fractional Chern insulators. II. First-principles calculations and continuum models of rhombohedral graphene superlattices}",
      journal = {\prb},
         year = 2024,
        month = may,
       volume = {109},
       number = {20},
          eid = {205122},
        pages = {205122},
          doi = {10.1103/PhysRevB.109.205122},
archivePrefix = {arXiv},
       eprint = {2311.12920},
 primaryClass = {cond-mat.mes-hall},
       adsurl = {https://ui.adsabs.harvard.edu/abs/2024PhRvB.109t5122H}
}

@article{kim2023imaging,
  title={Imaging Inter-Valley Coherent Order in Magic-Angle Twisted Trilayer Graphene},
  author={Kim, Hyunjin and Choi, Youngjoon and Lantagne-Hurtubise, {\'E}tienne and Lewandowski, Cyprian and Thomson, Alex and Kong, Lingyuan and Zhou, Haoxin and Baum, Eli and Zhang, Yiran and Holleis, Ludwig and others},
  journal={Nature},
  volume={623},
  number={7989},
  pages={942--948},
  year={2023},
  publisher={Nature Publishing Group UK London}
}

@article{kwan2021kekule,
  title={{{Kekul{\'e} Spiral Order at All Nonzero Integer Fillings in Twisted Bilayer Graphene}}},
  author={Kwan, Yves H and Wagner, Glenn and Soejima, Tomohiro and Zaletel, Michael P and Simon, Steven H and Parameswaran, Siddharth A and Bultinck, Nick},
  journal={Physical Review X},
  volume={11},
  number={4},
  pages={041063},
  year={2021},
  publisher={APS}
}

@article{arp2024intervalley,
  title={Intervalley Coherence and Intrinsic Spin--Orbit Coupling in Rhombohedral Trilayer Graphene},
  author={Arp, Trevor and Sheekey, Owen and Zhou, Haoxin and Tschirhart, CL and Patterson, Caitlin L and Yoo, HM and Holleis, Ludwig and Redekop, Evgeny and Babikyan, Grigory and Xie, Tian and others},
  journal={Nature Physics},
  volume={20},
  number={9},
  pages={1413--1420},
  year={2024},
  publisher={Nature Publishing Group UK London}
}

@misc{liu2024visualizingincommensurateintervalleycoherent,
      title={Visualizing Incommensurate Inter-valley Coherent States in Rhombohedral Trilayer Graphene}, 
      author={Yiwen Liu and Ambikesh Gupta and Youngjoon Choi and Yaar Vituri and Hari Stoyanov and Jiewen Xiao and Yanzhen Wang and Haibiao Zhou and Barun Barick and Takashi Taniguchi and Kenji Watanabe and Binghai Yan and Erez Berg and Andrea F. Young and Haim Beidenkopf and Nurit Avraham},
      year={2024},
      eprint={2411.11163},
      archivePrefix={arXiv},
      primaryClass={cond-mat.mes-hall},
      url={https://arxiv.org/abs/2411.11163}, 
}

@misc{han2026evidencemetallicwignercrystal,
      title={{{Evidence of Metallic Wigner Crystal in Rhombohedral Graphene}}}, 
      author={Tonghang Han and Jackson P. Butler and Shenyong Ye and Zhenqi Hua and Surajit Dutta and Zach Hadjri and Zhenghan Wu and Jixiang Yang and Junseok Seo and Phatthanon Pattanakanvijit and Emily Aitken and Kenji Watanabe and Takashi Taniguchi and Peng Xiong and Eli Zeldov and Zhengguang Lu and Raymond Ashoori and Long Ju},
      year={2026},
      eprint={2604.00113},
      archivePrefix={arXiv},
      primaryClass={cond-mat.mes-hall},
      url={https://arxiv.org/abs/2604.00113}, 
}

@misc{dong2026crystalscaughtdopingmetallic,
      title={{{Crystals Caught Doping: Metallic Wigner Crystals in Rhombohedral Graphene}}}, 
      author={Junkai Dong and Tomohiro Soejima and Daniel E. Parker and Ashvin Vishwanath},
      year={2026},
      eprint={2604.00114},
      archivePrefix={arXiv},
      primaryClass={cond-mat.str-el},
      url={https://arxiv.org/abs/2604.00114}, 
}

@misc{feng2026selfdopedcrystalpreemptedbandinversion,
      title={{{Self-Doped Crystal from Preempted Band-inversion Transitions}}}, 
      author={Jiechao Feng and Zhaoyu Han and Michael P. Zaletel and Zhihuan Dong},
      year={2026},
      eprint={2604.09820},
      archivePrefix={arXiv},
      primaryClass={cond-mat.str-el},
      url={https://arxiv.org/abs/2604.09820}, 
}

@article{ghazaryan2023platform,
  title = {Multilayer Graphenes as a Platform for Interaction-Driven Physics and Topological Superconductivity},
  author = {Ghazaryan, Areg and Holder, Tobias and Berg, Erez and Serbyn, Maksym},
  journal = {Phys. Rev. B},
  volume = {107},
  issue = {10},
  pages = {104502},
  numpages = {16},
  year = {2023},
  month = {Mar},
  publisher = {American Physical Society},
  doi = {10.1103/PhysRevB.107.104502},
  url = {https://link.aps.org/doi/10.1103/PhysRevB.107.104502}
}

@article{shtyk2017monkey,
  title = {Electrons at the Monkey Saddle: A Multicritical {{Lifshitz}} Point},
  author = {Shtyk, A. and Goldstein, G. and Chamon, C.},
  journal = {Phys. Rev. B},
  volume = {95},
  issue = {3},
  pages = {035137},
  numpages = {10},
  year = {2017},
  month = {Jan},
  publisher = {American Physical Society},
  doi = {10.1103/PhysRevB.95.035137},
  url = {https://link.aps.org/doi/10.1103/PhysRevB.95.035137}
}

@misc{qin2026extremeanisotropymetallicsuperconducting,
      title={{{Extreme Anisotropy in the Metallic and Superconducting Phases of Rhombohedral Hexalayer Graphene}}}, 
      author={Peiyu Qin and Hai-Tian Wu and Ron Q. Nguyen and Erin Morissette and Naiyuan J. Zhang and K. Watanabe and T. Taniguchi and J. I. A. Li},
      year={2026},
      eprint={2504.05129},
      archivePrefix={arXiv},
      primaryClass={cond-mat.mes-hall},
      url={https://arxiv.org/abs/2504.05129}, 
}

@misc{zhu2026microscopicoriginorbitalmagnetization,
      title={Microscopic Origin of Orbital Magnetization in Chiral Superconductors}, 
      author={Jihang Zhu and Chunli Huang},
      year={2026},
      eprint={2601.12387},
      archivePrefix={arXiv},
      primaryClass={cond-mat.supr-con},
      url={https://arxiv.org/abs/2601.12387}, 
}

@misc{bernevig2025berrytrashcanmodelinteracting,
      title={{{"Berry Trashcan" Model of Interacting Electrons in Rhombohedral Graphene}}}, 
      author={B. Andrei Bernevig and Yves H. Kwan},
      year={2025},
      eprint={2503.09692},
      archivePrefix={arXiv},
      primaryClass={cond-mat.str-el},
      url={https://arxiv.org/abs/2503.09692}, 
}

@article{lifshitsTheoryShubnikovHaas1958,
  title = {Theory of the {{Shubnikov}}---de {{Haas}} Effect},
  author = {Lifshits, E. M. and Kosevich, A. M.},
  year = 1958,
  month = jan,
  journal = {Journal of Physics and Chemistry of Solids},
  volume = {4},
  number = {1},
  pages = {1--10},
  issn = {0022-3697},
  doi = {10.1016/0022-3697(58)90189-6},
  urldate = {2026-05-30},
}

@article{leeb_field_2025,
    title = {A {Field} {Guide} to {Non}-{Onsager} {Quantum} {Oscillations} in {Metals}},
    volume = {4},
    copyright = {© 2025 The Author(s). Advanced Physics Research published by Wiley-VCH GmbH},
    issn = {2751-1200},
    url = {https://onlinelibrary.wiley.com/doi/abs/10.1002/apxr.202400134},
    doi = {10.1002/apxr.202400134},
    number = {4},
    urldate = {2026-05-30},
    journal = {Advanced Physics Research},
    author = {Leeb, Valentin and Huber, Nico and Pfleiderer, Christian and Knolle, Johannes and Wilde, Marc A.},
    year = {2025},
    pages = {2400134},
}

@article{wang2024account,
  title = {{{Phase Diagram of Twisted Bilayer ${\mathrm{MoTe}}_{2}$ in a Magnetic Field with an Account for the Electron-Electron interaction}}},
  author = {Wang, Minxuan and Wang, Xiaoyu and Vafek, Oskar},
  journal = {Phys. Rev. B},
  volume = {110},
  issue = {20},
  pages = {L201107},
  numpages = {6},
  year = {2024},
  month = {Nov},
  publisher = {American Physical Society},
  doi = {10.1103/PhysRevB.110.L201107},
  url = {https://link.aps.org/doi/10.1103/PhysRevB.110.L201107}
}

@misc{nguyen2025hierarchysuperconductivitytopologicalcharge,
      title={{{A Hierarchy of Superconductivity and Topological Charge Density Wave States in Rhombohedral Graphene}}}, 
      author={Ron Q. Nguyen and Hai-Tian Wu and Erin Morissette and Naiyuan J. Zhang and Peiyu Qin and Kenji Watanabe and Takashi Taniguchi and Aaron W. Hui and Dima E. Feldman and J. I. A. Li},
      year={2025},
      eprint={2507.22026},
      archivePrefix={arXiv},
      primaryClass={cond-mat.mes-hall},
      url={https://arxiv.org/abs/2507.22026}, 
}

@misc{dutta2026reconfigurablechiralsuperconductivity,
      title={Reconfigurable Chiral Superconductivity}, 
      author={Surajit Dutta and Nadav Auerbach and Tonghang Han and Yaozhang Zhou and Gal Shavit and Niladri-Sekhar Kander and Yuri Myasoedov and Martin E. Huber and Kenji Watanabe and Takashi Taniguchi and Long Ju and Eli Zeldov},
      year={2026},
      eprint={2605.13303},
      archivePrefix={arXiv},
      primaryClass={cond-mat.mes-hall},
      url={https://arxiv.org/abs/2605.13303}, 
}

@misc{sheekey2026visualizingorbitalmagnetismelectron,
      title={Visualizing Orbital Magnetism in Electron Doped Rhombohedral Multilayer Graphene}, 
      author={Owen I. Sheekey and Trevor B. Arp and Benjamin A. Foutty and Ruoxi Zhang and Tixuan Tan and Ludwig F. W. Holleis and Yi Guo and Sandesh S. Kalantre and Canxun Zhang and Mark Zakharyan and David Gong and Aidan Keough and Youngjoon Choi and Ysun Choi and Siyuan Xu and Tian Xie and Ben Hodder Alexander and Marisa Hocking and Qingrui Cao and Martin E. Huber and Takashi Taniguchi and Kenji Watanabe and Chenhao Jin and Etienne Lantagne-Hurtubise and Aaron Sharpe and Trithep Devakul and Andrea F. Young},
      year={2026},
      eprint={2605.30316},
      archivePrefix={arXiv},
      primaryClass={cond-mat.mes-hall},
      url={https://arxiv.org/abs/2605.30316}, 
}

@article{zhouSuperconductivityRhombohedralTrilayer2021a,
  title={Superconductivity in Rhombohedral Trilayer Graphene},
  author={Zhou, Haoxin and Xie, Tian and Taniguchi, Takashi and Watanabe, Kenji and Young, Andrea F},
  journal={Nature},
  volume={598},
  number={7881},
  pages={434--438},
  year={2021},
  publisher={Nature Publishing Group UK London}
}

@article{PhysRevB.108.235128,
  title = {Ground-state order in magic-angle graphene at filling $\nu=-3$: A full-scale density matrix renormalization group study},
  author = {Wang, Tianle and Parker, Daniel E. and Soejima, Tomohiro and Hauschild, Johannes and Anand, Sajant and Bultinck, Nick and Zaletel, Michael P.},
  journal = {Phys. Rev. B},
  volume = {108},
  issue = {23},
  pages = {235128},
  numpages = {8},
  year = {2023},
  month = {Dec},
  publisher = {American Physical Society},
  doi = {10.1103/PhysRevB.108.235128},
  url = {https://link.aps.org/doi/10.1103/PhysRevB.108.235128}
}

@article{chirolliDiodeEffectFraunhofer2025,
    title = {Diode effect in the Fraunhofer pattern of disordered planar Josephson junctions},
    volume = {8},
    copyright = {2025 The Author(s)},
    issn = {2399-3650},
    url = {https://www.nature.com/articles/s42005-025-02364-y},
    doi = {10.1038/s42005-025-02364-y},
    number = {1},
    urldate = {2026-06-01},
    journal = {Communications Physics},
    publisher = {Nature Publishing Group},
    author = {Chirolli, Luca and Greco, Angelo and Crippa, Alessandro and Strambini, Elia and Cuoco, Mario and Amico, Luigi and Giazotto, Francesco},
    month = nov,
    year = {2025},
    pages = {483},
}

@article{rashidiSelfFieldInducedJosephsonDiode2025,
    title = {Self-Field-Induced Josephson Diode Effect},
    volume = {25},
    issn = {1530-6984},
    url = {https://doi.org/10.1021/acs.nanolett.5c02198},
    doi = {10.1021/acs.nanolett.5c02198},
    number = {26},
    urldate = {2026-06-01},
    journal = {Nano Letters},
    publisher = {American Chemical Society},
    author = {Rashidi, Arman and Ahadi, Sina and Stemmer, Susanne},
    month = jul,
    year = {2025},
    pages = {10544--10548},
}

@article{thonhauser2005orbital,
  title={Orbital magnetization in periodic insulators},
  author={Thonhauser, Timo and Ceresoli, Davide and Vanderbilt, David and Resta, Raffaele},
  journal={Physical review letters},
  volume={95},
  number={13},
  pages={137205},
  year={2005},
  publisher={APS}
}
